\pdfoutput=1

\documentclass[11pt,twoside,a4paper,cmspaper,final,collab]{cms-tdr}
\def\svnVersion{455181P}\def\cmsCernNoTag{CERN-EP-2017-145}\def\cmsCernDate{\today}\def\cmsMessage{Published in Physical Review D as \href{http://dx.doi.org/10.1103/PhysRevD.97.072008}{\doi{10.1103/PhysRevD.97.072008.}}}

\begin{document}\cmsNoteHeader{B2G-12-016}

\hyphenation{had-ron-i-za-tion}
\hyphenation{cal-or-i-me-ter}
\hyphenation{de-vices}
\RCS$Revision: 440271 $
\RCS$HeadURL: svn+ssh://svn.cern.ch/reps/tdr2/papers/B2G-12-016/trunk/B2G-12-016.tex $
\RCS$Id: B2G-12-016.tex 440271 2017-12-27 15:55:17Z alverson $
\newlength\cmsTableSkip\setlength\cmsTableSkip{1.8ex}
\newlength\cmsFigWidth
\ifthenelse{\boolean{cms@external}}{\setlength\cmsFigWidth{0.95\columnwidth}}{\setlength\cmsFigWidth{0.6\textwidth}}
\ifthenelse{\boolean{cms@external}}{\providecommand{\cmsLeft}{upper}\xspace}{\providecommand{\cmsLeft}{left}\xspace}
\ifthenelse{\boolean{cms@external}}{\providecommand{\cmsRight}{lower}\xspace}{\providecommand{\cmsRight}{right}\xspace}
\ifthenelse{\boolean{cms@external}}{\providecommand{\suppMaterial}{the supplemental material [URL will be inserted by publisher]}}{\providecommand{\suppMaterial}{Appendix~\ref{app:suppMat}}}
\ifthenelse{\boolean{cms@external}}{\providecommand{\cmsTableResize}[1]{#1}}{\providecommand{\cmsTableResize}[1]{\resizebox{\textwidth}{!}{#1}}}
\ifthenelse{\boolean{cms@external}}{\providecommand{\CL}{C.L.\xspace}}{\providecommand{\CL}{CL\xspace}}
\providecommand{\Q}{\ensuremath{\cmsSymbolFace{Q}}\xspace}
\providecommand{\cPV}{\ensuremath{\cmsSymbolFace{V}}\xspace}
\providecommand{\QD}{\ensuremath{\cmsSymbolFace{D}}\xspace}
\providecommand{\QQbar}{\ensuremath{\Q\overline{\Q}}\xspace}
\providecommand{\q}{\PQq}
\providecommand{\qbar}{\PAQq}
\providecommand{\qW}{\ensuremath{\PW\PQq}\xspace}
\providecommand{\qZ}{\ensuremath{\PZ\PQq}\xspace}
\providecommand{\qH}{\ensuremath{\PH\PQq}\xspace}
\providecommand{\Mfit}{\ensuremath{m_\text{fit}}\xspace}

\cmsNoteHeader{B2G-12-016}
\title{Search for vector-like light-flavor quark partners in proton-proton collisions at \texorpdfstring{$\sqrt{s}= 8\TeV$}{sqrt(s)=8 TeV}}

\date{\today}

\abstract{
A search is presented for heavy vector-like quarks (VLQs) that couple only to light quarks
in proton-proton collisions at $\sqrt{s} = 8\TeV$ at the LHC.
The data were collected by the CMS experiment during 2012 and correspond
to an integrated luminosity of 19.7\fbinv.
Both single and pair production of VLQs are considered.
The single-production search is performed for down-type VLQs (electric charge of magnitude $1/3$),
while the pair-production search is
sensitive to up-type (charge of magnitude $2/3$) and down-type VLQs.
Final states with at least one muon or one electron are considered.
No significant excess over standard model expectations is observed,
and lower limits on the mass of VLQs are derived.
The lower limits range from
400 to 1800\GeV, depending on the single-production cross section and
the VLQ branching fractions $\mathcal{B}$ to \PW, \PZ, and Higgs bosons.
When considering pair production alone,
VLQs with masses below 845\GeV are excluded for $\mathcal{B}(\PW) = 1.0$, and below 685\GeV
for $\mathcal{B}(\PW) = 0.5$, $\mathcal{B}(\PZ) = \mathcal{B}(\PH) = 0.25$.
The results are more stringent than those previously obtained
for single and pair production of VLQs coupled to light quarks.
}

\hypersetup{%
pdfauthor={CMS Collaboration},%
pdftitle={Search for vector-like light-flavor quark partners in proton-proton collisions at sqrt(s)=8 TeV},%
pdfsubject={CMS},%
pdfkeywords={CMS, physics, vector-like quarks}}

\maketitle

\section{Introduction}
\label{sec:introduction}
Vector-like quarks (VLQs) are hypothetical spin-$1/2$ fermions, whose left-
and right-handed chiral components transform in the same way under the standard
model (SM) symmetries, and hence have vector couplings to gauge bosons.
Such VLQs appear in a number of models that extend the SM to address open
questions in particle physics. These models include:
beautiful mirrors~\cite{beautiful},
little-Higgs models~\cite{littlesthiggs,Schmaltz200340,littlehiggsreview},
composite-Higgs models~\cite{compositehiggs}, theories invoking
extra dimensions~\cite{extradimensionsVLQ}, grand unified theories~\cite{GUTVLQ},
and models providing insights into the SM flavor structure~\cite{gaugeflavor}.

Owing to the possible role of third-generation quarks in the solution of
problems in electroweak symmetry breaking,
the VLQs in many of the aforementioned models mix predominantly with third generation quarks.
In addition, indirect experimental constraints on the quark couplings of the lighter generations
from precision electroweak measurements are typically stronger
than those on third-generation couplings~\cite{VLQ}.
However, the coupling corrections from several different VLQs may cancel,
which can significantly relax constraints on the mixing of VLQs with the
first and second generations.
In this paper, we consider the pair production of heavy VLQs, denoted by \Q,
with electric charge of magnitude $1/3$ or $2/3$, that are partners of the
first-generation SM quarks. We also consider the electroweak single production
of vector-like down-type quarks with electric charge of magnitude $1/3$,
which we denote by \QD in this context.

\begin{figure}[htbp]
\centering
\includegraphics[width=0.4\textwidth]{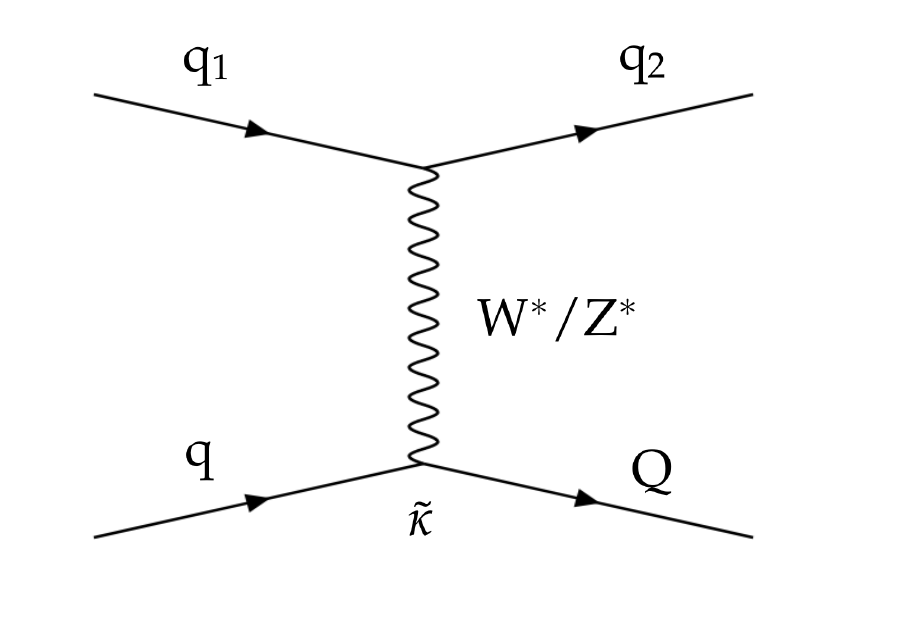}
\includegraphics[width=0.4\textwidth]{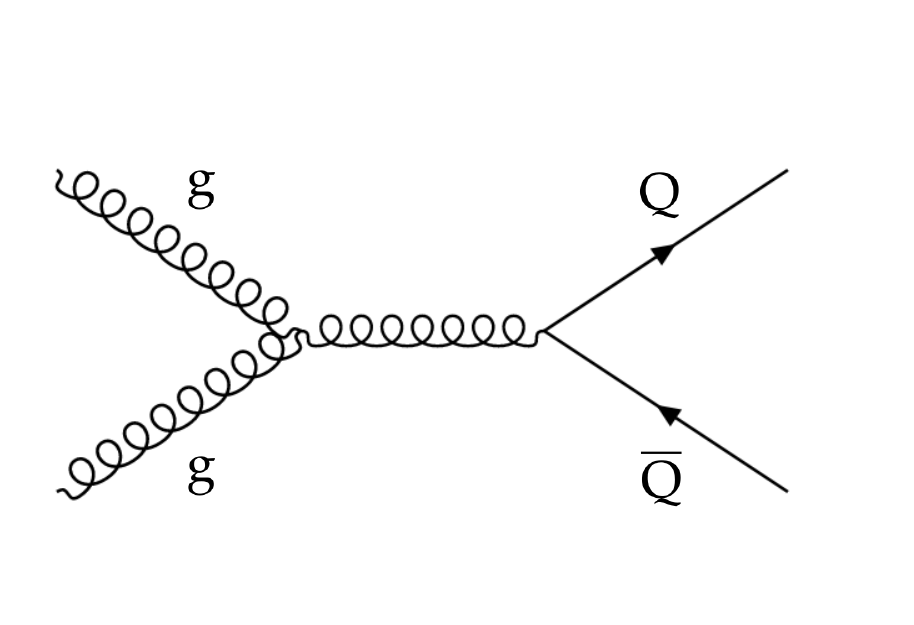}
        \caption{Vector-like quarks (denoted \Q) can be produced in proton-proton
  collisions either singly through electroweak interactions (the
  $t$ channel mode (\cmsLeft) is shown as an example), or in pairs via the strong
  interaction (\cmsRight). For single production we consider in the present work only
  vector-like quarks with electric charge of magnitude $1/3$ (denoted D).
}
      \label{fig:vlqdiagram}
\end{figure}

Figure~\ref{fig:vlqdiagram} shows examples of Feynman diagrams for the leading-order electroweak
single production and strong pair production of VLQs coupled to first-generation quarks.
In order to describe the production processes, new couplings of the VLQs to light-flavor quarks
via \PW, \PZ, and Higgs bosons (\PH) are introduced, whereas no new coupling to gluons is considered.
Assuming a short enough lifetime, the new quarks do not hadronize before decaying to
$\PW\PQq$, $\PZ\PQq$, or $\PH\PQq$, where $\PQq$ indicates a SM quark.
The branching fractions for the different
decay modes depend on the multiplet in which the VLQ
resides~\cite{handbookVLQ}. In most models, the neutral-current branching
fractions $\mathcal{B}(\Q \to \PZ\PQq)$ and
$\mathcal{B}(\Q \to\PH\PQq)$ are roughly the same size,
and the charged-current branching fraction $\mathcal{B}(\Q \to \PW\PQq)$
can vary between 0 and 1.
Other decay modes are assumed to be negligible, so the sum of the three branching fractions is one.

The cross section for the charged-current (neutral-current) production
of single VLQs is proportional to
$\tilde{\kappa}_\PW^2$ ($\tilde{\kappa}_\PZ^2$), where $\tilde{\kappa}$ is
a scaled coupling parameter defined in Section~\ref{sec:single_strateg}.
The pair-production cross section does not depend on these parameters as
it proceeds via the strong interaction.
Because the Q quark isosinglet is the simplest model having
$\mathcal{B}(\Q \to \PW\PQq) = 0.5$ and $\mathcal{B}(\Q \to \PZ\PQq) = \mathcal{B}(\Q \to
\PH\PQq) = 0.25$, implied by the equivalence theorem~\cite{equivalencetheorem}, it is chosen as a benchmark point in the signal model parameter space.

Previous searches for single and pair production of such VLQs have been
performed by the ATLAS experiment at $\sqrt{s} = 7$ and 8\TeV~\cite{ATLASlightVLQ2011,ATLASlightVLQ_pair_8}.
These searches exclude singly produced VLQs with masses below 900\,(760\GeV),
with $\Q\PQq \to \PW\PQq\PQq$ ($\Q\PQq \to \PZ\PQq\PQq)$,
and pair-produced VLQs with masses below 690\GeV, with $\mathcal{B}(\Q \to \PW\PQq) = 1$, at 95\% confidence level.

For high VLQ masses above 1 TeV, the kinematically favored single-production mode may be the dominant production mode,
since the pair-production cross section via the strong interaction drops rapidly as a function of the VLQ mass.
Nevertheless, since the single-production cross section depends on unknown model-dependent parameters, the pair-production mode may be dominant for sub-TeV VLQ masses.
Furthermore, the VLQs may decay to \PW, \PZ, and Higgs bosons with unknown decay branching fractions.
These considerations motivate searches for VLQs over a wide
mass range with search methods optimized for both singly and pair
produced VLQs, decaying in a variety of modes.

In this paper we report results of a search for VLQs in proton-proton
collisions at a center-of-mass energy of 8\TeV using the CMS detector at the CERN LHC.
The data set analyzed corresponds to an integrated luminosity of about 19.7\fbinv.
Events with one or more isolated leptons are used for the search.
The signal channels considered are listed in Table~\ref{tab:vlqmodes}.
The processes $ \QD \PQq \to \PH\PQq\PQq$ and $\QQbar \to \PH\PQq\PH\PQq$
have not been considered because of the low efficiency for selecting isolated leptons in such decay modes.
\begin{table}
\centering
\topcaption{Decay channels of vector-like quarks considered in the analysis.}
\label{tab:vlqmodes}
\begin{scotch}{ll}
  Production &  Channel \\
\hline
\multirow{2}{*}{Single (electroweak)}  &  $ \QD \PQq \to \PW\PQq\PQq$ \\
                                    &  $ \QD \PQq \to \PZ\PQq\PQq$ \\[\cmsTableSkip]
\multirow{6}{*}{Pair (strong)}  & \rule{0pt}{2.3ex}$\QQbar \to \PW\PQq\PW\PQq$ \\
                                  & $\QQbar \to \PW\PQq\PZ\PQq$ \\
                                  & $\QQbar \to \PW\PQq\PH\PQq$ \\
                                  & $\QQbar \to \PZ\PQq\PZ\PQq$ \\
                                  & $\QQbar \to \PZ\PQq\PH\PQq$ \\
\end{scotch}
\end{table}
The search for singly produced VLQs is performed only for vector-like down-type quarks.
The search for pair-produced VLQs is also applicable to up-type quarks,
as their decay products are experimentally indistinguishable from those of down-type VLQs.

This is the first search for VLQs coupled to light-flavor quarks that simultaneously considers
the single and pair production modes, in a scan over the branching fractions of the VLQs to
$\PW$, $\PZ$, and Higgs bosons.
Furthermore, for the first time in these topologies, kinematic fits using boosted jet substructure techniques in single-lepton events
are applied to improve the VLQ mass reconstruction, and events with at least two leptons are analyzed
to retain sensitivity to VLQs that have a high probability of decaying to a Z boson.

\section{Analysis strategy}
\label{sec:analysis_strategy}
In this analysis, the search for singly produced vector-like \QD quarks involves the reconstruction of a VLQ resonance
in final states with exactly one or two leptons and two or three jets.
In the search for pair produced VLQs, in the final state with one lepton, missing transverse momentum,
and four jets, a kinematic fit is performed to reconstruct the VLQ mass.
Final states with two, three or four leptons and at least two jets are also considered, using reconstructed observables sensitive to the VLQ mass.
The results of all channels, which are mutually exclusive, are combined in the calculation of the limits
on the VLQ masses and the production cross sections.

The searches are performed without assuming that the hypothetical quark
belongs to a particular SU(2) multiplet structure.
Therefore the analysis is not optimized for a combined search for all
quarks in a given multiplet. As such, the exclusion limits presented in this analysis
are expected to be more conservative than those that would be obtained in a dedicated
model-dependent search combining the signal from all quarks within a multiplet.
On the other hand, the approach used here allows a more model-independent interpretation.

\subsection{Search for single production}
\label{sec:single_strateg}

We consider the electroweak charged-current and neutral-current  modes of the single production of vector-like \QD quarks.
The interaction Lagrangian density for the vector-like \QD quarks contains three unknown parameters,
corresponding to the couplings to the three bosons, $\kappa_\PW$,
$\kappa_\PZ$, and $\kappa_\PH$~\cite{VLQ,vlqbuchkremer}:
\ifthenelse{\boolean{cms@external}}{
\begin{multline}
\label{eq:interactiontermsVLQdown}
   \mathcal{L}_{\mathrm{interaction}, \QD }  = \frac{g_\PW}{\sqrt{2}} \kappa_\PW \PWp_\mu \PAQu_\mathrm{R}\gamma^\mu
    \QD _\mathrm{R}\\
   + \frac{g_\PW}{2\mathrm{cos}\theta_\PW} \kappa_\PZ \PZ_\mu \PAQd_\mathrm{R} \gamma^\mu  \QD _\mathrm{R}
   - \frac{m_\Q}{v} \kappa_\PH \PH \PAQd_\mathrm{R}  \QD _\mathrm{L} + \text{h.c.}
\end{multline}
}{
\begin{equation}
\label{eq:interactiontermsVLQdown}
   \mathcal{L}_{\mathrm{interaction}, \QD }  = \frac{g_\PW}{\sqrt{2}} \kappa_\PW \PWp_\mu \PAQu_\mathrm{R}\gamma^\mu
    \QD _\mathrm{R}
   + \frac{g_\PW}{2\mathrm{cos}\theta_\PW} \kappa_\PZ \PZ_\mu \PAQd_\mathrm{R} \gamma^\mu  \QD _\mathrm{R}
   - \frac{m_\Q}{v} \kappa_\PH \PH \PAQd_\mathrm{R}  \QD _\mathrm{L} + \text{h.c.}
\end{equation}
}
Here $v \approx 246$\GeV is the Higgs field vacuum expectation value, $m_\Q$ is the VLQ mass,
$\theta_\PW$ is the weak mixing angle and $g_\PW$ is the coupling
strength of the weak interaction. In Eq.~(\ref{eq:interactiontermsVLQdown})
the terms for just one chirality are given (the $\mathrm{R}$ and $\mathrm{L}$ field indices
refer to right- and left-handed helicities, respectively), but there are
equivalent terms for the other helicities.

The coupling parameters, $\kappa$, are model dependent, and originate from
the mixing between SM quarks and VLQs.
These couplings can be re-parametrized as
$\kappa = v \tilde{\kappa} / \sqrt{2}m_\Q$, with the new parameter
$\tilde{\kappa}$ being naturally of order unity in a weakly coupled theory~\cite{VLQ}.

In the particular scenario where the VLQ couples only to
the first-generation quarks, it can be shown~\cite{vlqbuchkremer} that the
neutral-current coupling strength parameter, $\tilde{\kappa}_\PZ$, may
be expressed approximately through the charged-current coupling strength parameter,
$\tilde{\kappa}_\PW$, and the branching fractions of
the decays of the VLQ to $\PW$ and $\PZ$ bosons,
$\mathcal{B}_\PW = \mathcal{B}(\Q \to \PW\PQq)$ and
$\mathcal{B}_\PZ = \mathcal{B}(\Q \to \PZ\PQq)$, via:
\begin{equation}
\label{eq:kappaWZrelation}
  \tilde{\kappa}_\PZ \approx \sqrt{2\frac{\mathcal{B}_\PZ}{\mathcal{B}_\PW}}\tilde{\kappa}_\PW,
\end{equation}
if $\mathcal{B}_\PW \neq 0$.
It is therefore sufficient to determine limits on the cross section and mass
as a function of the three free parameters,
$\tilde{\kappa}_\PW$, $\mathcal{B}_\PW$ and $\mathcal{B}_\PZ$,
producing cross section and mass limits that then depend only on these parameters.
If $\mathcal{B}_\PW$ approaches 0, with $\tilde{\kappa}_\PW$ fixed to a non-zero value,
Eq.~(\ref{eq:kappaWZrelation}) implies that $\tilde{\kappa}_\PZ$ diverges, and when
$\mathcal{B}_\PW$ is exactly zero, Eq.~(\ref{eq:kappaWZrelation})
is no longer applicable. Results for an alternative single-production coupling parametrization that does not exhibit
divergent behavior throughout the parameter scan are available in \suppMaterial.

The expected signal topologies are listed in the upper two rows of Table~\ref{tab:vlqmodes}.
It should be noted that singly produced VLQs are produced in
association with a forward-going first-generation quark.
As will be explained in Section~\ref{sec:single}, we define two event categories corresponding to these two topologies,
based on whether one or two isolated leptons are present in the final state.
In these event categories we employ the reconstructed mass of the D quark
decaying into a $\PW$ or $\PZ$ boson and a quark to search for a signal.

\subsection{Search for pair production}
\label{sec:pair_strateg}

In the search for strongly produced VLQ pairs, $\QQbar$,
several event categories are defined that are
optimized for the decay modes of pair produced VLQs listed in Table~\ref{tab:vlqmodes}.
Signal events do not often contain \PQb jets, except in the cases where a Higgs boson is produced.

The single-lepton event categories are optimized for the following decay modes of VLQ pairs:
\begin{align}
\label{eq:eqn01}
\QQbar &\to \qW \qW \to \ell\nu ~{\q}_\ell ~ \q \qbar' ~{\q}_{\mathrm{h}};  \\
\label{eq:eqn02}
\QQbar &\to \qW \qZ ~\to \ell\nu ~{\q}_\ell ~ \q \qbar ~{\q}_{\mathrm{h}};  \\
\label{eq:eqn03}
\QQbar &\to \qW \qH \to \ell\nu ~{\q}_\ell ~ \bbbar ~{\q}_{\mathrm{h}}.
\end{align}

In these events the $\PW$ boson decays leptonically into a muon or an electron plus a neutrino
and the other boson ($\PW$, $\PZ$, or $\PH$) decays into a pair of quarks.
These events are classified as either $\Pgm$+jets or $\Pe$+jets events.
A light quark, ${\q}_\ell$, is produced in association with the
leptonically decaying $\PW$ boson, and ${\q}_{\mathrm{h}}$ is the equivalent for the
hadronically decaying boson.
We perform a constrained kinematic fit for each event to reconstruct the mass of the VLQ.
The full kinematic distributions of the final state are reconstructed, and the mass
of the Q quark, $m_{\text{fit}}$, is obtained, as detailed in
Section~\ref{sec:pairsinglelep}.
In addition, the $S_\mathrm{T}$ variable is defined as the scalar sum of the transverse momenta
$\pt^{\ell}$ of the charged lepton, the transverse momenta $\pt^{\text{jet}}$
of the jets, and the $\ptmiss$ value:
\begin{equation} \label{eq:STvlq}
  S_\mathrm{T} = \sum \pt^{\ell} + \sum \pt^{\text{jet}} + \ptmiss .
\end{equation}
The variable $\ptmiss$, referred to as the missing transverse momentum,
is defined as the magnitude of the missing transverse momentum vector,
which is the projection on the plane perpendicular to the beams
of the negative vector sum of the momenta of all reconstructed particles
in the event.
The $S_\mathrm{T}$ variable, calculated after the fit, is used to
define a phase space region where the signal-to-background ratio is enhanced.

In the dilepton event categories we employ two variables to search for a VLQ signal, as will be discussed in Section~\ref{sec:pairmultilep}.
The first variable is the reconstructed mass of the Q quark decaying into
a $\PZ$ boson and a quark, the second one is the $S_\mathrm{T}$
variable defined in Eq.~(\ref{eq:STvlq}).

In the multilepton event category, in this analysis defined as containing three or four leptons, the number of expected events is small.
Here, rather than using a kinematic variable to identify a possible signal, events are counted after imposing the selection criteria.

\section{The CMS detector}
\label{sec:detector}
The central feature of the CMS apparatus is a superconducting solenoid of
6\unit{m} internal diameter, providing a magnetic field of 3.8\unit{T}.
Within the solenoid volume are a silicon pixel and strip
tracker, a lead tungstate crystal electromagnetic calorimeter (ECAL), and
a brass and scintillator hadron calorimeter (HCAL), each composed of a barrel
and two endcap sections. Forward calorimeters extend the
pseudorapidity coverage provided by the barrel
and endcap detectors. Muons are measured in gas-ionization detectors embedded
in the steel flux-return yoke outside the solenoid.

The CMS detector is nearly hermetic, allowing momentum balance measurements to be made
in the plane transverse to the beam direction.
A more detailed description of the CMS detector, together with a definition
of the coordinate system used and the relevant kinematic variables, can be
found in Ref.~\cite{Chatrchyan:2008zzk}.

\section{Event samples}
\label{sec:data}
The data used for this analysis were recorded during the 2012 data taking
period, at a proton-proton center-of-mass energy of 8\TeV. The total
integrated luminosity of the data sample is 19.7\fbinv (19.6\fbinv
in the categories optimized for single VLQ production and those requiring at least two leptons optimized for pair production).
The trigger used to select the muon data sample
is based on the presence of at least one muon with a pseudorapidity
satisfying $\abs{\eta}<2.1$ and transverse momentum $\pt > 40$\GeV (in the single-lepton pair-production category),
or at least one isolated muon with $\pt > 24$\GeV
(in all other categories). For the electron data sample, events must
pass a trigger requiring the presence of one isolated electron with
$\pt > 27$\GeV.

Simulated samples are used to estimate signal efficiencies and
background contributions. The processes $\Pp\Pp\to  \QD \PQq$ and $\Pp\Pp\to \QQbar$
are simulated using the {\MADGRAPH~5.1.5.3} event generator~\cite{Madgraph5}
with CTEQ6L1 parton distribution functions (PDFs)~\cite{cteq6l},
with a decay width of 1\% of the VLQ mass and without extra partons,
and then passed to \PYTHIA~6.424~\cite{pythia} with the Z2* tune~\cite{UEtunes1,UEtunes2} for hadronization.
The following SM background processes are simulated: $\ttbar$ production
(including $\ttbar$ production in association with a vector boson and one
or more jets, denoted $\ttbar \cPZ$+jets and $\ttbar \PW$+jets); single
top quark production via the $\PQt\PW$, $s$-channel, and $t$-channel
processes; single-boson and diboson production ($\PW$+jets, $\cPZ$+jets, $\PW\PW$,
$\PW\cPZ$, and $\cPZ\cPZ$), triboson processes ($\PW\PW\PW$, $\PW\PW\cPZ$,
$\PW\cPZ\cPZ$, $\cPZ\cPZ\cPZ$), and multijet events.

Samples of the SM background processes, $\ttbar$+jets, and single top quark
production via $\PQt\PW$, $s\text{-}$, and $t$-channels, are simulated
using the {\POWHEG~1.0}~\cite{powheg1,powheg2,powheg3} event generator.
The diboson processes ($\PW\PW$, $\PW\PZ$, and $\PZ\PZ$)
and multijet events are generated using the {\PYTHIA} event
generator. The $\ttbar \cPZ$+jets, $\ttbar \PW$+jets, $\PW$+jets,
$\PZ$+jets and triboson samples are simulated using the {\MADGRAPH} event
generator. The {\PYTHIA} generator is used for parton shower development and
hadronization, for all simulated background processes.
The CTEQ6M PDFs are used for {\POWHEG}, while for the other generators the CTEQ6L1 PDFs are used.

The VLQ single-production cross sections are calculated
at leading order (LO) with the {\MADGRAPH} generator, and the pair-production cross sections,
at next-to-next-to-LO (NNLO)~\cite{Czakon_Mitov}.
The production cross sections for the background processes
are taken from the corresponding cross section measurements made by the CMS
experiment~\cite{cms_ttbar, cms_single_tW, cms_WW, cms_WZ_ZZ}: $\ttbar$+jets,
single top quark production in the $\PQt\PW$ mode, $\PW\PW$, $\PW\cPZ$, and $\cPZ\cPZ$;
and are in agreement with theoretical calculations at next-to-LO (NLO) or NNLO accuracy.
The cross section for multijet processes is calculated at leading order by
\PYTHIA. The cross sections of the remaining processes mentioned above are calculated
either at NLO or at NNLO.

All simulated events are processed through the CMS detector simulation
based on \GEANTfour~\cite{geant4}.
To simulate the effect of additional proton-proton collisions within the same or adjacent bunch crossings (pileup),
additional inelastic events are generated using \PYTHIA and superimposed on the hard-scattering events.
The Monte Carlo (MC) simulated events are weighted to reproduce
the distribution of the number of pileup interactions observed in data,
with an average of 21 reconstructed collisions per beam crossing.

\section{Event reconstruction}
\label{sec:event_reco}
The event reconstruction uses the particle flow (PF) algorithm~\cite{CMS-PRF-14-001}
which reconstructs and identifies each individual particle
with an optimized combination of all subdetector information.
In this process,
the identification of the particle type (photon, muon, electron, charged
hadron, neutral hadron) plays an important role in the determination of
the particle direction and energy.
Muons are identified by tracks or hits in the muon system that are associated with the extrapolated
trajectories of charged particles reconstructed in the inner tracker and have small energy deposits in the traversed calorimeter cells.
Electrons are identified as charged-particle tracks that are associated with potentially several ECAL clusters that result from the showering
of the primary particles and from secondary bremsstrahlung photons produced in the tracker material~\cite{Khachatryan:2015hwa}.
Charged hadrons are identified as charged-particle tracks associated with energy deposits in the HCAL, and identified
as neither electrons nor muons.
Finally, neutral hadrons are identified as HCAL
energy clusters not linked to any charged hadron trajectory, or as ECAL and
HCAL energy excesses with respect to the expected charged-hadron energy deposit.

The energy of each muon is obtained from the corresponding track
momentum.
The energy of each electron is determined from a
combination of the track momentum at the interaction vertex, the corresponding
ECAL cluster energy, and the energy sum of all bremsstrahlung photons attached
to the track.
The energy of each charged hadron is determined from a combination
of the track momentum and the corresponding ECAL and HCAL energies, corrected
for the response function of the calorimeters
to hadronic showers. Finally, the energy of each neutral hadron is obtained from
the corresponding corrected ECAL and HCAL energies.

Particles found using the PF algorithm are clustered into jets using the
direction of each particle at the interaction vertex. Charged hadrons
that are associated with pileup vertices are excluded, using a method
referred to as charged-hadron subtraction. Particles that are identified
as charged leptons, isolated according to criteria discussed later,
are removed from the jet clustering procedure.
In the analysis, two types of jets are used: jets reconstructed with the
infrared- and collinear-safe anti-\kt algorithm~\cite{antikt} operated
with a distance parameter $R = 0.5$ (AK5 jets) and jets reconstructed with the
Cambridge--Aachen algorithm~\cite{ca8_1} using a distance parameter
$R = 0.8$ (CA8 jets), as implemented in {\FASTJET} version 3.0.1~\cite{fastjet, hep-ph/0512210}.
An event-by-event jet-area-based correction~\cite{Chatrchyan:2011ds, fastjet1, fastjet2}
is applied to remove, on a statistical basis, pileup contributions that have
not already been removed by the charged-hadron subtraction procedure.

The momentum of each jet is determined from the vector
sum of all particle momenta in the jet, and is found from simulation to
be within 5 to 10\% of the true momentum for all values of \pt and over the whole
detector acceptance.
Jet energy corrections varying with \pt and $\eta$ are applied to each jet
to account for the combined response function of the calorimeters.
They are derived from simulation,
and are confirmed with in situ measurements of the energy balance of dijet
and photon+jet events~\cite{Khachatryan:2016kdb}. The jet energy resolution
amounts typically to 15--20\% at 30\GeV, 10\% at 100\GeV, and 5\% at 1\TeV.

As the mass of the heavy VLQ increases, the Lorentz boosts of the decay
products also increase.
The quark pairs from the hadronic decays of $\PW$, $\PZ$, or Higgs
bosons become increasingly collimated and eventually the resulting
hadronic showers cannot be resolved as separate jets.
The CA8 jets are used to identify these merged hadronic boson decays and
a jet pruning algorithm, which removes soft/wide-angle radiation~\cite{pruning,pruning1,cms_pruning} is then applied to resolve
the merged subjets.

Charged leptons originating from decays of heavy VLQs
are expected to be isolated from nearby jets. Therefore, a relative isolation
($I_\text{rel}$) criterion is used to suppress backgrounds from non-prompt leptons or hadrons misidentified as leptons inside jets.
Relative isolation is calculated as the sum
of the \pt of the charged hadrons, neutral hadrons, and photons
in a cone of $\Delta R=\sqrt{\smash[b]{(\Delta \phi)^2+(\Delta \eta)^2}}$ around the
lepton, with the lepton track itself removed from the sum, divided by the lepton \pt. Here $\Delta\phi$
and $\Delta\eta$ are the azimuthal angle and pseudorapidity differences with respect
to the lepton direction. In the calculation of $I_\text{rel}$ using PF reconstruction,
the isolation cone size is taken to be $\Delta R = 0.4$ for muons and $\Delta R = 0.3$
for electrons. In the calculation of $I_\text{rel}$, pileup corrections are applied.

Charged leptons are categorized by the stringency of their selection criteria in two types, namely ``tight'' and
``loose'' leptons, as defined in Table~\ref{tab:leptonbaseselvlq}.
In the analysis, events with at least one tight muon or electron are selected,
while the loose lepton criteria are used to identify and exclude the presence of additional
leptons in the event.
Additional requirements for tight and loose leptons used in
the single-lepton channel optimized for VLQ pair production are described in Section~\ref{sec:pairsinglelep}.

To identify jets as originating from a \PQb quark (b-tagged jets), the combined
secondary vertex (CSV) algorithm is used~\cite{btagCMS7TeV,CMS-PAS-BTV-13-001}.
This algorithm combines variables that distinguish \PQb jets from
non-\PQb jets, such as the track impact parameter significance and
properties of the secondary vertex. The algorithm uses a
likelihood ratio technique to compute a \PQb tagging discriminator.
We use two operating points (with different thresholds applied to the \PQb tagging
discriminator): medium and loose, which are designated as CSVM and CSVL,
respectively~\cite{CMS-PAS-BTV-13-001}. The medium (loose) CSV discriminant
operating point corresponds to a light-quark or gluon mistag rate of about
1\%\,(10\%) and a \PQb tagging efficiency of about 70\%\,(84\%).
B-tagging is applied to AK5 jets and to subjets of CA8 jets.

Data-to-simulation \PQb tagging efficiency and mistag rate scale
factors correct for the small differences between the efficiencies observed in data and in simulation.
We use scale factors that depend on both jet \pt and
$\eta$~\cite{CMS-PAS-BTV-13-001}.

\begin{table}
\centering
\topcaption{Initial selection requirements for tight and loose leptons.
}
\label{tab:leptonbaseselvlq}
\begin{scotch}{ll{c}@{\hspace*{5pt}}ll}
 \multicolumn{2}{c}{Muons} && \multicolumn{2}{c}{Electrons} \\
 Tight & Loose &&  Tight & Loose \\\cline{1-2}\cline{4-5}\\[-1.5ex]
 $\pt > 20$\GeV    & $\pt > 10$\GeV    &&  $\pt > 20$\GeV    & $\pt > 15$\GeV \\
 $\abs{\eta} < 2.1$    & $\abs{\eta} < 2.5$    &&  $\abs{\eta} < 2.5$    & $\abs{\eta} < 2.5$ \\
 $I_\text{rel} < 0.12$   & $I_\text{rel} < 0.2$    &&  $I_\text{rel} < 0.1$
   & $I_\text{rel} < 0.15$ \\
\end{scotch}
\end{table}

\section{Analysis}
\label{sec:analysis}
\subsection{Search for single production}
\label{sec:single}

We use two collections of AK5 jets with $\pt > 30$\GeV. The first collection
consists of all jets that satisfy $\abs{\eta}<2.4$; these jets are referred
to as \textit{selected central jets}. The second collection contains all jets that
satisfy $2.4<\abs{\eta}<5.0$; these jets are referred to as \textit{selected
forward jets}.
In order to exploit the presence of
first-generation quarks in the final state of VLQ processes, we
require the presence of a number of selected central jets for which the b-tag CSV discriminant
lies below the CSVL threshold. These jets are referred to as ``anti-tagged'' jets.

Events with one or two tight muons or electrons are selected.
The leptons (jets) in each event are ordered by transverse momentum.
The lepton (jet) with the largest $\pt$ is labelled as the leading lepton (jet)
and the others are labelled as subleading leptons (jets).
We define two event categories that are sensitive to the single production topologies presented in Table~\ref{tab:vlqmodes},
$\PWm\PQq\PQq$ and $\PZ\PQq\PQq$.
In order to enhance the signal sensitivity to the $ \QD \PQq \to \PW\PQq\PQq$ mode,
we require the lepton charge in the corresponding category, indicated as $\PWm\PQq\PQq$, to be negative.
For a \QD mass of 1100\GeV, this choice approximately doubles the signal-to-background ratio.
The production rate for  \QD
quarks is higher than that for $\overline{\QD}$ quarks~\cite{VLQ}
because of the proton PDFs. The production of $\PW$ bosons in the SM
is also charge asymmetric for the same reason,
with more $\PWp$ bosons produced than $\PWm$ bosons.
We therefore use only the $\PWm\PQq\PQq$ category in this search,
and do not consider the corresponding category with a positively charged lepton, $\PWp\PQq\PQq$, to search for a signal.
The definition of the event categories used to search for single production of VLQs
is summarized in Table~\ref{tab:vlqsubsamples1}.

The leptonically decaying $\PW$ and $\cPZ$ boson candidates are reconstructed and
thresholds are imposed on their transverse momenta, $\pt(\PW)$ or $\pt(\cPZ)$.
A $\PW$ boson candidate is reconstructed as follows.
The $z$ component of the neutrino momentum is obtained by imposing the $\PW$ boson mass constraint on the lepton-neutrino system,
resulting in a quadratic equation in the neutrino $p_z$.
If the solution is complex, the real part is taken as the $z$ component.
If both solutions are real we take the one where the total reconstructed
neutrino momentum has the largest difference in $\eta$ with respect to the
leading central jet in the event.
We require the separation between the lepton and the reconstructed neutrino to satisfy
$\Delta R < 1.5$, because these two particles, when produced in the decay of a boosted $\PW$
boson, are expected to be close to each other. A requirement on the transverse
mass $M_T = \sqrt{\smash[b]{2\pt^{\ell} \ptmiss \{1-\cos[\Delta\phi(\ell, \ptmiss )]\}} }> 40$\GeV is imposed to suppress the multijet background.
A $\cPZ$ boson candidate is reconstructed from two same-flavor opposite-sign dileptons, and requirements on the mass,
$m_{\ell\ell}$, of the dilepton system are imposed, as described in Table~\ref{tab:vlqsubsamples1}.

\begin{table*}[htb]
\centering
 \topcaption{The event categories as optimized for the VLQ single production.
 The categories are based on the number of tight muons
 or electrons present in the event, along with additional criteria optimized for specific VLQ
 topologies. Events containing any additional loose leptons are excluded.
\label{tab:vlqsubsamples1}}
\begin{scotch}{lll}
 Event category & Tight leptons ($\Pgm$,$\Pe$)  &  Additional selection criteria\\
\hline
            &                               & 1 or 2 selected central jets, all anti-tagged\\
            &                               &  \quad leading $\pt > 200$\GeV\\
$\PWm\PQq\PQq$           & 1 with $\pt > 30$\GeV      &  1 selected forward jet\\
            &      \quad  negative charge      &  $\pt(\PW\to \ell\nu) > 150$\GeV\\
           &             &   $\Delta R (\ell,\nu) < 1.5$    \\
           &             &  $\ptmiss > 60$\GeV,  $M_T > 40$\GeV\\[\cmsTableSkip]
  \multirow{5}{*}{$\Z\PQq\PQq$} &         &   1 or 2 selected central jets, all anti-tagged\\
           &   2 opposite-sign same-flavor       &   \quad leading $\pt > 200$\GeV\\
           &   \quad leading $\pt > 30$\GeV      &     1 selected forward jet\\
           &   \quad subleading  $\pt > 20$\GeV  & $\abs{m_{\ell\ell} - m_\PZ} < 7.5$\GeV\\
           &                                     & $\pt(\PZ \to \ell\ell) > 150$\GeV\\
\end{scotch}

\end{table*}

The event yields for the observed data as well as for the expected SM
backgrounds are shown in Table~\ref{tab:yieldsubsamples_single} for the muon channel
and the electron channel.
The respective normalizations of the simulated $\PW$ and $\cPZ$ boson production processes
in association with either light-flavor jets or heavy-flavor jets are derived
from data by fitting the CSVL b-tagged jet multiplicity distribution in control samples.
A deficit of data events compared to simulation is observed in both
the signal-depleted $\PWp\PQq\PQq$ and the signal-enriched $\PWm\PQq\PQq$ categories,
motivating a dedicated background prediction in the $\PWm\PQq\PQq$ category as described below.

\begin{table*}[hbtp]
\centering
\addtolength{\tabcolsep}{-2pt}
\topcaption{Event yields in the muon and electron channels for the event categories optimized for the single production search.
The $\PWp\PQq\PQq$ event category is not used in the search, but is shown for comparison, in order to demonstrate the expected
lepton charge asymmetry. For the separate background components the indicated uncertainties are statistical only,
originating from the limited number of MC events, while for the total background yield the combined statistical and systematic uncertainty is given.
The prediction for the signals is shown assuming branching
fractions of $\mathcal{B}_\PW = 0.5$ and
$\mathcal{B}_\PZ = \mathcal{B}_\PH = 0.25$.
The label `Other' designates the background originating
from $\ttbar \PW$, $\ttbar \PZ$ and triboson processes.}
\label{tab:yieldsubsamples_single}
\newcolumntype{x}{D{,}{\,\pm\,}{4.2}}
\cmsTableResize{\begin{scotch}{lxxxxxx}
 &\multicolumn{2}{c}{ $\PWp\PQq\PQq$ }& \multicolumn{2}{c}{$\PWm\PQq\PQq$} &\multicolumn{2}{c}{  $\Z\PQq\PQq$ } \\
\hline
Channel       & \multicolumn{1}{c}{muon} & \multicolumn{1}{c}{electron} & \multicolumn{1}{c}{muon} & \multicolumn{1}{c}{electron} & \multicolumn{1}{c}{muon} & \multicolumn{1}{c}{electron} \\[\cmsTableSkip]

 Estimated backgrounds &   &   &    &   &     &\\[\cmsTableSkip]
  $\ttbar$+jets & 26,2 & 23,2 & 28,3 & 24,2 & \multicolumn{1}{c}{$<$1} & \multicolumn{1}{c}{$<$1}    \\ [0.5ex]
  $\PW$+jets & 2069,43 & 1906,41 & 1191,36 & 1082,32 & \multicolumn{1}{c}{$<$1} &\multicolumn{1}{c}{$<$1}    \\ [0.5ex]
  $\PZ$+jets & 17,3 & 10,3 & 22,4 & 8.7,1.9 & 541,20 & 428,18   \\ [0.5ex]
  Single top quark & 20,3 & 20,3 & 11,2 & 12,2 & \multicolumn{1}{c}{$<$1} & \multicolumn{1}{c}{$<$1}    \\ [0.5ex]
  VV        & 28,2 & 27,2  & 31,2 & 31,2  & 9.9,0.7  & 7.6,0.6  \\ [0.5ex]
  Multijet  & 3.9,0.9 & 8.5,2.5 & 2.8,0.8 & 5.7,2.0 & \multicolumn{1}{c}{$<$1} & \multicolumn{1}{c}{$<$1}  \\ [0.5ex]
  Other     & \multicolumn{1}{c}{$<$1} & \multicolumn{1}{c}{$<$1}   & \multicolumn{1}{c}{$<$1} & \multicolumn{1}{c}{$<$1} & \multicolumn{1}{c}{$<$1}   & \multicolumn{1}{c}{$<$1}  \\[\cmsTableSkip]
  Total background & 2170,440 & 2000,400 & 1290,240 & 1160,230 & 550,110 & 436,87   \\ [0.5ex]
  Observed & \multicolumn{1}{c}{2082} & \multicolumn{1}{c}{1838} & \multicolumn{1}{c}{1112} & \multicolumn{1}{c}{1027}& \multicolumn{1}{c}{527}& \multicolumn{1}{c}{421} \\[\cmsTableSkip]
  Signal ($m_\Q = 600$\GeV, $\tilde{\kappa}_\PW = 0.1$) & \multicolumn{1}{c}{1.8} & \multicolumn{1}{c}{1.5} & \multicolumn{1}{c}{4.6}&\multicolumn{1}{c}{4.1}& \multicolumn{1}{c}{1.5}&\multicolumn{1}{c}{1.2} \\ [0.5ex]
  Signal ($m_\Q = 1100$\GeV, $\tilde{\kappa}_\PW = 1$) & \multicolumn{1}{c}{8.9}& \multicolumn{1}{c}{6.7}& \multicolumn{1}{c}{44.4}& \multicolumn{1}{c}{43.6}& \multicolumn{1}{c}{12.1}& \multicolumn{1}{c}{11.4} \\ [0.5ex]
\end{scotch}}
\end{table*}

In each of the event categories we reconstruct the mass of the VLQ candidate from the $\PW$ or $\cPZ$ boson candidates and the leading central jet in the event.
The reconstructed mass can be used to efficiently discriminate between the SM background and the VLQ processes.

\begin{figure*}[hbtp]
  \centering
\includegraphics[width=0.48\textwidth]{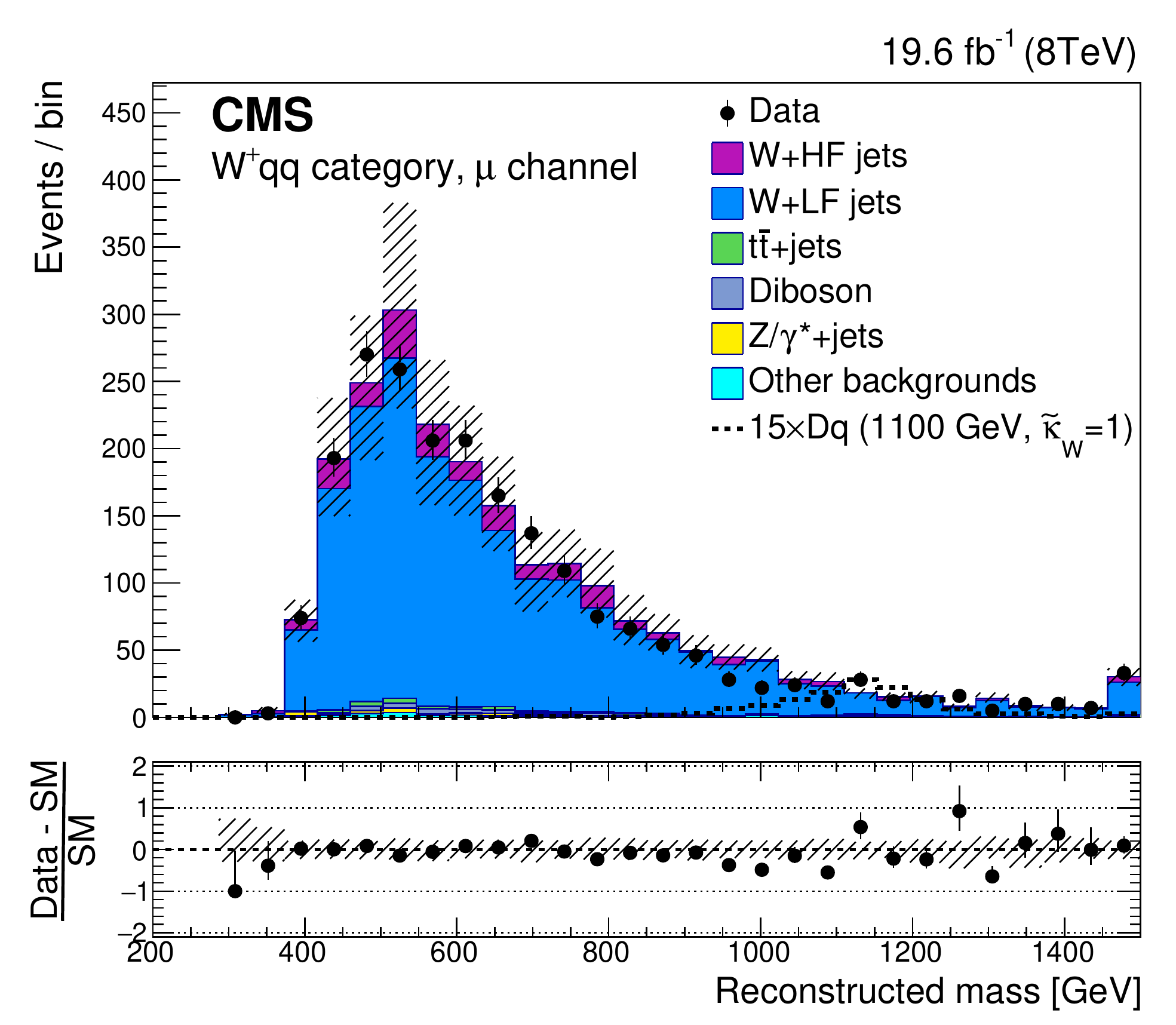}
\includegraphics[width=0.48\textwidth]{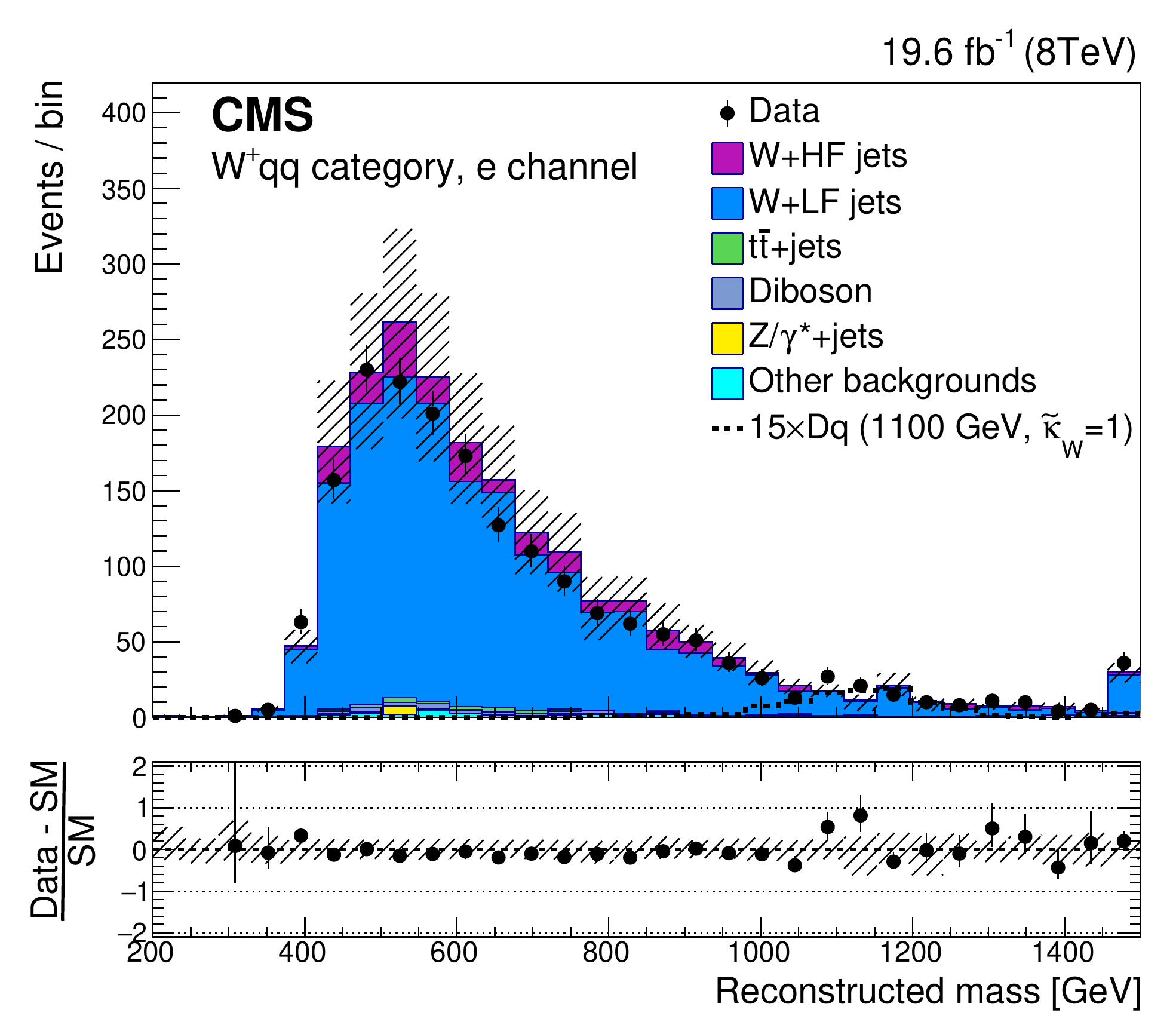}
\includegraphics[width=0.48\textwidth]{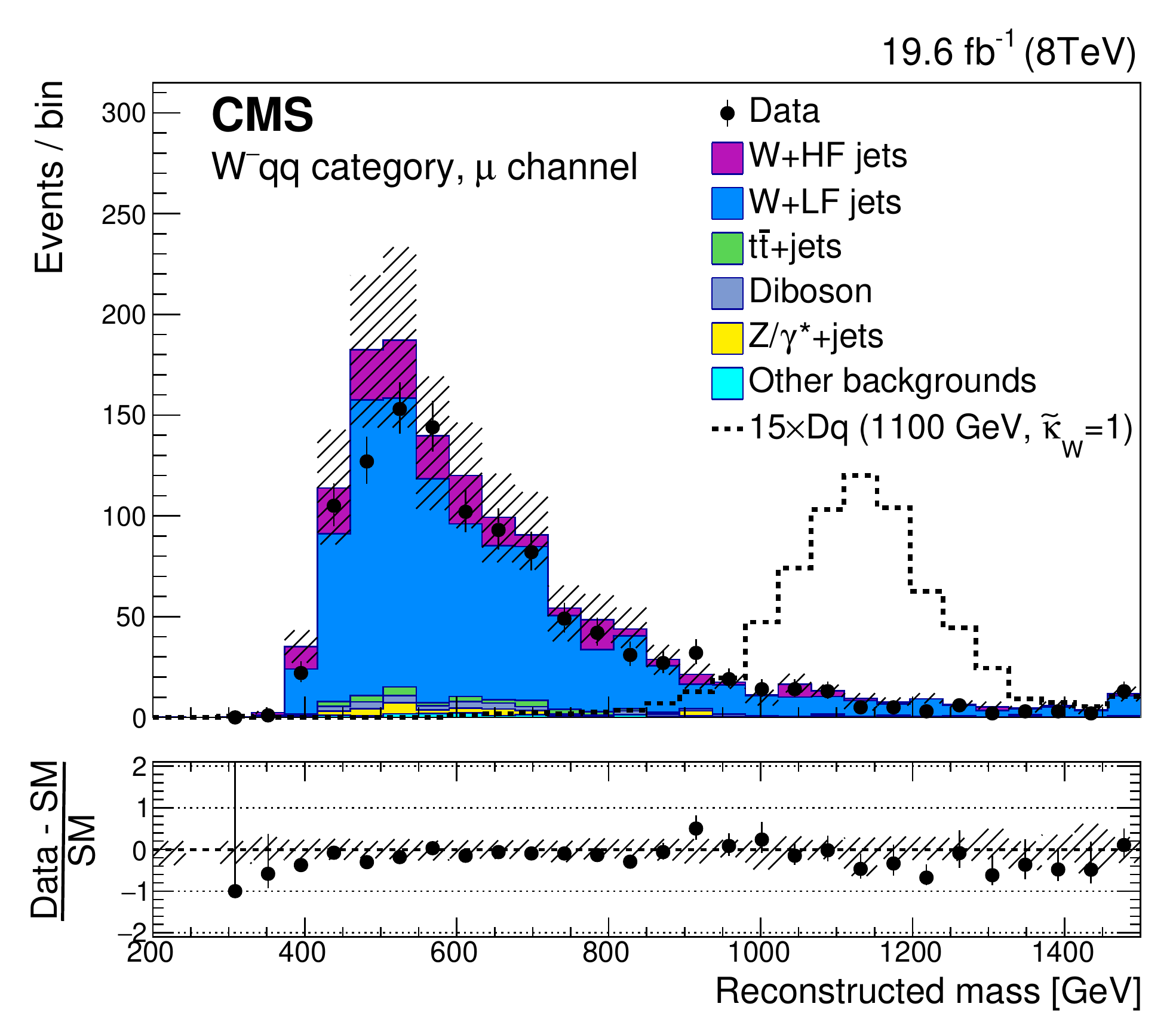}
\includegraphics[width=0.48\textwidth]{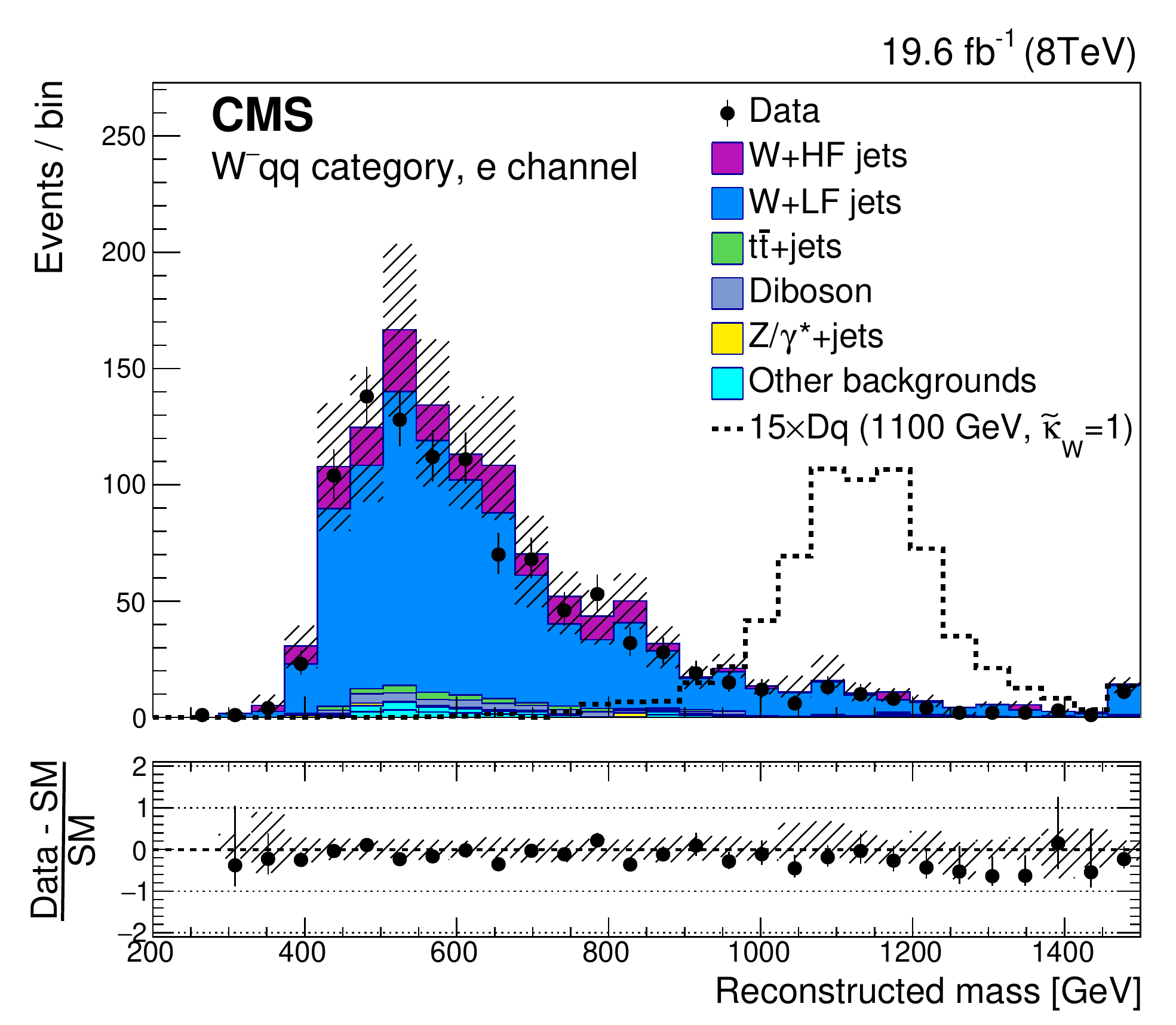}
\caption{The reconstructed mass of the VLQ candidate in the $\PWp\PQq\PQq$
event category (upper) and the $\PWm\PQq\PQq$ event category (lower), in the muon channel
(left) and the electron channel (right). The contributions of simulated events where the $\PW$ boson is produced in association
with light-flavor (LF) jets and heavy-flavor (HF) jets are shown separately.
The distribution for a heavy VLQ signal (indicated as D$\PQq$ representing a down-type VLQ produced
in association with a SM quark) of mass 1100\GeV and $\tilde{\kappa}_\PW = 1$ (for $\mathcal{B}_\PW = 0.5$
and $\mathcal{B}_\PZ = \mathcal{B}_\PH = 0.25$) is scaled up
by a factor of 15 for visibility.
The enhanced \QD quark signal contribution in the $\PWm\PQq\PQq$ event category in comparison to the $\PWp\PQq\PQq$
event category is clearly shown.
The hatched bands represent
the combined statistical and systematic uncertainties, and the highest bin contains the overflow.
\label{fig:VLQmassWqqplusminus}}
\end{figure*}

\begin{figure*}
  \centering
\includegraphics[width=0.48\textwidth]{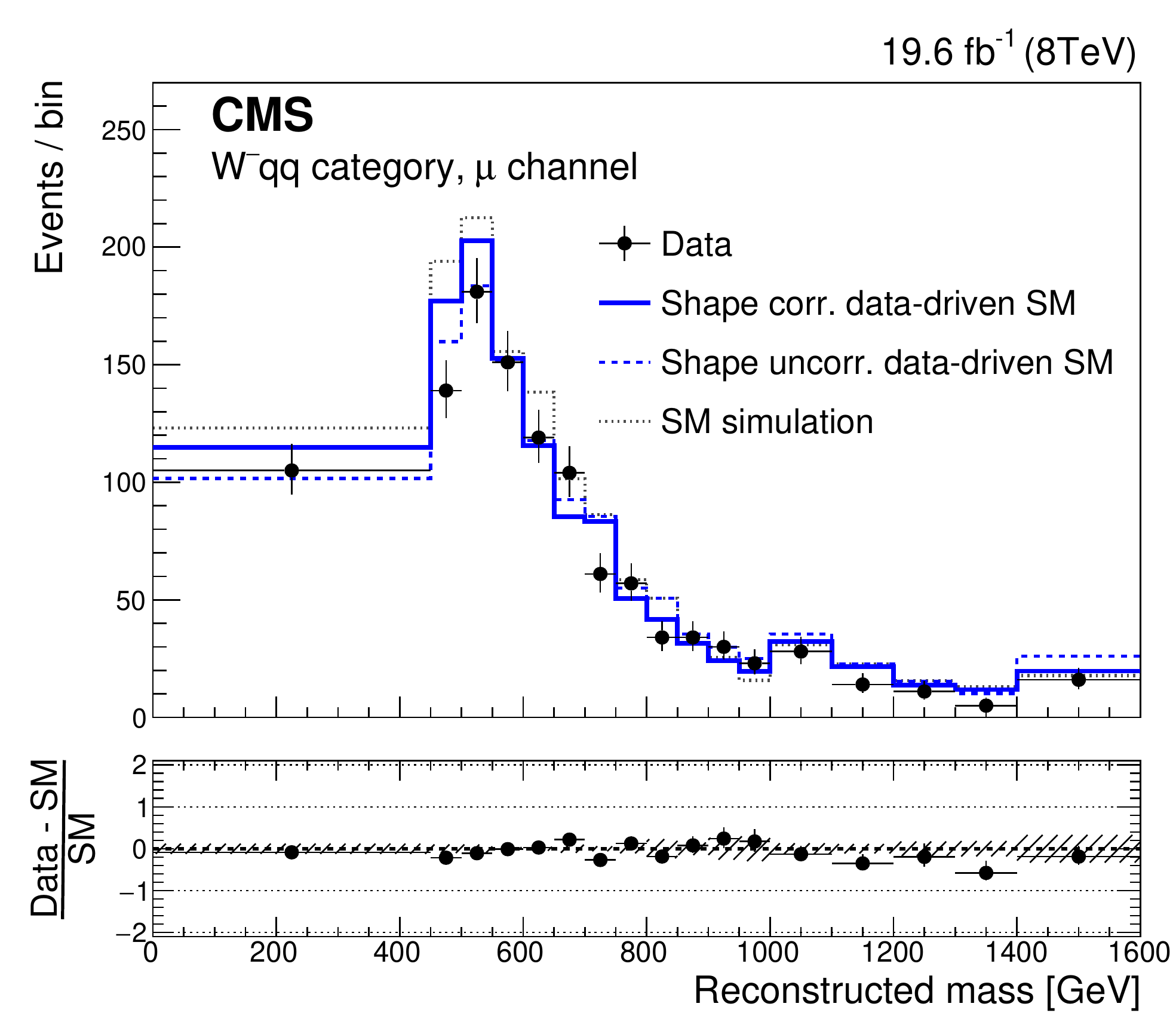}
 \includegraphics[width=0.48\textwidth]{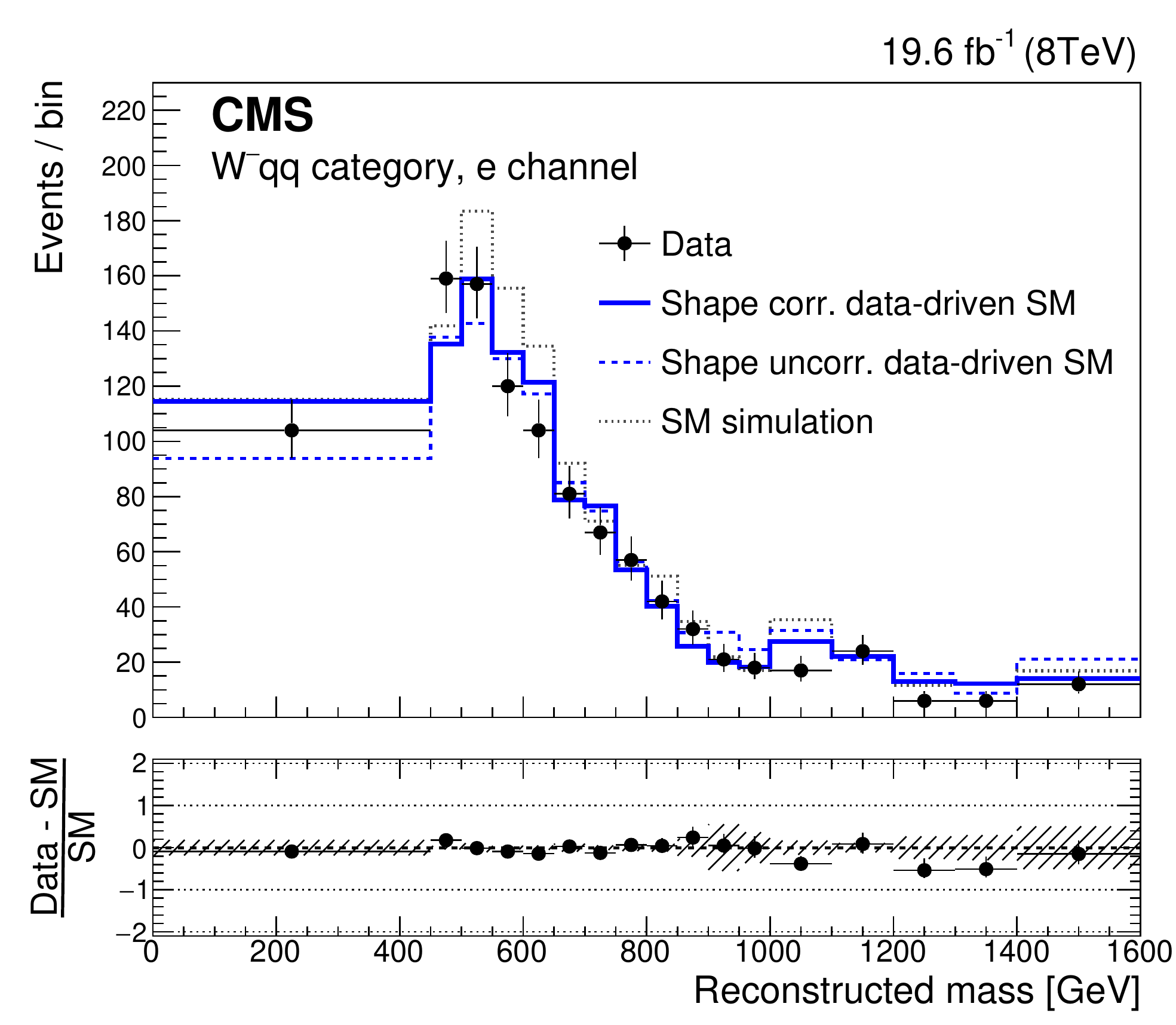}
  \caption{
The reconstructed VLQ candidate mass in the $\PWm\PQq\PQq$ category
for the muon channel (left) and the electron channel (right), for the background prediction and the data.
The solid bold (blue) line is the background distribution estimated from data, with a final shape
correction that accounts for the difference between the \PW+jets simulation in the control region
and the $\PWm\PQq\PQq$ signal region. The dashed (blue) line is the same, but without the shape correction.
The dotted (grey) line represents the SM prediction from simulation. The lower panel shows the ratio of the data to
the data-driven background distribution with shape corrections. For bins from 1000\GeV onwards,
a wider bin width is chosen to reduce statistical uncertainties in the background estimation from the data control region.
The horizontal error bars on the data points indicate the bin width.
\label{fig:VLQmass_Wminusqqdatadrivenbkg}}
\end{figure*}

Figure~\ref{fig:VLQmassWqqplusminus} shows the reconstructed mass of the VLQ candidate
for the $\PWp\PQq\PQq$ category (\cmsLeft) and the
$\PWm\PQq\PQq$ category (\cmsRight), comparing data to simulation.
The distributions of the reconstructed VLQ candidate mass comparing data to the prediction derived from a control region in data
are shown in Fig.~\ref{fig:VLQmass_Wminusqqdatadrivenbkg}
for the muon channel (\cmsLeft) and the electron channel (\cmsRight).
The predicted reconstructed mass distributions for the W+jets and multijet backgrounds
in the $\PWm\PQq\PQq$ category are obtained using a control region in data in the following way.
The control region is defined with the same $\PWm\PQq\PQq$ selection
requirements as outlined in Table~\ref{tab:vlqsubsamples1}, but with the
selection of a lepton with positive charge instead of a negative charge,
and with a forward-jet veto instead of requiring the presence of a forward jet.
The contribution of a potential signal in this control region is negligible because of these inverted requirements.
In order to obtain the predicted distribution in the $\PWm\PQq\PQq$ category,
the observed distribution in the control region is scaled with the ratio,
calculated from simulation, of negatively charged $\PW$ boson events to positively charged $\PW$ boson events.
Finally, we apply a shape correction to account for the difference observed in the
$\PW$+jets simulation between the control region and the $\PWm\PQq\PQq$
signal region, which originates from the different forward jet and lepton charge requirements.

\begin{figure}[hbtp]
  \centering
\includegraphics[width=0.48\textwidth]{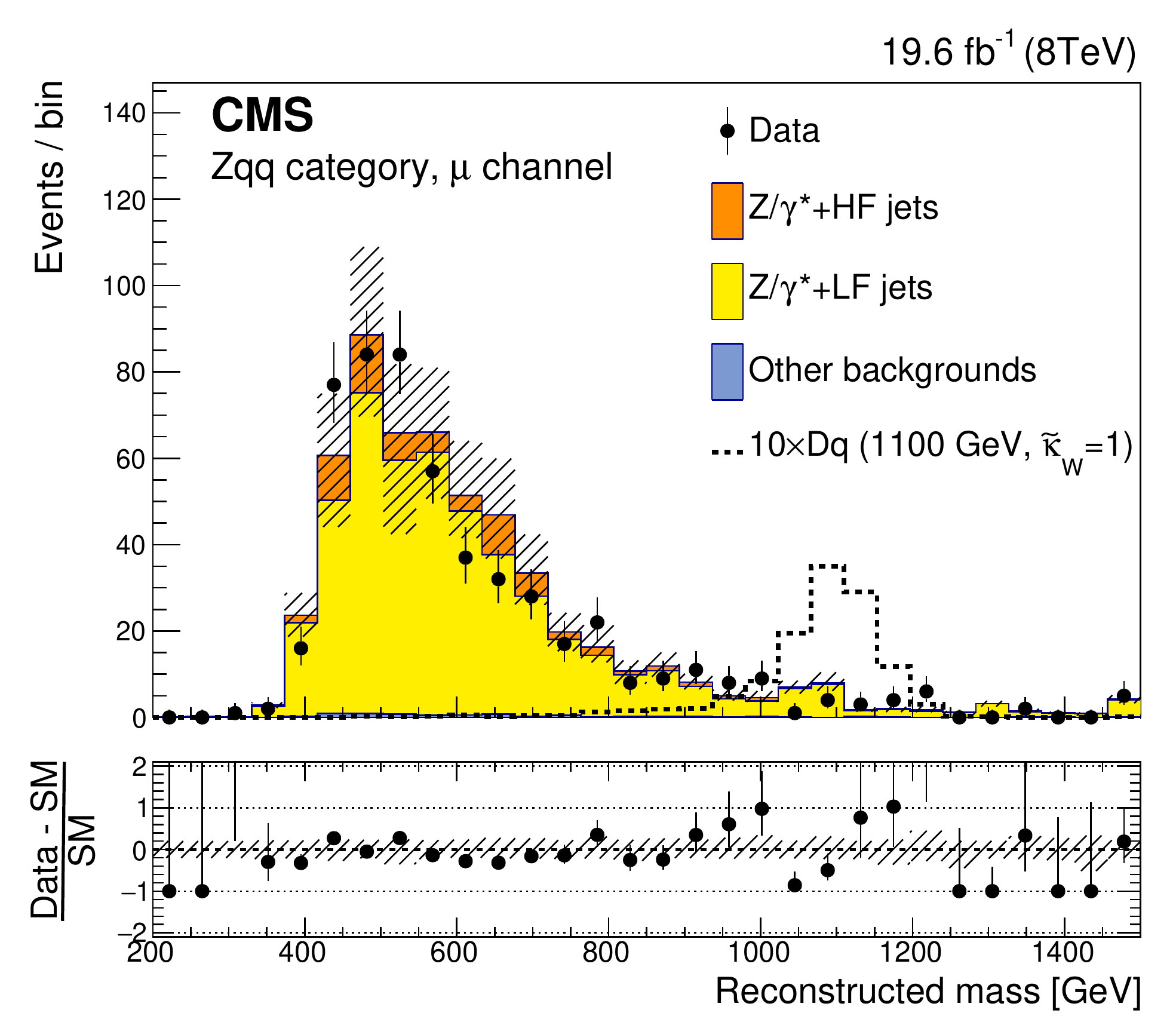}
\includegraphics[width=0.48\textwidth]{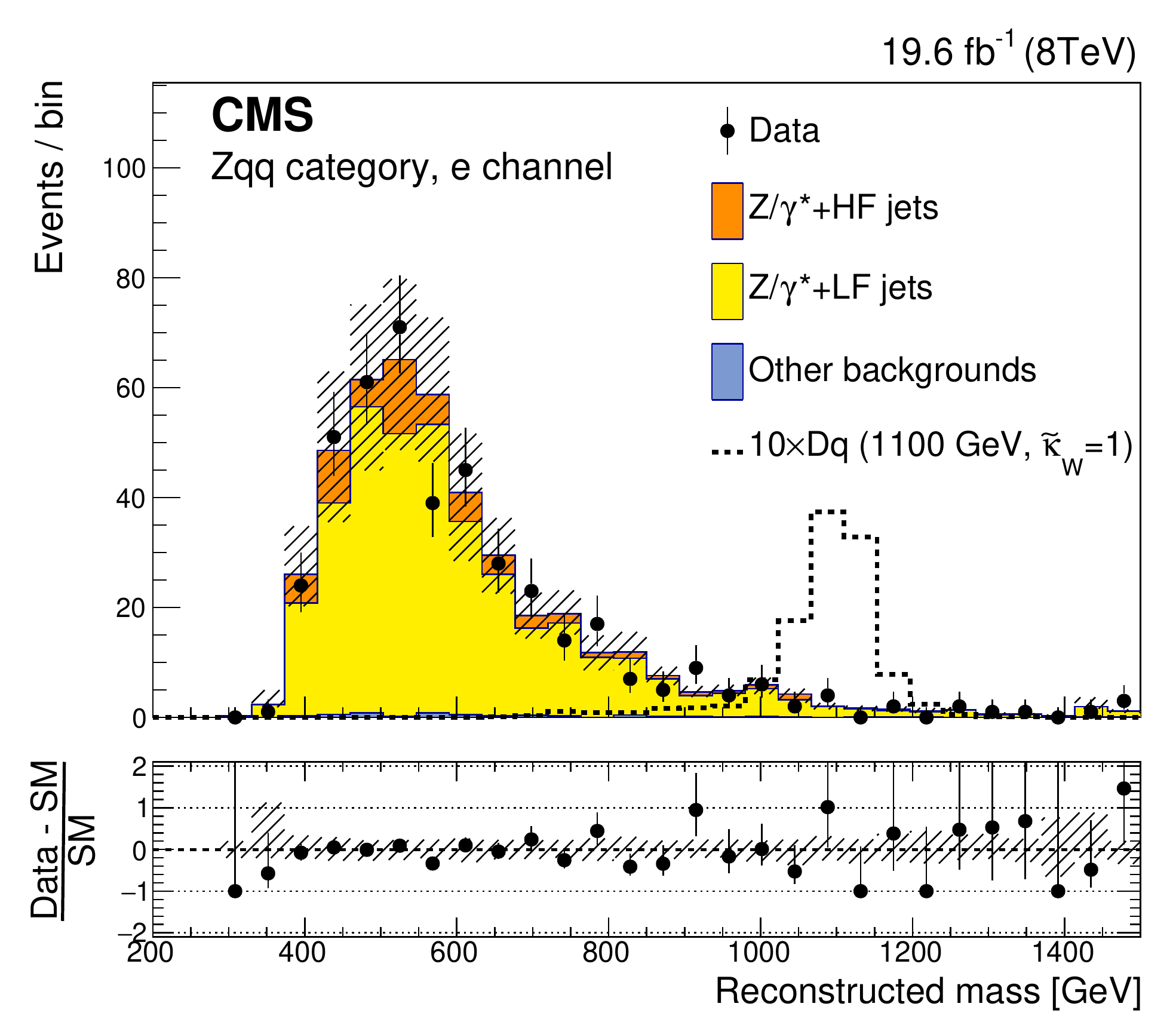}
\caption{The reconstructed mass of the VLQ candidate in the Zqq event category,
in the muon channel (\cmsLeft) and the electron channel (\cmsRight).
The contributions of simulated events where the $\PZ$ boson is produced in association
with light-flavor (LF) jets and heavy-flavor (HF) jets are shown separately.
The distribution for a heavy VLQ signal (indicated as D$\PQq$ representing a down-type VLQ produced
in association with a SM quark) of mass 1100\GeV and $\tilde{\kappa}_\PW = 1$
(for $\mathcal{B}_\PW = 0.5$ and $\mathcal{B}_\PZ = \mathcal{B}_\PH = 0.25$)
is scaled by a factor of 10 for better visibility. The hatched bands represent
the combined statistical and systematic uncertainties.
\label{fig:VLQmassZqq}}
\end{figure}

The reconstructed mass of the VLQ candidate in the Zqq category is shown
for data and the simulated signal sample in Fig.~\ref{fig:VLQmassZqq}, for the muon and electron channels.
The SM background is completely dominated by the $\PZ$+jets process.

\subsection{Search for pair production}
\label{sec:pair}

\subsubsection{Single-lepton channel}
\label{sec:pairsinglelep}
In the single-lepton event categories optimized for the search for pair produced VLQs, each of the selected events must contain exactly
one charged lepton (muon or electron) and at least four jets. The jet
multiplicity requirement ensures that there is no overlap with the
single-lepton $\PWm\PQq\PQq$ category selection outlined in Section~\ref{sec:single},
which selects events with at most three jets.
The jet collection may consist of AK5 jets or also of the subjets of a
V-tagged CA8 jet, where \cPV indicates a $\PW$, $\PZ$, or Higgs boson.

For heavy VLQs the quark pair from the hadronic decay
of the V boson may become so collimated that the overlapping hadronic showers
cannot be resolved as separate jets.
This means it is not possible to perform a kinematic fit
to the final state and therefore the signal reconstruction efficiency drops.
The CA8 jets with $\pt > 200$\GeV are used to identify the merged hadronic V boson decays by applying
a jet pruning algorithm, which resolves
the merged jets into subjets.
The efficiency drop caused by the jet merging at high VLQ masses
can be recovered by using the subjets in the kinematic fit.

A pruned CA8 wide-jet mass is equal to the invariant mass of the subjets.
A CA8 jet is considered to be: \PW-tagged if the pruned jet mass satisfies
$60 < M_\text{jet} < 100\GeV$, \Z-tagged if it satisfies
$65 < M_\text{jet} < 115\GeV$, or \PH-tagged if
$100 < M_\text{jet} < 140\GeV$.
If two subjets cannot be resolved, no \cPV tagging is done.
The three different \cPV tagging selections overlap, such that the same event
can be selected in different categories. As explained at the end of this
section, the overlap is removed in the final distributions and each event
is only counted once.

If the \cPV-tagged jet overlaps with any AK5 jets, the AK5 jet is
replaced with the two subjets of the matched CA8 jet.
Jets are considered as overlapping if $\Delta R < 0.04$,
where $\Delta R$ is constructed using the directions of the CA8 and AK5 jets.
The \PQb tagging of subjets is used in case of H-tagged CA8 jets.

Muon (electron) candidates in selected events contain tight muons (electrons) with $\pt > 45\,(30)\GeV$.
Events in the $\Pgm$+jets ($\Pe$+jets) channel must satisfy $\ptmiss > 20\,(30)\GeV$.
Events having a loose muon or electron in addition to a tight lepton
are vetoed. For this selection, loose leptons are defined as in
Table~\ref{tab:leptonbaseselvlq}, except that loose electrons
have relative isolation $I_\text{rel} < 0.2$ and $\pt > 20 \GeV$.
The jet collection described previously is used in a kinematic fit after
the following additional selection requirements. Selected AK5 jets
must have $\pt > 30$\GeV, while CA8 jets should have $\pt > 200$\GeV.
All jets should satisfy $\abs{\eta} < 2.4$.
We require the presence of at least four jets, and the highest four
\pt-ordered jets in the collection must satisfy
$\pt > 120$, 90, 50, and 30\GeV, respectively.

We perform constrained kinematic fits of the selected events to the hypotheses
described by Eqs.~(\ref{eq:eqn01}), (\ref{eq:eqn02}) and (\ref{eq:eqn03}).
The kinematic reconstruction of events is performed using the HitFit
package~\cite{hitfit_web}, which was developed by the D0 experiment at
Fermilab~\cite{snyder} for the measurement of the top quark mass in the
lepton+jets channel.

The fit is performed by minimizing a $\chi^2$ quantity constructed from
the differences between the measured value of each momentum component for each reconstructed object and the
fitted value of the same quantity divided by the corresponding uncertainties.
The four-momenta of the final-state particles are subject to the following constraints:
\begin{align}
m(\ell\nu)&= m_\PW, \\
m(\q\qbar')&= m_\PW,\ \text{or}\ m(\q\qbar)= m_\PZ,\ \text{or}\  m(\bbbar)= m_\PH, \\
m(\ell\nu{\q}_\ell) &= m_\text{hadr} = m_{\text{fit}},
\end{align}
where $m_\PW$ denotes the $\PW$ boson mass, $m_\PZ$ the $\PZ$ boson mass, and $m_\PH$ the
Higgs boson mass, with the values taken from the PDG~\cite{PDG}. The $m_\text{hadr}$ variable represents the mass of the three quarks on the hadronic
side of the decay ($m(\q\qbar'{\q}_{\mathrm{h}})$, $m(\q\qbar{\q}_{\mathrm{h}})$ or $m(\bbbar{\q}_{\mathrm{h}})$).
The kinematic fit is performed for each \cPV hypothesis in parallel.

The $z$ component of the neutrino momentum is estimated from one of
the two constraints given above that contain the neutrino momentum, with a
two-fold quadratic ambiguity.
The solutions found for the $z$ component of the neutrino momentum
are used as starting values for the fit.
If there are two real
solutions, they are both taken in turn, doubling the number of fitted
combinations. In the case of complex solutions, the real part is taken as
a starting value.
Using one constraint for calculation of $z$ component of the neutrino
momentum leaves only two constraints for the kinematic fit.
Only the combinations for which the $\chi^2$ probability of the fit exceeds
0.1\% are accepted. If the jet collection contains more than four jets,
then the five highest \pt jets are considered, and all possible combinations
of four jets are checked.

In order to distinguish between jets originating from quarks and from gluons,
we use the quark-gluon likelihood discrimination tagger (QGT)~\cite{quark_gluon}.
To reduce the combinatorial background in the assignment of jets to final-state
quarks, \cPV tagging, QGT tagging, and \PQb tagging information is used.
If a \cPV tag is present, only combinations where the subjets of the \cPV-jet
match decay products of the corresponding boson are considered.
The QGT tag requirements are then applied to those jets which are assigned
to the $\{\PQq_\ell, \PQq_{\mathrm{h}}\}$ quark pair. To suppress jets that may
have originated from gluons we require the QGT discriminant values to satisfy
the requirements $\mathrm{QGT}_{\q_\ell} > 0.4$ or $\mathrm{QGT}_{\q_{\mathrm{h}}}> 0.4$.
This excludes combinations in which both light quark jets have discriminant
values favoring gluons.

A b-tagged jet veto is applied to the jets that have been assigned
to the quark pair $\{\PQq_\ell, \PQq_{\mathrm{h}}\}$.
Since the \cPV-tagged events have a better signal-to-background
ratio, we apply softer b-tag selection requirements for this event category,
as described in Table~\ref{tab:b-veto}.
A more stringent requirement is applied on events without a \cPV tag.
\begin{table}
\centering
 \topcaption{Combinations of pairs of jets that have not been identified
as \cPV-jet matches, which can be accepted for matching to the quark pair
$\{\PQq_\ell, \PQq_{\mathrm{h}}\}$.
In the left column, the group with the lowest available b-tag content is
chosen, and within that group, the combination with the lowest $\chi^2$
is selected. In the right column, only the anti-tagged category is accepted.}
\label{tab:b-veto}
\begin{scotch}{ll}
Events with \cPV-tag & Events without \cPV-tag \\
\hline
0 CSVL \PQb tags                        &  0 CSVL \PQb tags  \\
1 CSVL \PQb tag only; no CSVM \PQb tags & \\
2 CSVL \PQb tags; no CSVM \PQb tags     & \\
\end{scotch}
\end{table}

Additional \PQb tagging requirements are applied to the jets associated with a Higgs boson decay.
For H-tagged events, at least one jet from the Higgs boson decay must have a CSVL
b tag, and for non-H-tagged events, at least one jet must have a CSVM \PQb tag.

After applying the kinematic fit we impose an additional threshold on
$S_\mathrm{T}$: $S_\mathrm{T} > 1000 \GeV$, where $S_\mathrm{T}$ is calculated
from jets selected during the kinematic fit, using post-fit transverse
momentum values. The $S_\mathrm{T}$ requirement strongly suppresses
the remaining background.

Table~\ref{tab:events} presents the event yields obtained after applying
the selections described above. There is good agreement between data and
the expected SM background. The number of expected signal events is also
presented.

\begin{table*}[htbp]
\centering
\addtolength{\tabcolsep}{-2pt}
\topcaption{Numbers of expected background events from simulation and of data
events in the $\PW\PQq\PW\PQq$,  $\PW\PQq\Z\PQq$ , and $\PW\PQq\PH\PQq$ channels after applying the single-lepton event selection in the search for pair produced VLQs.
For the separate background components the indicated uncertainties are statistical only,
originating from the limited number of MC events, while for the total background yield the combined statistical and systematic uncertainty is given.}
\label{tab:events}
\newcolumntype{z}{D{,}{\,\pm\,}{4.2}}\newcolumntype{x}{D{,}{\,\pm\,}{4.2}}
\begin{scotch}{lzzzzzz}
   & \multicolumn{2}{c}{$\qW\qW$} & \multicolumn{2}{c}{$\qW\qZ$} & \multicolumn{2}{c}{$\qW\qH$} \\
\hline
Channel       & \multicolumn{1}{c}{$\Pgm$+jets} & \multicolumn{1}{c}{$\Pe$+jets} & \multicolumn{1}{c}{$\Pgm$+jets} & \multicolumn{1}{c}{$\Pe$+jets} & \multicolumn{1}{c}{$\Pgm$+jets} & \multicolumn{1}{c}{$\Pe$+jets} \\[\cmsTableSkip]
Background process  &              &            &            &        &         &       \\[\cmsTableSkip]
$\ttbar$+jets  & 257,5&269,5&295,6&304,7 &224,6&241,6\\
$\PW + \ge 3$jets & 396,13   & 462,14 & 426,12 & 497,14& 42,4   & 42,4\\
Single top quark    & 13,2     & 25,3   & 13,2   & 30,4  & 11,2   & 17,3\\
Z/$\gamma^\ast+\ge3$jets&27,2& 27,2 & 30,2   & 30,2  & 2.8,0.5 & 2.9,0.5\\
$\PW\PW$, $\PW\PZ$, $\PZ\PZ$     & 10,1     & \multicolumn{1}{c}{$<$1}   & 10,1   & \multicolumn{1}{c}{$<$1}  & 1.7,0.6 & \multicolumn{1}{c}{$<$1}\\
Multijet       & \multicolumn{1}{c}{$<$1}     & 59,4   & \multicolumn{1}{c}{$<$1}   & 59,4  & \multicolumn{1}{c}{$<$1}    & 11,2\\[\cmsTableSkip]
Total background & 703,80 & 840,100 & 773,86 & 920,110 & 282,37  & 314,41\\
Observed         &\multicolumn{1}{c}{741}&\multicolumn{1}{c}{896}&\multicolumn{1}{c}{793}&\multicolumn{1}{c}{943}&\multicolumn{1}{c}{292}& \multicolumn{1}{c}{313} \\[\cmsTableSkip]
Signal ($m_\Q = 600$\GeV) &\multicolumn{1}{c}{112}&\multicolumn{1}{c}{117}&\multicolumn{1}{c}{63}&\multicolumn{1}{c}{64}&\multicolumn{1}{c}{36}& \multicolumn{1}{c}{35} \\
Signal ($m_\Q = 800$\GeV) &\multicolumn{1}{c}{20}&\multicolumn{1}{c}{20}&\multicolumn{1}{c}{11}&\multicolumn{1}{c}{11}&\multicolumn{1}{c}{6.5}& \multicolumn{1}{c}{5.7} \\
Signal ($m_\Q = 1000$\GeV)&\multicolumn{1}{c}{3.3}&\multicolumn{1}{c}{3.3}&\multicolumn{1}{c}{1.8}&\multicolumn{1}{c}{2.0}&\multicolumn{1}{c}{1.1}& \multicolumn{1}{c}{0.8} \\
\end{scotch}
\end{table*}

\begin{figure*}[htbp]
\includegraphics[width=0.48\textwidth]{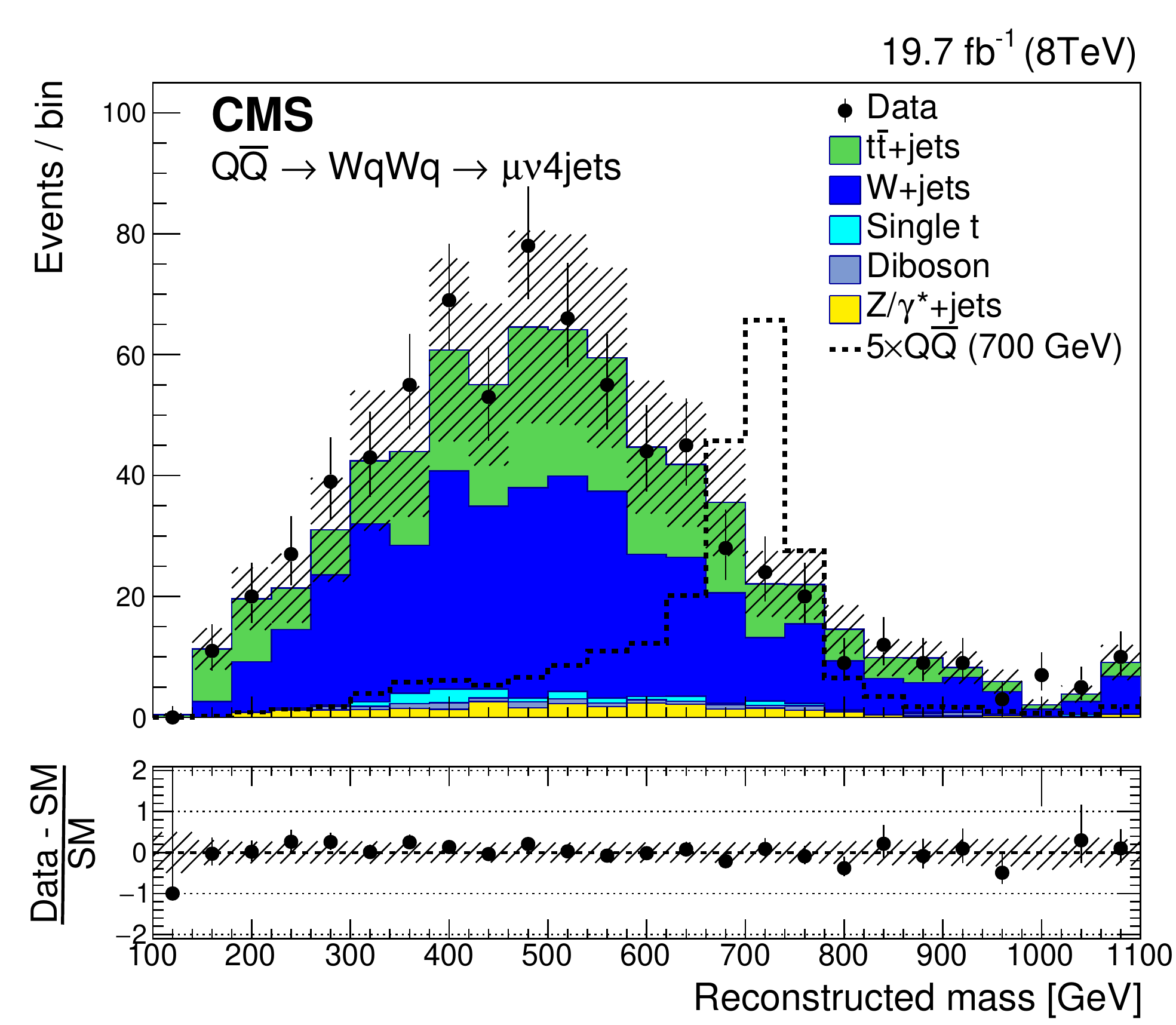}
\includegraphics[width=0.48\textwidth]{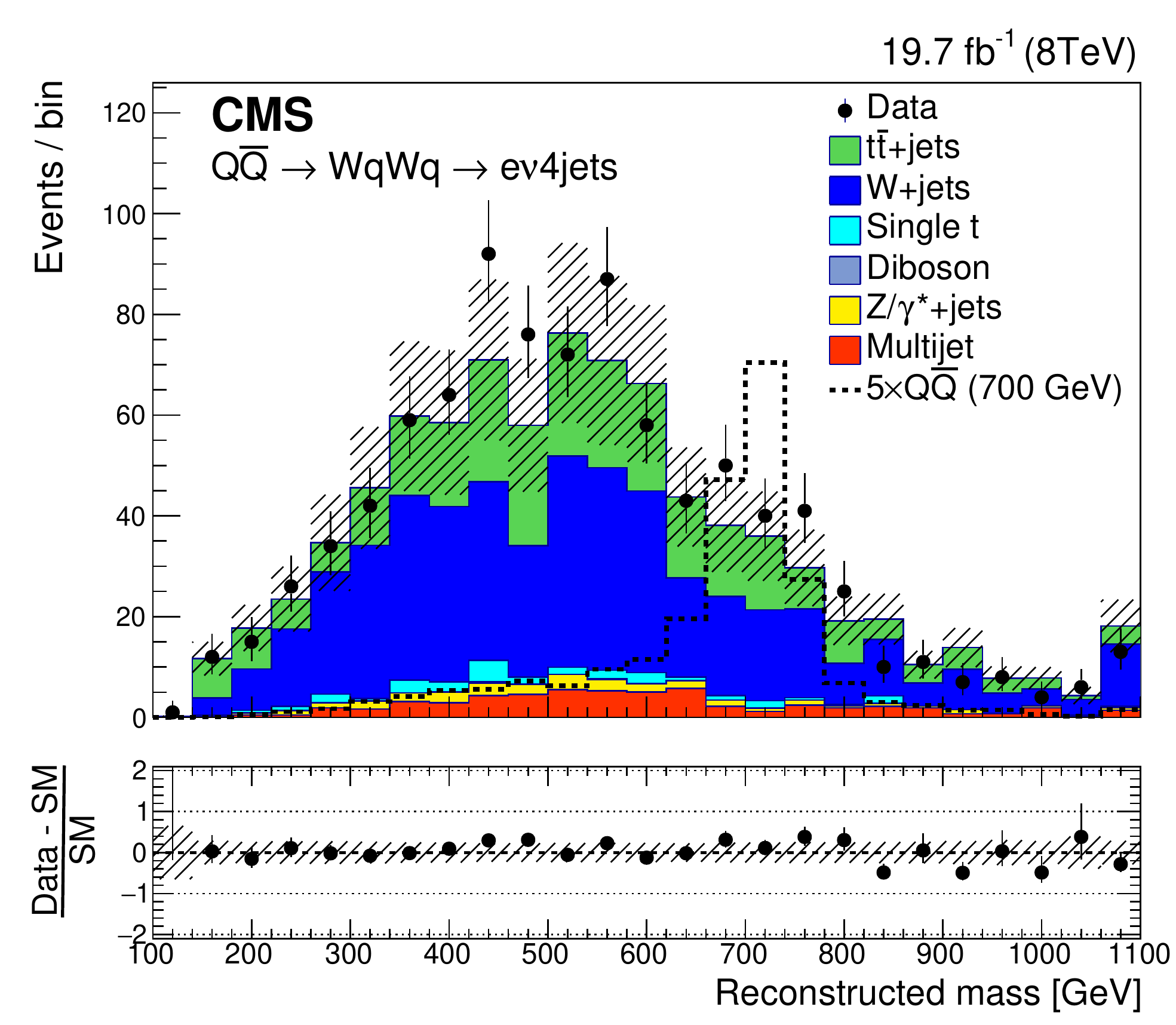}\\
\includegraphics[width=0.48\textwidth]{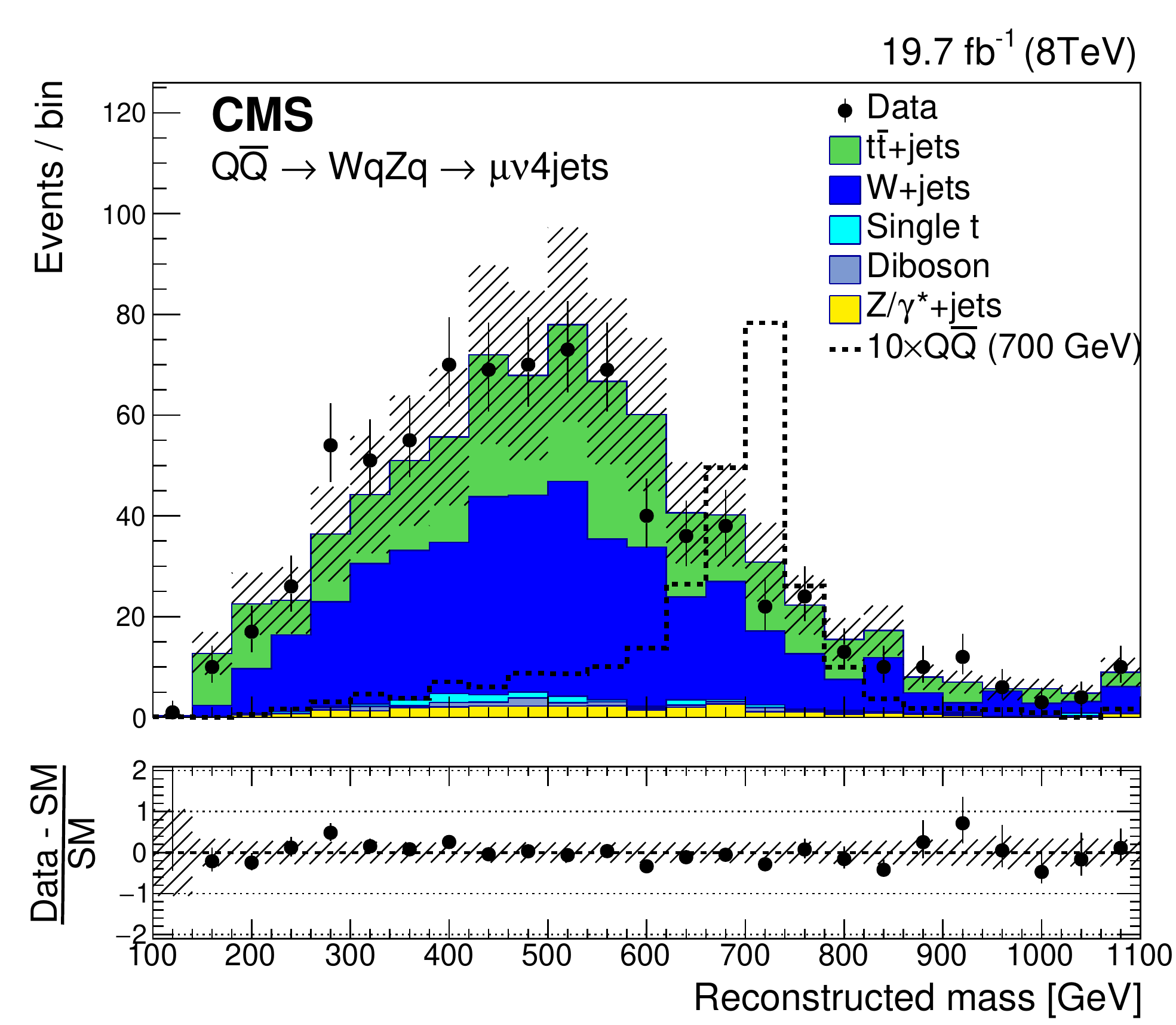}
\includegraphics[width=0.48\textwidth]{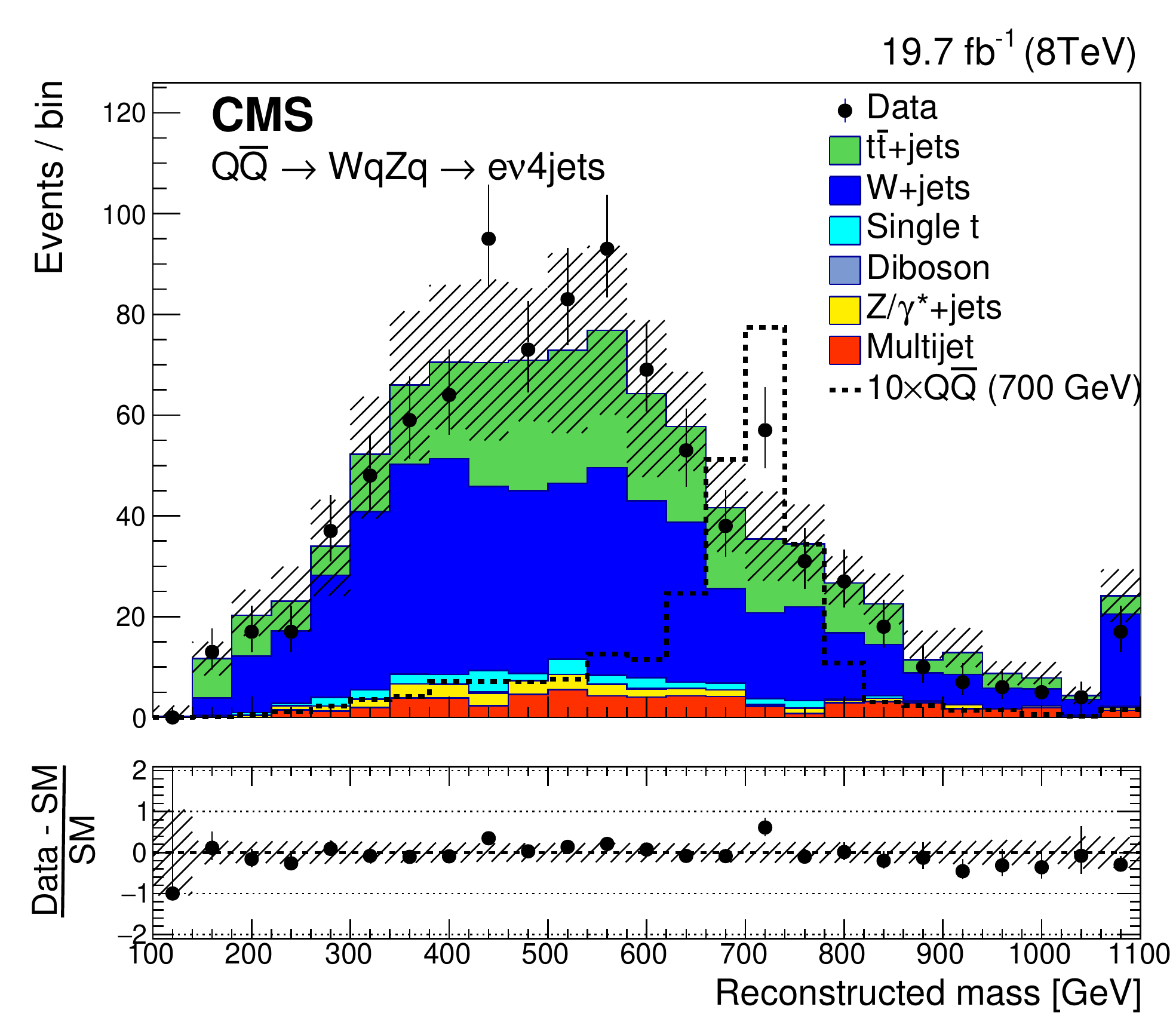}\\
\includegraphics[width=0.48\textwidth]{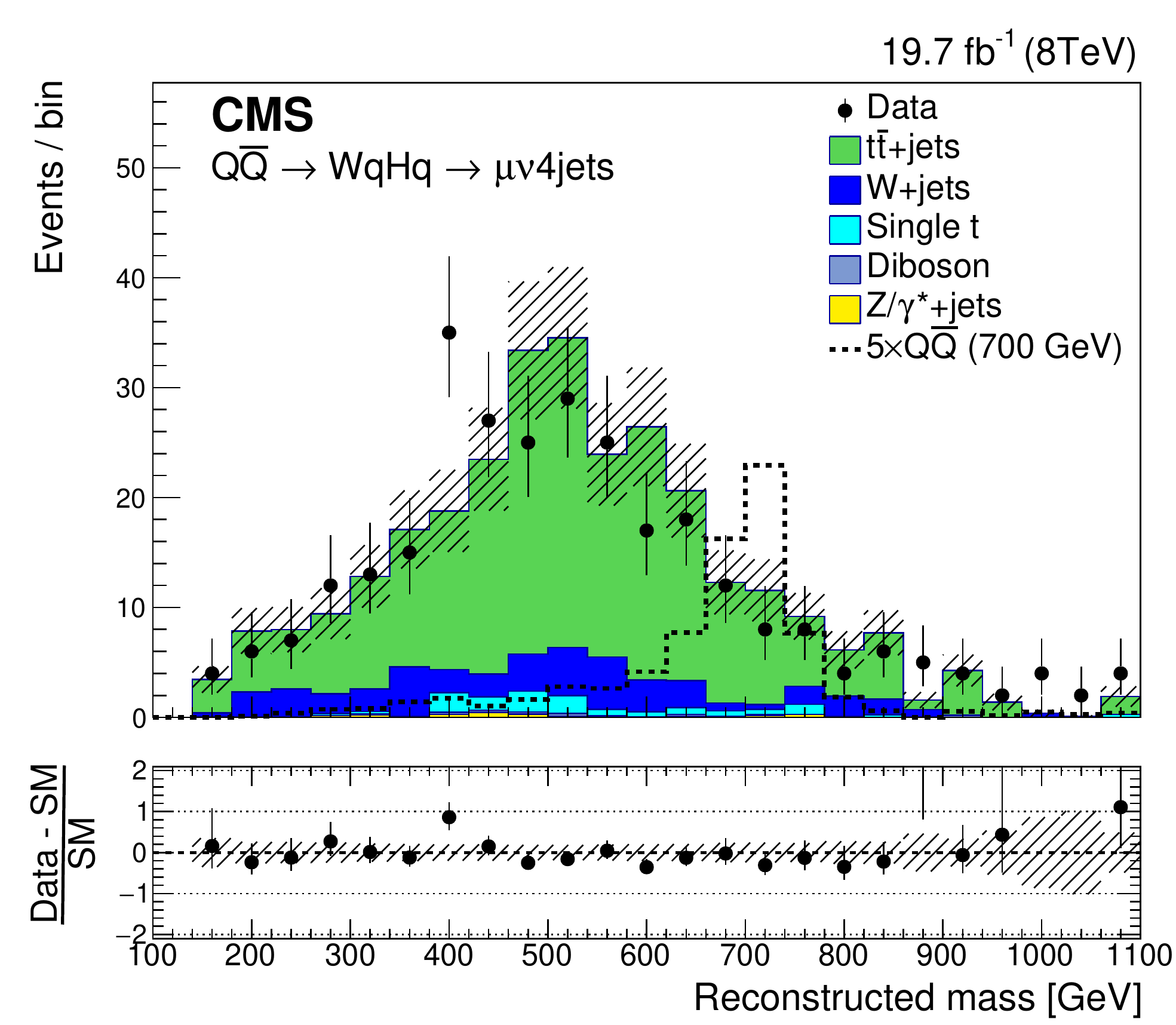}
\includegraphics[width=0.48\textwidth]{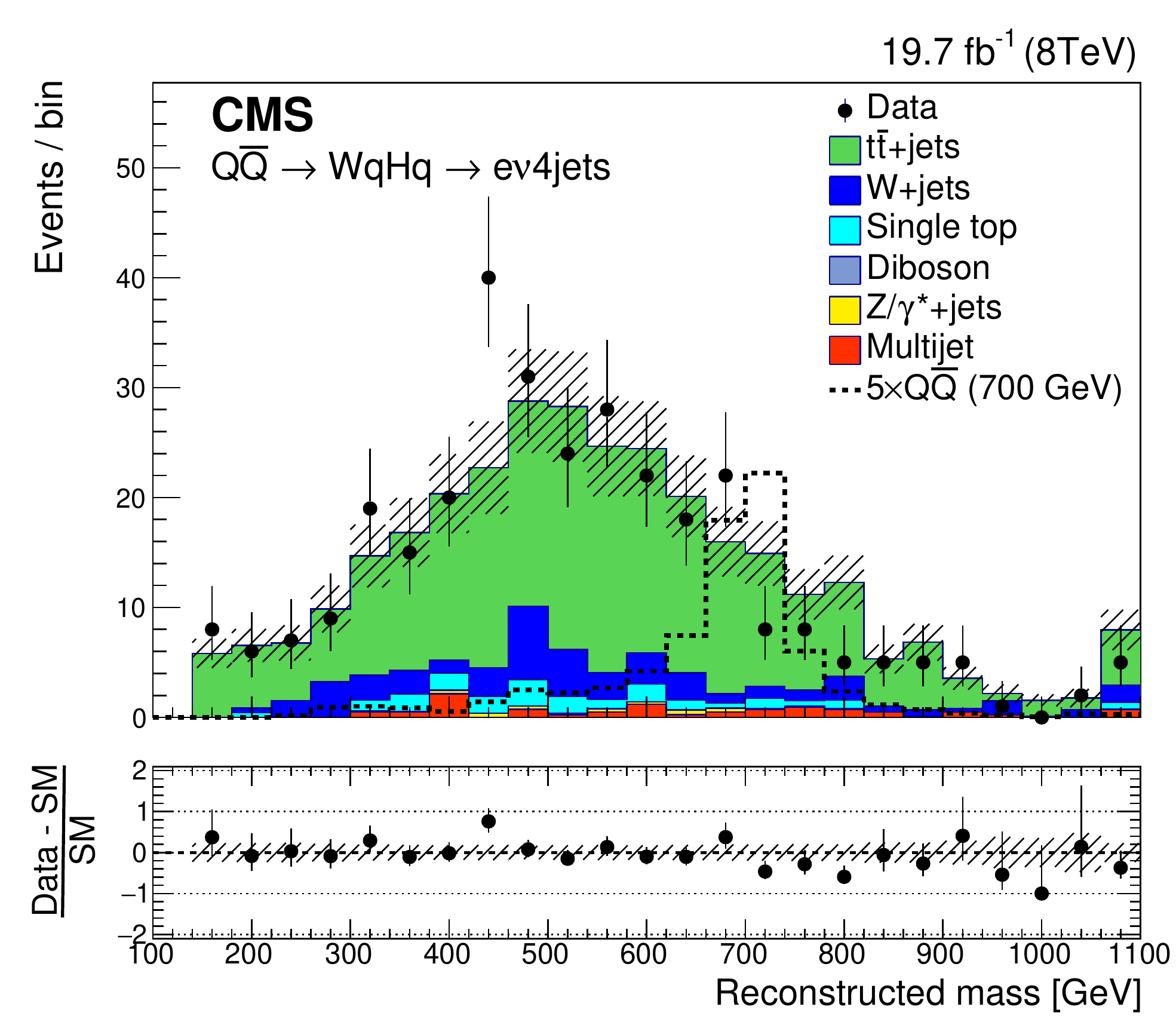}\\
  \caption{Reconstructed mass distributions for $\PW\PQq\PW\PQq$ (uppermost), $\PW\PQq\Z\PQq$  (middle), and
 $\PW\PQq\PH\PQq$ (lowest) reconstruction from a kinematic fit. Plots on the left are for the $\Pgm$+jets channel
and on the right, for the $\Pe$+jets channel.
The distribution for pair-produced VLQs of mass 700\GeV
for $\mathcal{B}_\PW = 1.0$ (uppermost), $\mathcal{B}_\PW = \mathcal{B}_\PZ = 0.5$ (middle) and
$\mathcal{B}_\PW = \mathcal{B}_\PH = 0.5$ (lowest) are scaled up for visibility by
a factor of 5, 10 and 5, respectively. The hatched bands represent
the combined statistical and systematic uncertainties.
}
\label{masses}
\end{figure*}

The result of the kinematic fit is one mass distribution per reconstruction hypothesis and lepton channel, as shown in Fig.~\ref{masses}.
The mass distributions are presented for the $\Pgm$+jets channel in the
plots on the left, and for the $\Pe$+jets channel in the plots on the right.
In the case of $\Pe$+jets events, the contribution from multijets is estimated
from control samples in data. Events are selected that pass the electron trigger,
but contain objects that satisfy inverted electron identification requirements.
The normalisation of the multijet contribution is determined from a maximum
likelihood fit of the observed $\ptmiss$ distribution. The shapes in this
fit are predicted by the MC simulation, where electroweak backgrounds are
constrained to their expected cross sections and float within uncertainties,
while the multijet normalization is allowed to float freely.

The uppermost row of distributions in Fig.~\ref{masses} are those associated with the
$\PW\PQq\PW\PQq$ reconstruction, while the middle row corresponds to the $\PW\PQq\Z\PQq$  reconstruction,
and the lowest row, to the $\PW\PQq\PH\PQq$ reconstruction. For both the $\PW\PQq\Z\PQq$  and $\PW\PQq\PH\PQq$ reconstruction, the expected pair-produced VLQ signals are shown
for $\mathcal{B}(\Q\to\qW$) = 0.5 and $\mathcal{B}(\Q\to\qZ$) = 0.5 or
$\mathcal{B}(\Q\to\qH$) = 0.5, respectively. These distributions show good agreement between data and the expected
SM background.

\begin{figure*}[htbp]
\centering
\includegraphics[width=0.48\textwidth]{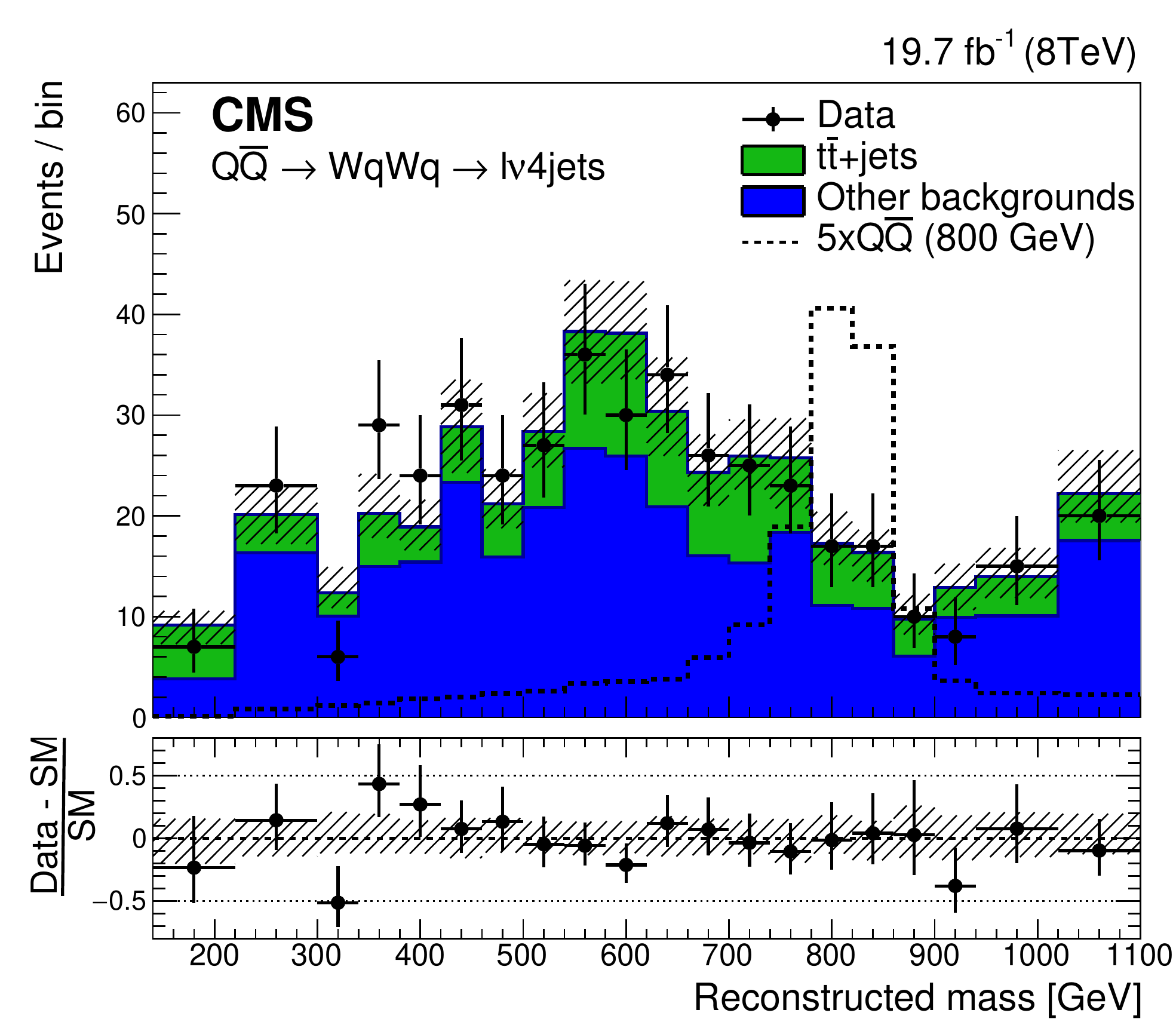}
\includegraphics[width=0.48\textwidth]{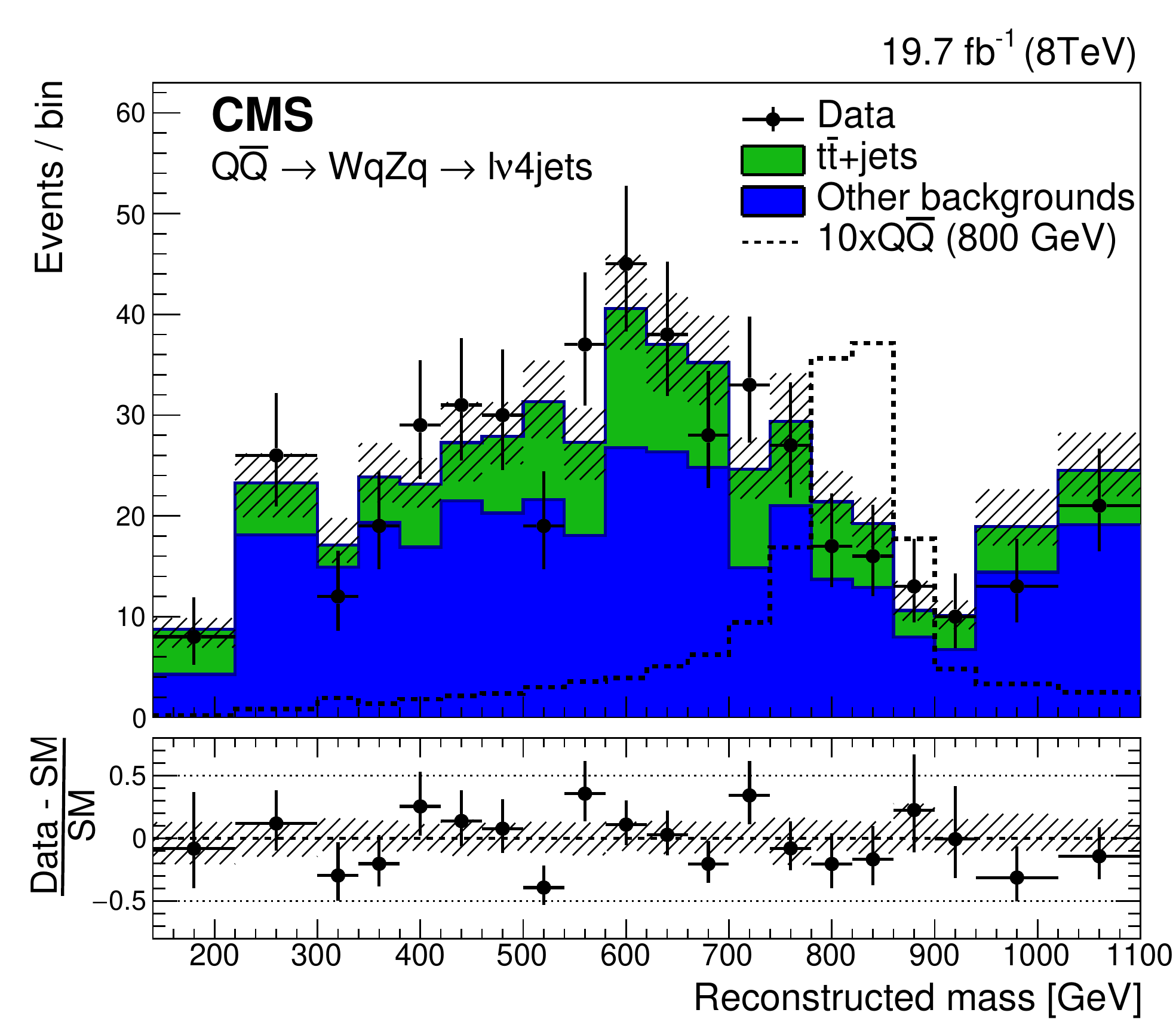}
\includegraphics[width=0.48\textwidth]{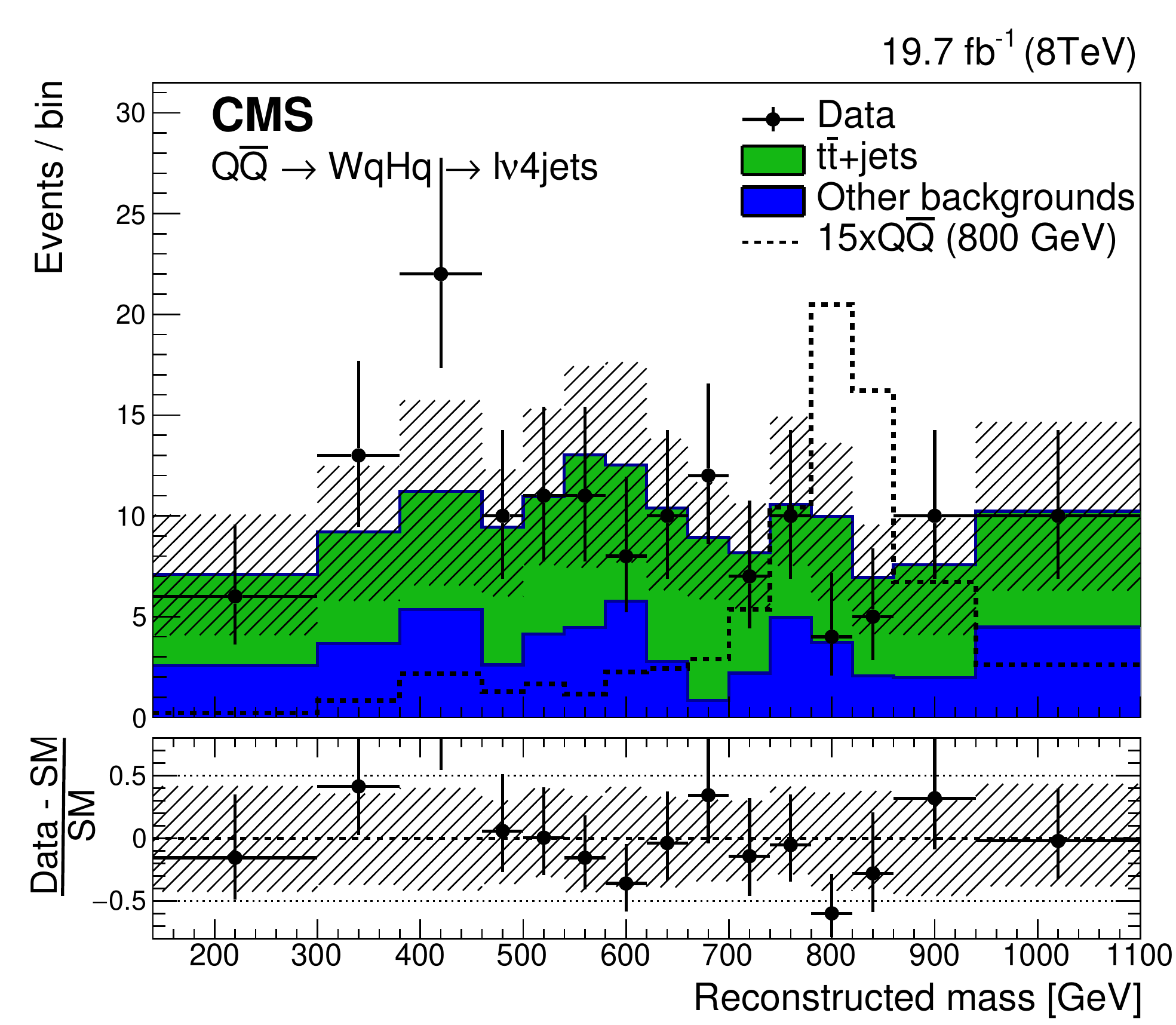}
  \caption{Mass distributions for the $\PW\PQq\PW\PQq$ (upper left), $\PW\PQq\Z\PQq$  (upper right), and $\PW\PQq\PH\PQq$ (lower)
reconstructions from a kinematic fit for the combination of the $\Pgm$+jets and $\Pe$+jets channel,
for events with $S_\mathrm{T} >$ 1240\GeV.
The distribution for pair-produced VLQs of mass 800\GeV
for $\mathcal{B}_\PW = 1.0$ (upper left), $\mathcal{B}_\PW = \mathcal{B}_\PZ = 0.5$ (upper right) and
$\mathcal{B}_\PW = \mathcal{B}_\PH = 0.5$ (lower) is scaled up for visibility by
a factor of 5, 10 and 15, respectively.
The hatched bands represent
the combined statistical and systematic uncertainties.
The horizontal error bars on the data points only indicate the bin width.
}
  \label{masses_1240}
\end{figure*}

Following the strategy described in Ref.~\cite{B2G-13-005}
we then further tighten the $S_\mathrm{T}$ requirement to $S_\mathrm{T} > 1240$\GeV.
This improves the signal-to-background ratio. At the same time we combine
the $\Pgm$+jets and $\Pe$+jets events, and use the resulting $\Mfit$ distributions
for the cross section limit calculations.  Figure~\ref{masses_1240} shows
these $\Mfit$ distributions for the $\PW\PQq\PW\PQq$ (uppermost), $\PW\PQq\Z\PQq$  (middle), and $\PW\PQq\PH\PQq$ (lowest)
reconstruction.

\begin{figure}[htbp]
\centering
\includegraphics[width=0.5\textwidth]{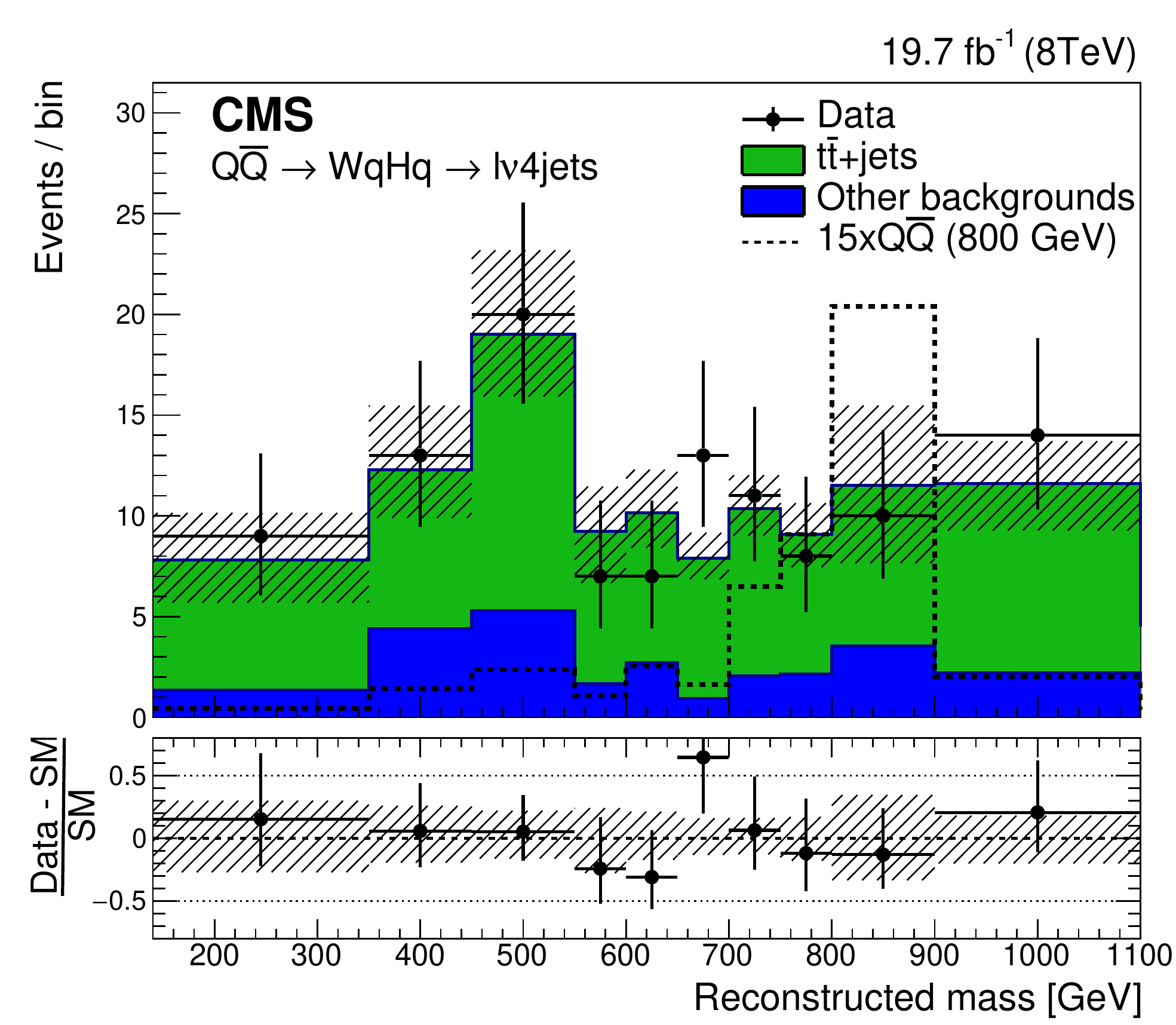}\\
  \caption{Mass distribution for the $\PW\PQq\PH\PQq$ reconstruction from a kinematic fit, for combined $\Pgm$+jets
and $\Pe$+jets channels and for events with $S_\mathrm{T} >$ 1240\GeV. Events appearing also in the $\PW\PQq\PW\PQq$ sample have been removed. The distribution for pair-produced VLQs of mass 800\GeV
for $\mathcal{B}_\PW = \mathcal{B}_\PH = 0.5$ is scaled up by a factor of 15 for visibility.
The hatched band represent
the combined statistical and systematic uncertainties.
The horizontal error bars on the data points only indicate the bin width.
}
  \label{qWqH_1240}
\end{figure}

We find that the $\PW\PQq\PW\PQq$ reconstruction gives a better expected mass
limit than the $\PW\PQq\Z\PQq$  reconstruction even for high values of $\mathcal{B}(\Q\to\qZ$).
The events selected and reconstructed for the $\PW\PQq\PW\PQq$ and $\PW\PQq\Z\PQq$  hypotheses are highly
correlated, with an 82\% overlap between the two. Furthermore, since the $\PW\PQq\PW\PQq$ reconstruction is more sensitive, we do not consider the $\PW\PQq\Z\PQq$
reconstruction further, and use only the $\PW\PQq\PW\PQq$ reconstruction for all branching
fraction combinations of the VLQ decaying to a $\PW$ boson or a
$\PZ$ boson. The $\PW\PQq\PH\PQq$ reconstruction improves the expected limits for large
decay branching fractions of the VLQ into a Higgs boson.
The events selected for the $\PW\PQq\PH\PQq$ reconstruction have a relatively small
correlation with those selected for the $\PW\PQq\PW\PQq$ channel events,
with only a 25\% event overlap. We therefore use $\PW\PQq\PH\PQq$ reconstructed events
and combine them with $\PW\PQq\PW\PQq$ events.
Events in the $\PW\PQq\PH\PQq$ selection that also appear in the $\PW\PQq\PW\PQq$ selection are removed, so that there is no double counting.
Figure~\ref{qWqH_1240} shows the reconstructed mass for $\PW\PQq\PH\PQq$ events where
events overlapping with the $\PW\PQq\PW\PQq$ reconstruction have been removed.
Table~\ref{tab:events_1240} shows the number of selected events
after applying the stricter $S_\mathrm{T}$ requirement for both the $\PW\PQq\PW\PQq$ reconstruction and the $\PW\PQq\PH\PQq$ reconstruction, excluding those events that
appear in both samples.

\begin{table*}[htbp]
\centering
\addtolength{\tabcolsep}{-2pt}
\topcaption{Numbers of expected background events from simulation and of data
events in the single-lepton $\PW\PQq\PW\PQq$ and $\PW\PQq\PH\PQq$ channels, after the application of the
$S_\mathrm{T} >$ 1240\GeV requirement. Events in the $\PW\PQq\PH\PQq$ channel that
also appear in the $\PW\PQq\PW\PQq$ channel are excluded.
For the separate background components the indicated uncertainties are statistical only,
originating from the limited number of MC events, while for the total background yield the combined statistical and systematic uncertainty is given.}
\label{tab:events_1240}
\newcolumntype{y}{D{,}{\,\pm\,}{3.2}}
\begin{scotch}{lyyyy}
  & \multicolumn{2}{c}{$\qW\qW$} & \multicolumn{2}{c}{$\qW\qH$} \\
\hline
Channel &\multicolumn{1}{c}{$\Pgm$+jets}& \multicolumn{1}{c}{$\Pe$+jets} &\multicolumn{1}{c}{$\Pgm$+jets}& \multicolumn{1}{c}{$\Pe$+jets} \\[\cmsTableSkip]
Backgr. process & \multicolumn{1}{c}{Events}  & \multicolumn{1}{c}{Events}  &  \multicolumn{1}{c}{Events}               &   \multicolumn{1}{c}{Events} \\[\cmsTableSkip]
$\ttbar$           &61,3   & 65,3& 34,3   & 46,3\\
$\PW + \ge 3$ jets     & 103,7      & 129,8    & 8,2  & 11,3\\
Single top quark    & 2,1    & 9,2  & 2,1  & 3,1\\
Z/$\gamma^\ast+\ge3$ jets & 7,1& 6,1  & \multicolumn{1}{c}{$<$1} & 1.0,0.4\\
$\PW\PW$, $\PW\PZ$, $\PZ\PZ$         & 3,1    & \multicolumn{1}{c}{$<$1}    & \multicolumn{1}{c}{$<$1} & \multicolumn{1}{c}{$<$1}\\
Multijets          & \multicolumn{1}{c}{$<$1}      & 15,2 & \multicolumn{1}{c}{$<$1} & 3,1\\[\cmsTableSkip]
Total backgr.      & 176,21  & 224,26 & 44,7  & 64,10\\
Observed           & \multicolumn{1}{c}{199}& \multicolumn{1}{c}{233}& \multicolumn{1}{c}{51}&  \multicolumn{1}{c}{61} \\[\cmsTableSkip]
Signal ($m_\Q = 600$\GeV)   & \multicolumn{1}{c}{53}& \multicolumn{1}{c}{54}& \multicolumn{1}{c}{5.7}& \multicolumn{1}{c}{5.7} \\
Signal ($m_\Q = 800$\GeV)   & \multicolumn{1}{c}{15}& \multicolumn{1}{c}{16}& \multicolumn{1}{c}{1.5}& \multicolumn{1}{c}{1.7} \\
Signal ($m_\Q = 1000$\GeV)  & \multicolumn{1}{c}{2.9}& \multicolumn{1}{c}{3.1}& \multicolumn{1}{c}{0.3}& \multicolumn{1}{c}{0.2} \\
\end{scotch}
\end{table*}

The distributions in Fig.~\ref{masses_1240} (upper left) and Fig.~\ref{qWqH_1240}
of the reconstructed mass are used in the rest of the analysis.
The binning in these distributions has been
chosen such that the statistical uncertainty on the background expectation
in each bin is less than 20\%.

\subsubsection{Dilepton and multilepton channels}
\label{sec:pairmultilep}
The event categories with at least two leptons optimized for the search for pair produced VLQs make use of the collections of central jets and anti-tagged jets
defined in Section~\ref{sec:single}, in addition to b-tagged jets, which are required
to have a \PQb tagging discriminant above the CSVM threshold.

We categorize the events according to the number of tight leptons along with
selection criteria applied to the jets and the missing transverse momentum.
Each of the event categories is designed to be particularly sensitive to one or
more of the pair production topologies presented in Table~\ref{tab:vlqmodes}. This is reflected in
the names used as identifiers for the categories: dileptonic $\PW\PQq\PW\PQq$, $\Z\PQq\PH\PQq$, dileptonic $\cPV\PQq\Z\PQq$,
and multileptonic $\cPV\PQq\Z\PQq$, where $\cPV$ indicates a $\PW$ or \Z boson.
For the decay channel $\QQbar \to \PW\PQq\PH\PQq$, no dedicated category has been defined,
to avoid an overlap of selected events with the single-lepton categories described in the previous section.

The definition of each event category optimized for pair production is summarized in Table~\ref{tab:vlqsubsamples2}.
In all event categories except dileptonic $\PW\PQq\PW\PQq$, a leptonically decaying $\cPZ$ boson candidate is reconstructed,
from two same-flavor opposite-sign dileptons, imposing a requirement on the dilepton mass $m_{\ell\ell}$,
as described in Table~\ref{tab:vlqsubsamples2}.
Thresholds are imposed on the transverse momentum $\pt(\cPZ)$ of the $\cPZ$ boson candidate.

\begin{table*}[htbp]
\centering
 \topcaption{The event categories as optimized for the VLQ pair production, with at least two leptons.
 The categories are based on the number of tight muons
 or electrons present in the event, along with additional criteria optimized for specific VLQ topologies.
 Events containing any additional loose leptons are excluded.
\label{tab:vlqsubsamples2}}
\begin{scotch}{lll}
    Event category    & Tight leptons ($\Pgm$,$\Pe$)  &  Additional selection criteria\\
\hline
                                   &                    & $\geq$2 selected central jets, all anti-tagged\\
 dileptonic      &  2 opposite-sign     & \quad leading $\pt > 200$\GeV\\
 $\PW\PQq\PW\PQq$      &  \quad leading $\pt > 30$\GeV & \quad subleading $\pt > 100$\GeV\\
       &  \quad subleading  $\pt > 20$\GeV             &  $\abs{m_{\ell\ell} - m_\PZ} > 7.5$\GeV (same flavor)\\
       &               &  $\ptmiss > 60$\GeV\\[\cmsTableSkip]
\multirow{7}{*}{$\Z\PQq\PH\PQq$}  &                  & $\geq$3 selected central jets, $\geq$2 anti-tagged\\
                       & 2 opposite-sign same-flavor      & \quad leading $\pt > 200$\GeV\\
                       & \quad leading $\pt > 30$\GeV     &   \quad subleading $\pt > 100$\GeV\\
                       & \quad subleading  $\pt > 20$\GeV & $\geq$1 b-tagged jet\\
                       &                  & $\abs{m_{\ell\ell} - m_\PZ} < 7.5$\GeV\\
                       &                  & $\pt(\PZ \to \ell\ell) > 150$\GeV\\[\cmsTableSkip]
                       &                  & $\geq$4 selected central jets, $\geq$2 anti-tagged\\
dileptonic           & 2 opposite-sign same-flavor               & \quad leading $\pt > 200$\GeV\\
$\cPV\PQq\Z\PQq$ & \quad leading $\pt > 30$\GeV      & \quad subleading $\pt > 100$\GeV\\
              & \quad subleading  $\pt > 20$\GeV  & veto events with b-tagged jets\\
              &                                   & $\abs{m_{\ell\ell} - m_\PZ} < 7.5$\GeV\\
              &                                   & $\pt(\PZ \to \ell\ell) > 150$\GeV\\[\cmsTableSkip]
              &                                   & $\geq$2 selected central jets, all anti-tagged\\
multileptonic           &                                   & \quad leading $\pt > 200$\GeV\\
$\cPV\PQq\Z\PQq$ &         3 or 4                    & \quad subleading $\pt > 100$\GeV\\
              &    \quad leading $\pt > 30$ Ge    & $|m_{\ell\ell} - m_\PZ| < 7.5$\GeV\\
              &    \quad others  $\pt > 20$\GeV   & $\pt(\PZ \to \ell\ell) > 150$\GeV\\
              &                                   & $\ptmiss > 60$\GeV (3 leptons)\\
              &                                   & $\Delta R(\ell,\ell) > 0.05$ (other flavor)\\
\end{scotch}
\end{table*}

The event yields for the observed data as well as for the expected SM
backgrounds are shown in Table~\ref{tab:yieldsubsamples_pair} for the muon channel
and the electron channel.
In the case of $\Pgm$-$\Pe$ dilepton events (for the dileptonic $\PW\PQq\PW\PQq$ event category only),
the event is assigned to the muon channel or the electron channel
depending on which trigger the event has passed online, with the priority given to the muon trigger.
If the event has passed the muon trigger, the selected muon has
$\pt > 30$\GeV and the electron has $\pt > 20$\GeV, then this
event will be assigned to the muon channel, even if the event also
passed the electron trigger. If the event has passed the electron trigger
as well as the muon trigger, the selected electron has $\pt > 30$\GeV
and the muon has $\pt$ in the range of 20--30\GeV, then the event will
be assigned to the electron channel. In the final case where the event only
passes the electron trigger, the selected electron has $\pt > 30$\GeV
and the muon has $\pt > 20$\GeV, the event will be assigned to the
electron channel.

\begin{table*}[hbtp]
\centering
\addtolength{\tabcolsep}{-2pt}
\topcaption{Event yields in the muon and electron channels for the event categories with at least two leptons, optimized for the pair production search.
For the separate background components the indicated uncertainties are statistical only,
originating from the limited number of MC events, while for the total background yield the combined statistical and systematic uncertainty is given.
The prediction for the signals is shown assuming branching
fractions of $\mathcal{B}_\PW = 0.5$ and
$\mathcal{B}_\PZ = \mathcal{B}_\PH = 0.25$.
The label `Other' designates the background originating
from $\ttbar \PW$, $\ttbar \PZ$ and triboson processes.}
\label{tab:yieldsubsamples_pair}
\newcolumntype{x}{D{,}{\,\pm\,}{4.2}}
\cmsTableResize{\begin{scotch}{lxxxxxx}
  &\multicolumn{2}{c}{ dileptonic $\PW\PQq\PW\PQq$ }& \multicolumn{2}{c}{ $\Z\PQq\PH\PQq$ } &\multicolumn{2}{c}{ dileptonic  $\cPV\PQq\Z\PQq$ } \\
\hline
 Channel       & \multicolumn{1}{c}{muon} & \multicolumn{1}{c}{electron} & \multicolumn{1}{c}{muon} & \multicolumn{1}{c}{electron} & \multicolumn{1}{c}{muon} & \multicolumn{1}{c}{electron} \\[\cmsTableSkip]

 Estimated backgrounds &   &   &    &   &     &\\[\cmsTableSkip]
  $\ttbar$+jets  & 62,4 & 22,2  & 2.1,0.7 &1.2,0.4 & \multicolumn{1}{c}{$<$1} & \multicolumn{1}{c}{$<$1} \\ [0.5ex]
  $\PW$+jets & \multicolumn{1}{c}{$<$1} & \multicolumn{1}{c}{$<$1} & \multicolumn{1}{c}{$<$1} & \multicolumn{1}{c}{$<$1} & \multicolumn{1}{c}{$<$1} & \multicolumn{1}{c}{$<$1} \\ [0.5ex]
  $\PZ$+jets & 79,6 & 55,5  & 53,3 & 41,2 & 238,5 & 202,4 \\ [0.5ex]
  Single top quark  & 4.6,1.5 & 1.7,0.8  & \multicolumn{1}{c}{$<$1} &\multicolumn{1}{c}{$<$1}  & \multicolumn{1}{c}{$<$1} &\multicolumn{1}{c}{$<$1}  \\ [0.5ex]
  VV         & 8.5,1.0 & 3.5,0.6 &1.0,0.2 & \multicolumn{1}{c}{$<$1} & 3.7,0.4 & 3.6,0.4  \\ [0.5ex]
  Multijet   & 14,2  & 9.2,2.6   & \multicolumn{1}{c}{$<$1} & \multicolumn{1}{c}{$<$1}   & \multicolumn{1}{c}{$<$1}  & \multicolumn{1}{c}{$<$1}\\ [0.5ex]
  Other     & 1.8,0.2  & \multicolumn{1}{c}{$<$1}  & 1.3,0.2  & \multicolumn{1}{c}{$<$1}& \multicolumn{1}{c}{$<$1} & \multicolumn{1}{c}{$<$1}\\[\cmsTableSkip]
  Total background & 170,21 & 92,17 & 58,14 & 43,10  & 243,45 & 207,37  \\ [0.5ex]
  Observed & \multicolumn{1}{c}{174}& 95 & \multicolumn{1}{c}{54} & \multicolumn{1}{c}{48}& \multicolumn{1}{c}{249}& \multicolumn{1}{c}{201}\\[\cmsTableSkip]
  Signal ($m_\Q = 600$\GeV, $\tilde{\kappa}_\PW = 0.1$) & \multicolumn{1}{c}{11.7}&\multicolumn{1}{c}{4.2}& \multicolumn{1}{c}{3.9}& \multicolumn{1}{c}{3.4}& \multicolumn{1}{c}{9.1}& \multicolumn{1}{c}{7.4}\\ [0.5ex]
  Signal ($m_\Q = 1100$\GeV, $\tilde{\kappa}_\PW = 1$) & \multicolumn{1}{c}{0.6}&\multicolumn{1}{c}{0.2}& \multicolumn{1}{c}{0.3}& \multicolumn{1}{c}{0.2}& \multicolumn{1}{c}{1.4}& \multicolumn{1}{c}{1.2}\\ [0.5ex]
\end{scotch}}
\end{table*}

In each of the mutually exclusive event categories an observable is constructed that
efficiently discriminates SM background events from VLQ processes.
In several of the event categories we reconstruct the mass of the VLQ candidate.
In other categories, where the mass of the VLQ candidate is poorly reconstructed,
or where the event yield is too low,
we use a simpler observable such as the $S_\mathrm{T}$ variable defined in
Eq.~(\ref{eq:STvlq}) or the event count.

\begin{figure}[hbtp]
  \centering
 \includegraphics[width=0.48\textwidth]{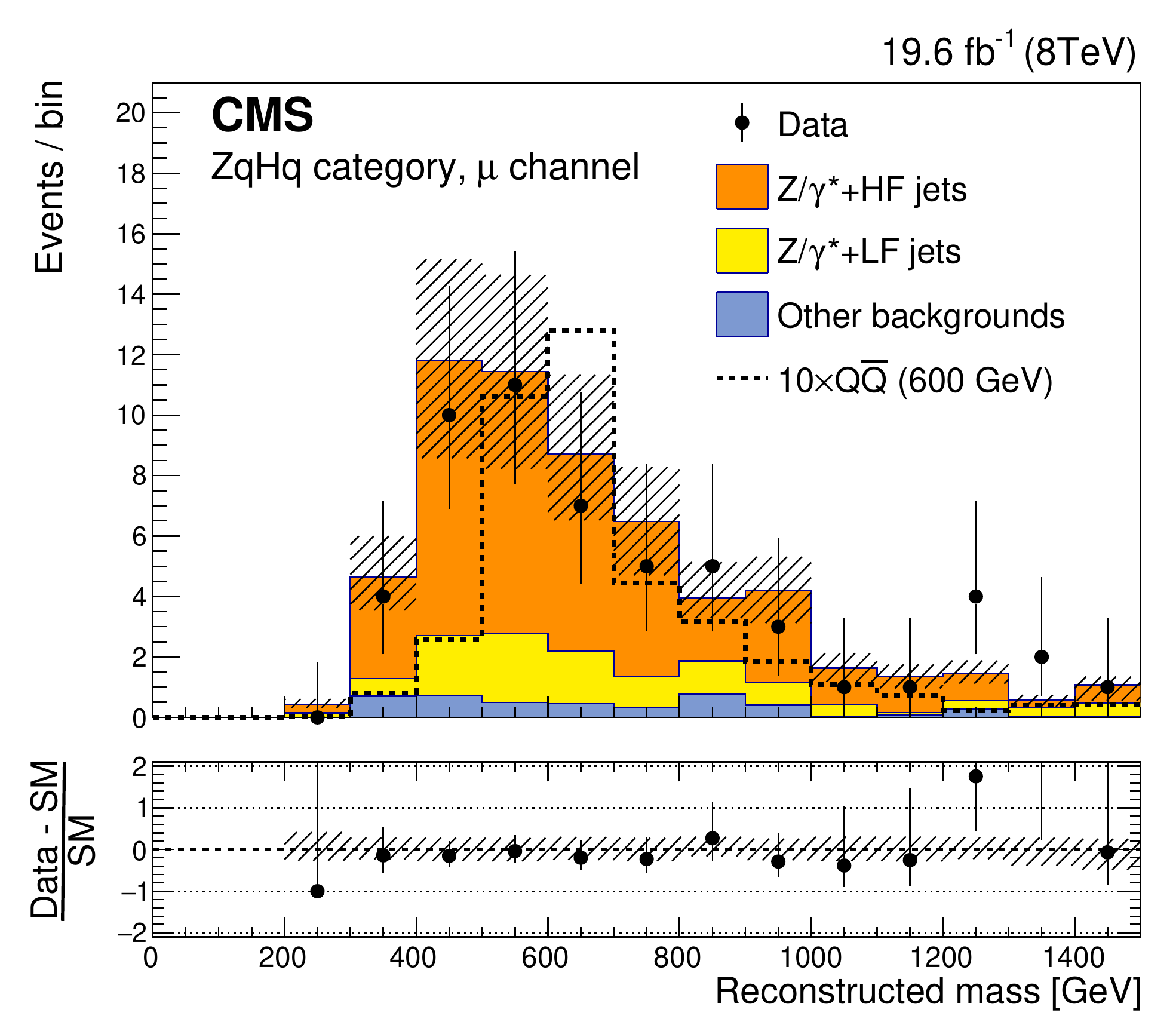}
 \includegraphics[width=0.48\textwidth]{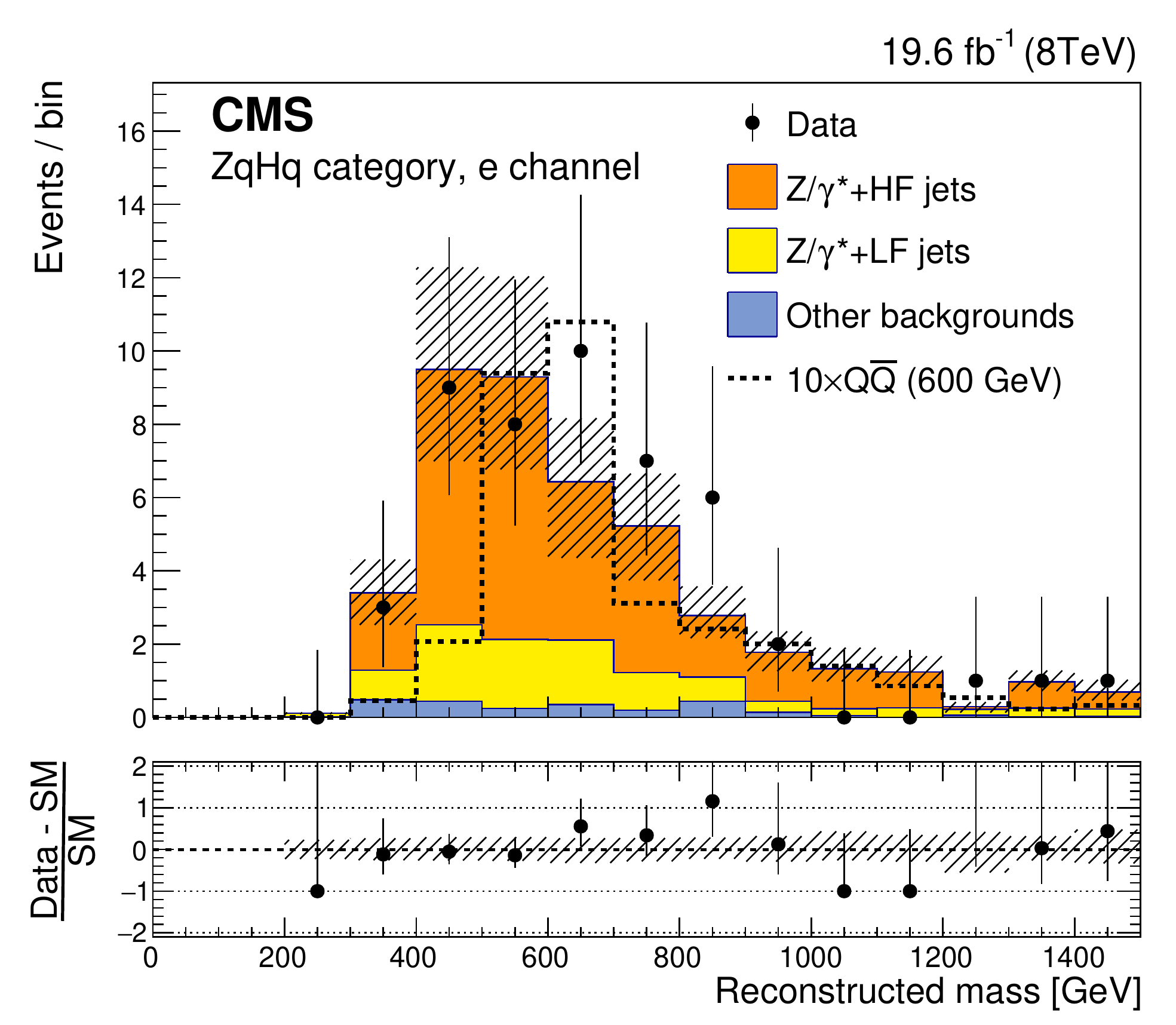}
\caption{The reconstructed mass of the VLQ candidate in the $\Z\PQq\PH\PQq$  event category, in the muon channel (\cmsLeft) and the electron channel (\cmsRight).
The contributions of simulated events where the $\PZ$ boson is produced in association
with light-flavor (LF) jets and heavy-flavor (HF) jets are shown separately.
The distribution for a heavy VLQ signal of mass 600\GeV and
$\tilde{\kappa}_\PW = 0.1$ (for $\mathcal{B}_\PW = 0.5$ and
$\mathcal{B}_\PZ = \mathcal{B}_\PH =0.25$) is scaled up by
a factor of 10 for visibility. The hatched bands represent
the combined statistical and systematic uncertainties.}
\label{fig:VLQmassZHqq}
\end{figure}

\begin{figure}[hbtp]
  \centering
 \includegraphics[width=0.48\textwidth]{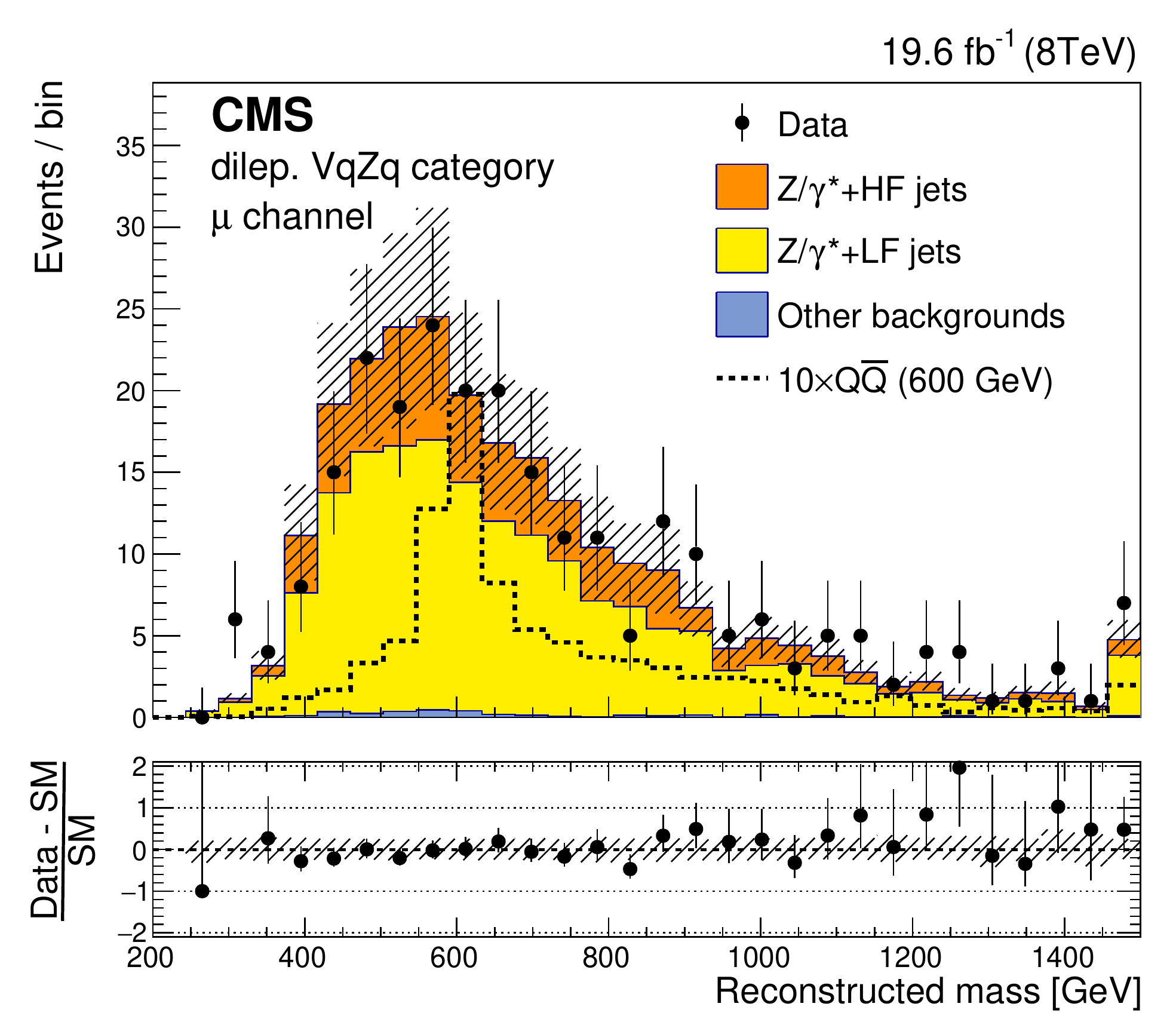}
 \includegraphics[width=0.48\textwidth]{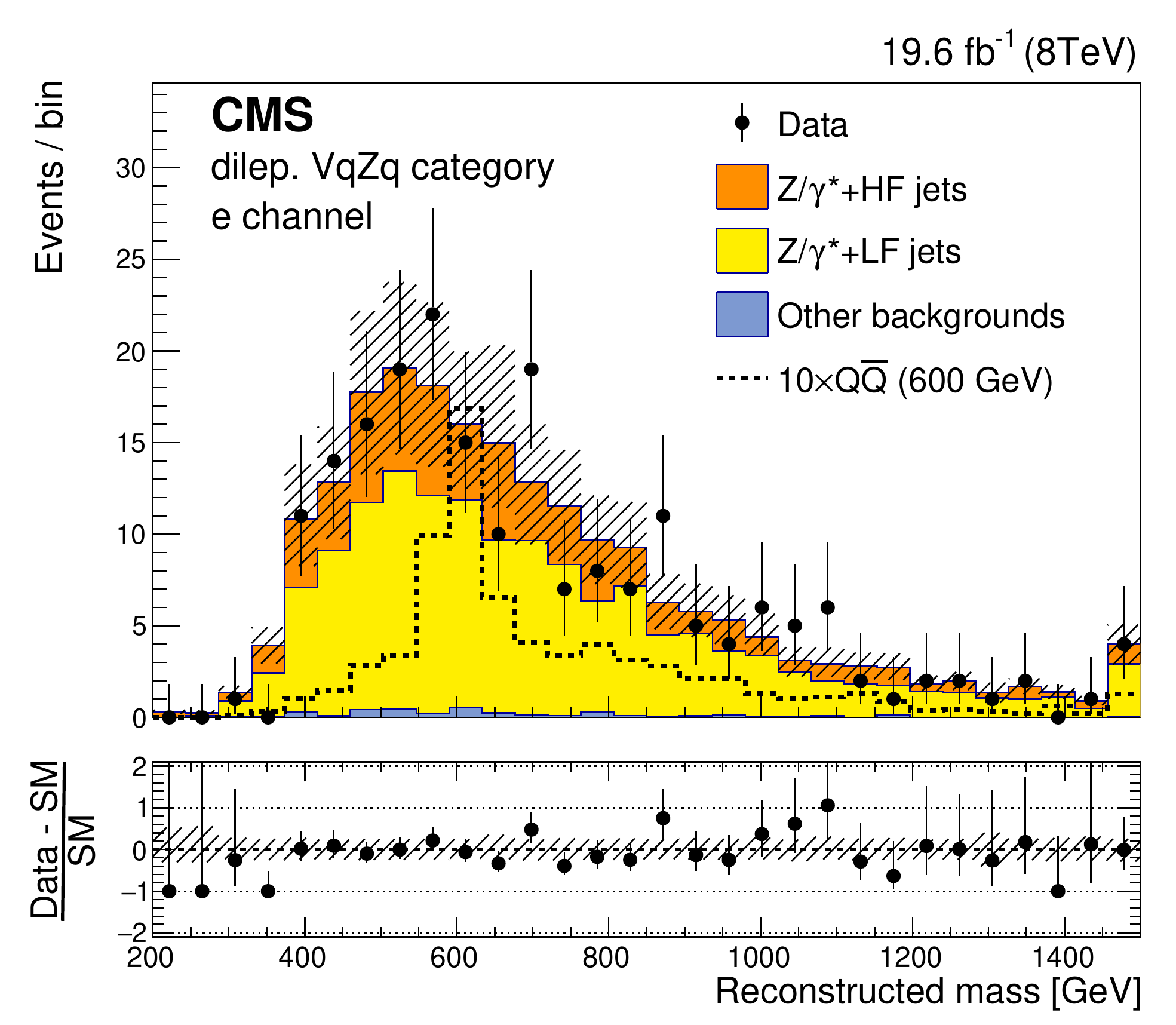}
  \caption{The reconstructed mass of the VLQ candidate in the
dileptonic $\cPV\PQq\Z\PQq$  event category, in the muon channel (\cmsLeft) and the electron
channel (\cmsRight). The contributions of simulated events where the $\PZ$ boson is produced in association
with light-flavor (LF) jets and heavy-flavor (HF) jets are shown separately.
The distribution for a heavy VLQ signal of
mass 600\GeV and $\tilde{\kappa}_\PW = 0.1$ (for $\mathcal{B}_\PW = 0.5$
and $\mathcal{B}_\PZ = \mathcal{B}_\PH = 0.25$) is scaled up
by a factor of 10 for visibility. The hatched bands represent
the combined statistical and systematic uncertainties.}
\label{fig:VLQmassSemilepVZqq}
\end{figure}

The VLQ candidate mass is reconstructed in the $\Z\PQq\PH\PQq$ and the dileptonic $\cPV\PQq\Z\PQq$ event categories
from two leptons forming a $\PZ$ boson candidate and a jet that potentially
corresponds to the light quark from the VLQ decay. For the latter,
we choose the highest $\pt$ anti-tagged jet with the largest $\Delta R$
separation from the $\PZ$ boson candidate.
The resulting mass distributions are shown in Fig.~\ref{fig:VLQmassZHqq} and Fig.~\ref{fig:VLQmassSemilepVZqq},
for the $\Z\PQq\PH\PQq$ and dileptonic $\cPV\PQq\Z\PQq$ categories, respectively.
The background
consists mainly of $\PZ$+jets events with a large contribution from those in which
the $\PZ$ boson is associated with heavy-flavor jets, because of the required presence of at least one b-tagged jet.

\begin{figure}[hbtp]
  \centering
\includegraphics[width=0.48\textwidth]{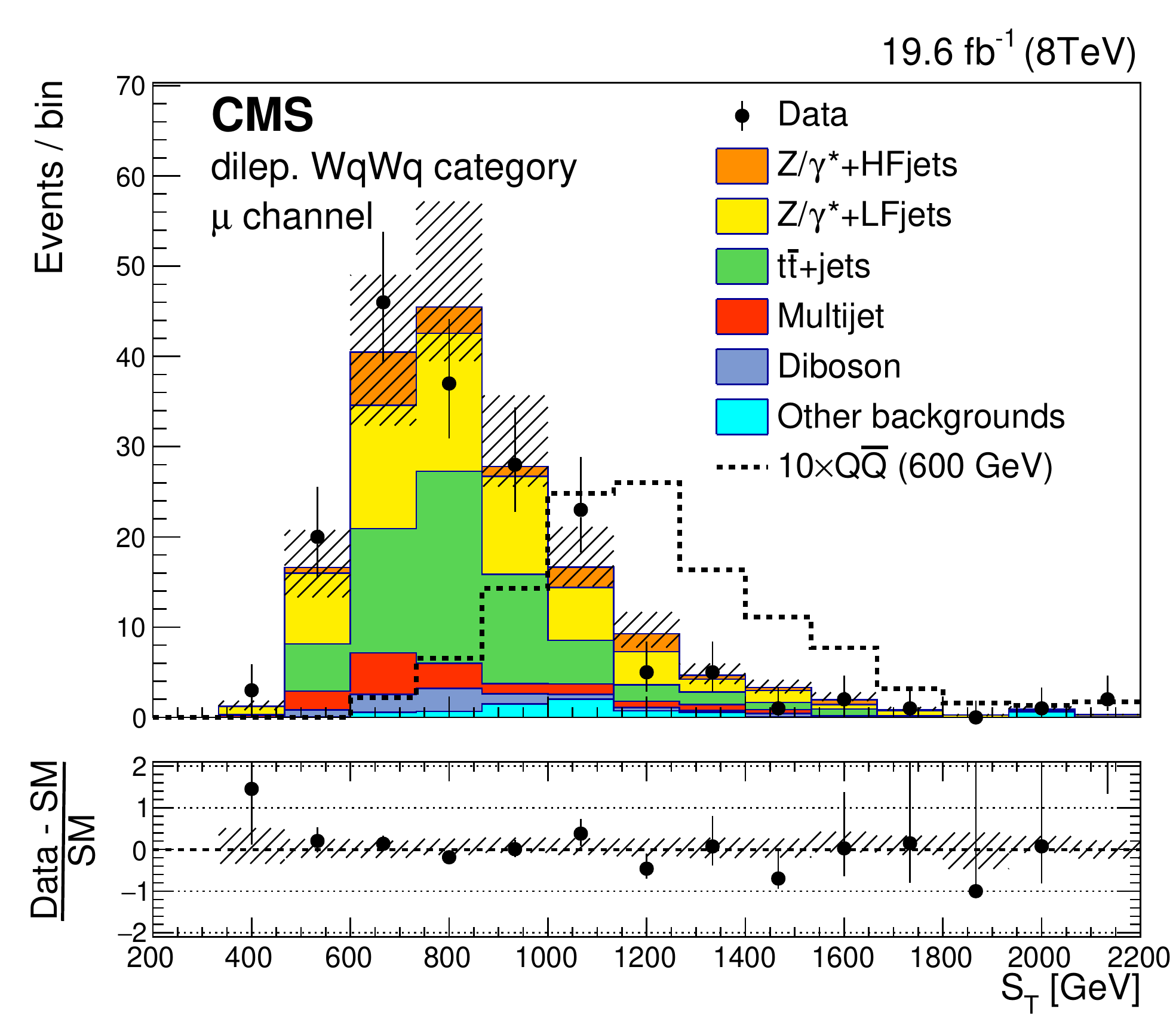}
\includegraphics[width=0.48\textwidth]{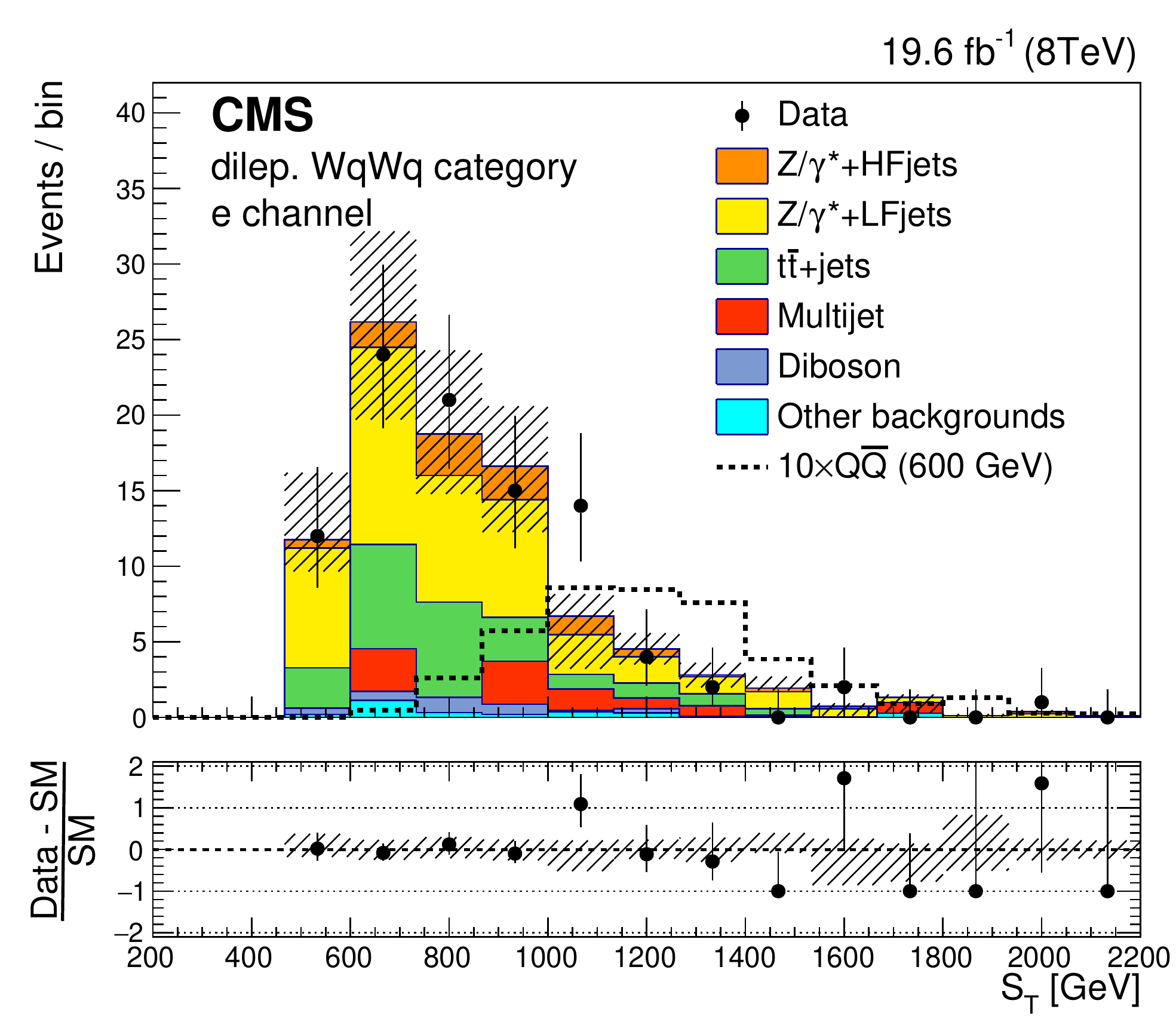}
\caption{The $S_T$ variable in the dileptonic $\PW\PQq\PW\PQq$ event category, in the muon channel
(\cmsLeft) and in the electron channel (\cmsRight).
The contributions of simulated events where the $\PZ$ boson is produced in association
with light-flavor (LF) jets and heavy-flavor (HF) jets are shown separately.
The distribution for a heavy VLQ signal of mass 600\GeV and $\tilde{\kappa}_\PW = 0.1$
(for $\mathcal{B}_\PW = 0.5$ and $\mathcal{B}_\PZ = \mathcal{B}_\PH = 0.25$)
is scaled up by a factor of 10 for visibility. The hatched bands represent
the combined statistical and systematic uncertainties.}
\label{fig:StWWqq}
\end{figure}

In the dileptonic $\PW\PQq\PW\PQq$ event category we use the $S_\mathrm{T}$ variable to discriminate
between SM and VLQ processes as shown in Fig.~\ref{fig:StWWqq}.
Since two neutrinos are present
in the topology of the dileptonic $\PW\PQq\PW\PQq$ event category, a full mass reconstruction is
not performed.

In the multileptonic $\cPV\PQq\Z\PQq$ event category (three or four leptons), the
number of events is too low to obtain a meaningful distribution.
Instead, the event count is used as observable.
The numbers of events observed and expected are summarized in
Table~\ref{tab:EventCountsFulllepVZqq}.
The main SM background originates from irreducible diboson and triboson
processes with three prompt charged leptons. We use control samples in data to
estimate the contribution from misidentified leptons passing the
tight-lepton selection criteria. This contribution is very small.

\begin{table*}[hbtp]
\centering
\topcaption{The total number of estimated background events compared
to the number of observed events, in the multileptonic $\cPV\PQq\Z\PQq$  event category, with either 3 or 4 leptons.
The numbers of expected signal events for two different signal hypotheses are shown.
The indicated uncertainties are statistical only, originating from the limited number of MC events.}
\label{tab:EventCountsFulllepVZqq}
\begin{scotch}{lc}
Irreducible background &	$0.4 \pm 0.1$   \\
Misidentified lepton background   &	$0.06 \pm 0.06$  \\[\cmsTableSkip]
Total background  &  $0.5 \pm 0.1$  \\
Observed & 2 \\[\cmsTableSkip]
Signal ($m_\Q = 600$\GeV, $\mathcal{B}_\PW = 0.5$, $\mathcal{B}_\PZ = 0.25$) & $2.1 $ \\
Signal ($m_\Q = 400$\GeV, $\mathcal{B}_\PZ = 1.0$) & $4.9 $  \\
\end{scotch} \end{table*}

\section{Combination}
\label{sec:combined}
We do not observe a significant excess of events over the
background prediction, and combine the results of the single and pair production searches by calculating upper
limits on the signal production cross sections and lower limits on the
mass of the VLQs.
The selection criteria defining the event categories optimized for single VLQ production and those optimized
for pair VLQ production, are orthogonal.
The discriminating observables for
the different event categories and the methods by which they are reconstructed are
summarized in Table~\ref{tab:discriminant}.
The distributions (templates) used in the limit calculation contain those of the observables in the single-lepton and dilepton event categories in the muon and the electron channel,
shown in Figs.~\ref{fig:VLQmass_Wminusqqdatadrivenbkg},~\ref{fig:VLQmassZqq},~\ref{fig:VLQmassZHqq},~\ref{fig:VLQmassSemilepVZqq}, and~\ref{fig:StWWqq}, where
the binning of the distributions is chosen in such a way that there are at
least 10 expected background events per bin.
In the single-lepton category optimized for VLQ pair production,
the distributions in Fig.~\ref{masses_1240} (upper left) and Fig.~\ref{qWqH_1240}
of the reconstructed mass are used.
In the event categories that require three and four leptons, we use the event counts of
Table~\ref{tab:EventCountsFulllepVZqq}.

\begin{table*}
\centering
 \topcaption{Discriminating variables used in the different event categories.
 The overlap of events in the $\PW\PQq\PW\PQq$ and $\PW\PQq\PH\PQq$ categories is removed, as explained in Section~\ref{sec:pairsinglelep}.}
\label{tab:discriminant}
\begin{scotch}{llll}
 Event category & Discriminating variable  &  Reconstructed using &  Shown in \\
\hline
$\PWm\PQq\PQq$  &  VLQ mass   & Lepton, neutrino, &
     Figs.~\ref{fig:VLQmassWqqplusminus},~\ref{fig:VLQmass_Wminusqqdatadrivenbkg}  \\
           &            & leading central jet &   \\[\cmsTableSkip]
$\Z\PQq\PQq$     &  VLQ mass & Two opposite-sign leptons, & Fig.~\ref{fig:VLQmassZqq} \\
        &               & leading central jet &   \\[\cmsTableSkip]
$\PW\PQq\PW\PQq$  & VLQ mass &  Kinematic fit,   & Figs.~\ref{masses},~\ref{masses_1240}  \\
                        &   &  see Section~\ref{sec:pairsinglelep}   &    \\[\cmsTableSkip]
$\PW\PQq\PH\PQq$ & VLQ mass &  Kinematic fit,   & Figs.~\ref{masses},~\ref{masses_1240},~\ref{qWqH_1240}  \\
                        &   &  see Section~\ref{sec:pairsinglelep}   &    \\[\cmsTableSkip]
$\Z\PQq\PH\PQq$   &  VLQ mass & Two opposite-sign leptons, & Fig.~\ref{fig:VLQmassZHqq} \\
        &               & high-$\pt$ anti-tagged,  &  \\
        &               &  jet with the largest  &  \\
        &               & $\Delta R$ separation from &  \\
        &               & the $\PZ$ boson candidate & \\[\cmsTableSkip]
dileptonic     &   VLQ mass    & Two opposite-sign leptons, & Fig.~\ref{fig:VLQmassSemilepVZqq} \\
$\cPV\PQq\Z\PQq$ &         & high-$\pt$ anti-tagged,  &  \\
        &               &  jet with the largest  &  \\
        &               & $\Delta R$ separation from &  \\
        &               & the $\PZ$ boson candidate & \\[\cmsTableSkip]
dileptonic   & $S_\mathrm{T}$ & See Section~\ref{sec:single_strateg}  & Fig.~\ref{fig:StWWqq} \\
$\PW\PQq\PW\PQq$ &         &     &  \\[\cmsTableSkip]
multileptonic   & Event count  & See Section~\ref{sec:pairmultilep} & Table~\ref{tab:EventCountsFulllepVZqq} \\
$\cPV\PQq\Z\PQq$ &   &  \\						
\end{scotch}
\end{table*}

The limit calculation is performed
using a Bayesian interpretation~\cite{PDG}.
Systematic uncertainties are taken into account as nuisance parameters.
For uncertainties affecting the shapes of the variables used in the search,
alternative templates are produced by varying each source of uncertainty within
$\pm$1 standard deviation, and associating the varied templates with
Gaussian prior constraints of the corresponding nuisance parameters.
Uncertainties affecting only the normalization
are included, using log-normal prior constraints.
A flat prior probability density function on the total signal yield is assumed.
The likelihood function is marginalized with respect to the nuisance
parameters representing the systematic uncertainties that arise from shape
and global normalization variations. The shapes of the background and signal templates
vary with the appropriate nuisance parameters.
Statistical uncertainties associated with the simulated distributions are also included
in this procedure using the Barlow-Beeston light method~\cite{barlow}.

\subsection{Systematic uncertainties}

The uncertainties in the $\ttbar$ total cross section, electroweak and
multijet background yields, integrated luminosity, lepton
efficiencies, the choice of PDFs, and constant
data-to-simulation scale factors affect only the normalization.
Uncertainties that affect the shape and normalization of the distributions include those in
the jet energy scale, jet energy resolution, $\ptmiss$
resolution, \PQb tagging efficiency, QGT tagging efficiency, number of additional $\Pp\Pp$
interactions per bunch crossing, and the factorization and renormalization scales assumed in the simulation.
Some of the uncertainties listed above have a negligible impact on the distributions
and are neglected in the limit calculation.

The main backgrounds are $\ttbar$, $\PW$+jets,
and $\PZ$+jets production. A 15\% uncertainty in the cross section for $\ttbar$
production is taken from the CMS measurement~\cite{cms_ttbar_diff}.
In the single-production event categories as well as the pair-production categories with multiple leptons,
we use values for the normalization
uncertainty in the $\PW$+jets and $\PZ$+jets background
contributions, which are obtained from estimates based on data. The values
are 20\% for the light-flavor component, and 30\% for the heavy-flavor component.
These uncertainties are estimated to cover the changes in the normalizations induced by modifying
the kinematic requirements that define the control samples.
The uncertainties corresponding to the normalization of the smaller single top quark, diboson,
$\ttbar \cPZ$+jets, $\ttbar \PW$+jets, and triboson backgrounds in these categories
are taken from the corresponding experimental measurements or the theoretical calculations.
In the single-lepton pair-production categories, in which a kinematic fit is performed, the normalization of the non-$\ttbar$ background
processes has been assigned an uncertainty of 50\%, reflecting the large uncertainty
in the heavy-flavor component of the $\PW$+jets process and in other background processes, in
the high-$S_\mathrm{T}$ signal region.

The integrated luminosity has an uncertainty of 2.6\%~\cite{cms_lumi_summer2013}.
Trigger efficiencies, lepton
identification efficiencies, and data-to-simulation scale factors are obtained from data
using the decays of \cPZ\ bosons to lepton pairs. The uncertainties associated
with all of these lepton related sources are included in the selection efficiency uncertainty,
and together they amount to a total uncertainty of 3\%.

The PDF uncertainties are estimated
by varying up and down by one standard deviation the CTEQ6 PDF set parameters.
Only the changes in acceptance caused by these uncertainties, not the change in total cross section, are propagated.
For each simulated event, the weight corresponding to each varied PDF parameter is calculated,
and an envelope for the distributions of the observables is created by taking the maximum and minimum of the variations bin by bin.
This results in a normalization uncertainty of 1.4\% for the signal and 8\% for the background,
with a negligible impact on the shape of the distributions.

The uncertainty in the jet energy scale is evaluated by
scaling the jet energy in the simulation by the $\eta$ and $\pt$ dependent uncertainties, ranging from 0.5\% to 2.3\%~\cite{Khachatryan:2016kdb}.
The $\eta$ dependent scale factors that smear the jet energy resolution
are varied within their uncertainty, changing the scale factors between 2.4\% and 3.8\%
depending on the absolute value of $\eta$.
Both  AK5 and CA8 jet collections are subject to these variations.
The systematic variations on the jet energy scale and resolution
are applied before the splitting of the CA8 jets in subjets.
The variations for subjets are done proportionally to the variations of their parent CA8 jet.

The changes in jet momentum resulting from the AK5 jet energy scale variations
are propagated to the $\ptmiss$. The effect of the unclustered energy uncertainty on
$\ptmiss$ is evaluated by varying the unclustered energy by $\pm$10\%,
and is found to be negligible.

The systematic uncertainty in the \PQb tagging efficiency
is estimated by varying the data-to-simulation scale factors, for both medium and loose working points,
within their uncertainty, separately for heavy-flavor (\PQb and \PQc) jets
and light-flavor jets.
The relative precision on the heavy-flavor scale factors is 2--4\% for $\pt$ below 120 GeV and about
5--9\% at the highest jet transverse momenta considered~\cite{CMS-PAS-BTV-13-001}.
The scale factors for light-flavor jets are measured with a precision of about 5--13\%.
For the evaluation of the systematic uncertainty originating from the QGT,
the QGT discriminant values of the jets in the simulation are smeared, depending on the flavor,
$\pt$, and $\eta$ of the jet~\cite{quark_gluon}.
The observed variations in the number of selected events, and in particular the variations in the signal
inefficiencies, are very small and neglected.

To evaluate the uncertainty related to the modeling of multiple
interactions in the same bunch crossing, the average number of interactions
in the simulation is varied by $\pm$5\% relative to the nominal value.
The impact of these variations on the distributions is found to be negligible.

The uncertainty in the factorization and renormalization scales assumed in the simulated $\ttbar$ sample
is estimated by varying the scales simultaneously by a factor of two, and a factor of one half, relative to the nominal value.

Several systematic uncertainties affect the backgrounds estimated
from control regions in data. The uncertainty in the estimated
misidentified lepton background is considered in the multileptonic $\cPV\PQq\Z\PQq$
category. In the $\PWm\PQq\PQq$ category, an uncertainty is assigned in the shape
correction applied to the reconstructed mass distribution, which accounts
for the different selection requirements between the control region and
the signal region. The templates modeling this uncertainty are
constructed by applying a shape correction twice the size of the nominal
correction, and not applying a shape correction.

The systematic uncertainties in the integrated luminosity, the lepton
efficiency scale factors, the jet energy scale, the jet energy resolution,
and the \PQb tag efficiency and mistag rate scale factor uncertainties are
considered as fully correlated across all channels.
The uncertainties in the normalization of the different background processes
are considered as uncorrelated among the event categories that make use of a kinematic fit and those that do not,
because of the different signal selection procedures.
The expected and observed mass limits change by
less than 5\GeV when treating the $\ttbar$+jets normalization uncertainty
as completely correlated across all categories.

\section{Results}
\label{sec:results}

\begin{figure}
  \centering
  \includegraphics[width=0.48\textwidth]{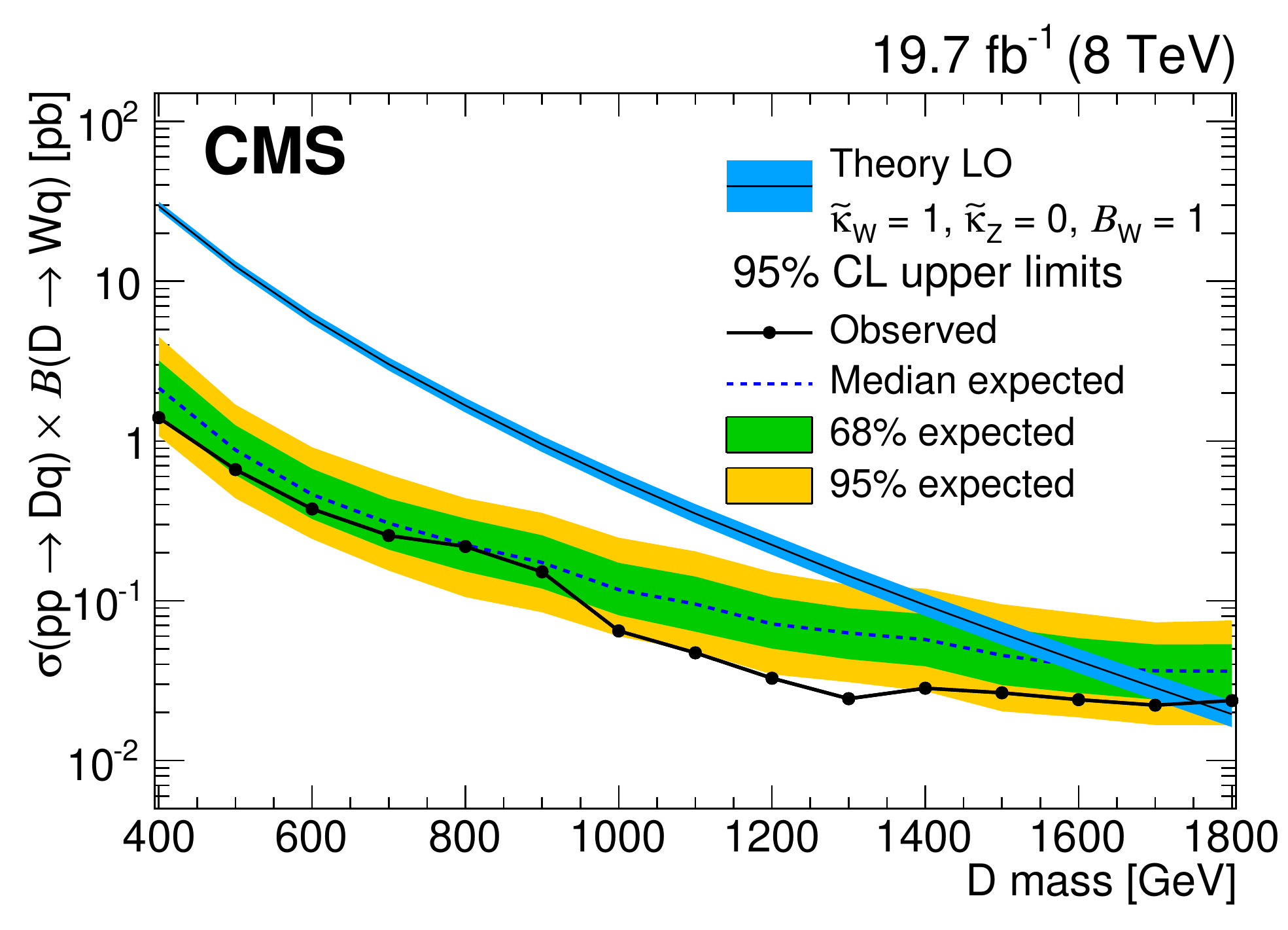}
  \includegraphics[width=0.48\textwidth]{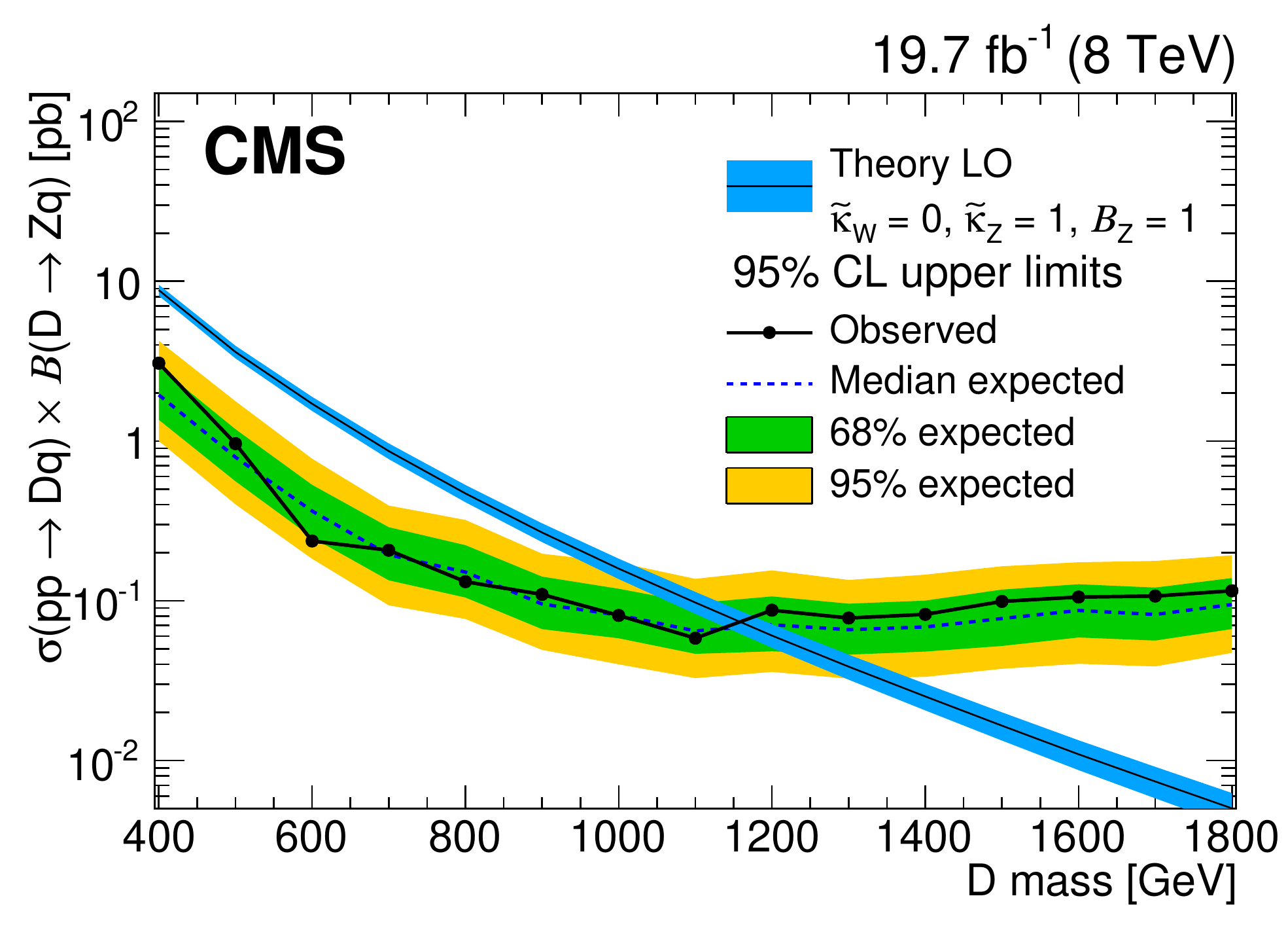}
  \caption{The 95\% \CL exclusion limits on the product of the production cross section
and the branching fraction, considering only single production of down-type VLQs,
and assuming a neutral current coupling of zero (\cmsLeft) or a charged current coupling of zero (\cmsRight).
The median expected and observed exclusion limits are indicated
  with a dashed and a solid line, respectively. The inner (green) band and the outer (yellow) band
  indicate the regions containing 68 and 95\%, respectively, of the distribution of limits expected under the background-only hypothesis.
The corresponding LO theory predictions are superimposed.
The predictions are represented by a solid black line centered within a blue band, which shows the uncertainty of the calculation.
The uncertainties are determined based on the choice of PDF set along with the renormalization and factorization scales.}
  \label{fig:xsection_vs_mass_singleonly}
\end{figure}

The 95\% confidence level (\CL) limit on the product of the production
cross section and the branching fraction as a function
of the VLQ mass, considering only single production of down-type VLQs and the corresponding optimized categories,
is shown in Fig.~\ref{fig:xsection_vs_mass_singleonly}.
The \cmsLeft{} (\cmsRight) plot shows the scenario
where a nonzero $\tilde{\kappa}_\PW$ ($\tilde{\kappa}_\PZ$)
is considered while setting $\tilde{\kappa}_\PZ = 0$
($\tilde{\kappa}_\PW = 0$) and including only the $\PWm\PQq\PQq$ (Zqq) event
category in the limit setting procedure. The LO theoretical predictions
for the cross section are superimposed.
The scale uncertainty in the prediction was estimated by comparing the effect of either doubling or halving the central value of the scale.
The PDF uncertainty is determined using the 44 eigenvectors of the CTEQ66 PDF
set~\cite{cteq66}.
A mass of 1755 (1620)\GeV is
observed (expected) to be excluded at the 95\% \CL for $\tilde{\kappa}_\PW = 1.0$
and $\mathcal{B}_\PW = 1$, and a mass of 1160 (1170)\GeV is observed
(expected) to be excluded at the 95\% \CL for $\tilde{\kappa}_\PZ = 1.0$ and
$\mathcal{B}_\PZ = 1$.

\begin{figure}[hbtp]
  \centering
   \includegraphics[width=0.48\textwidth]{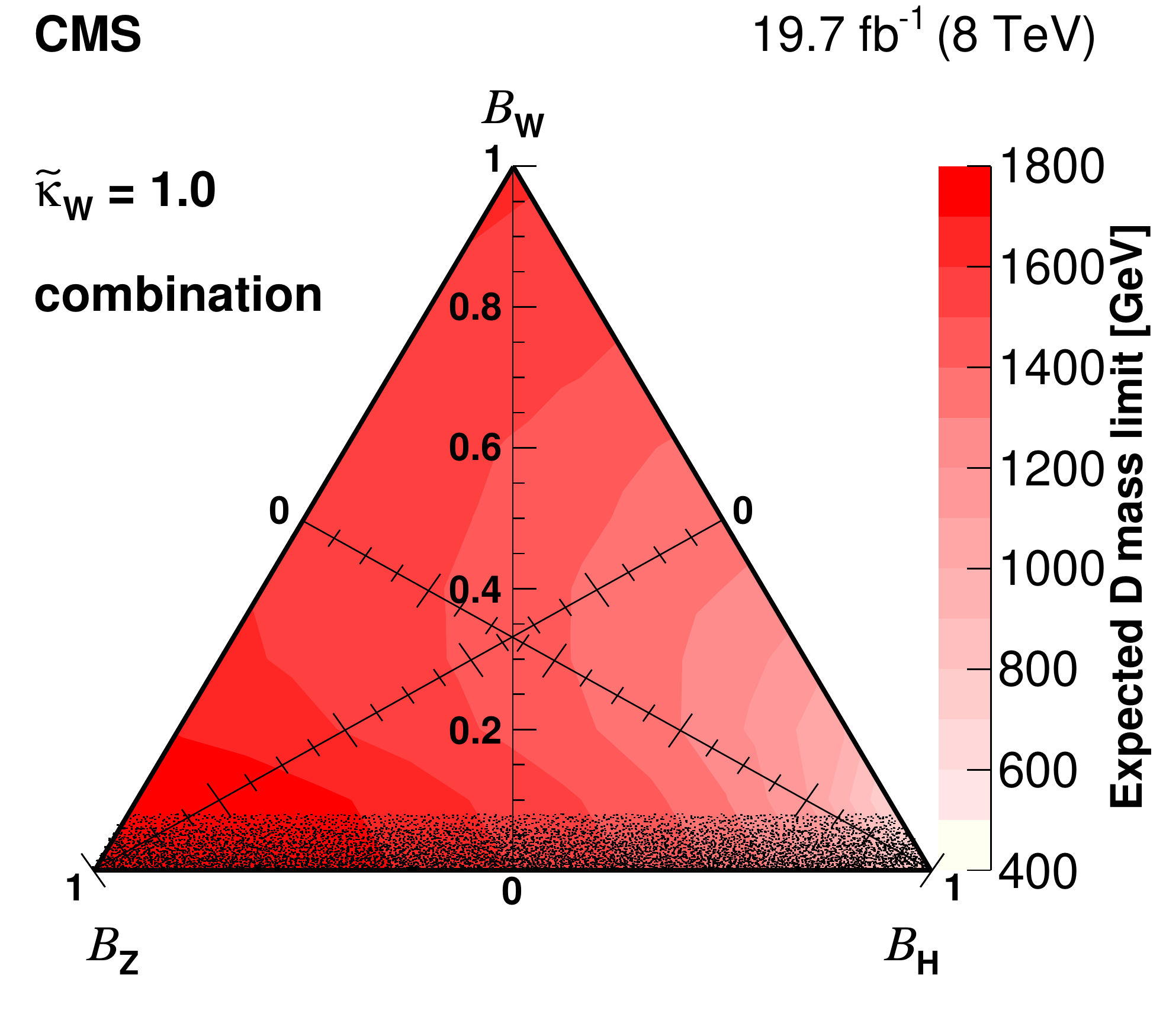}
   \includegraphics[width=0.48\textwidth]{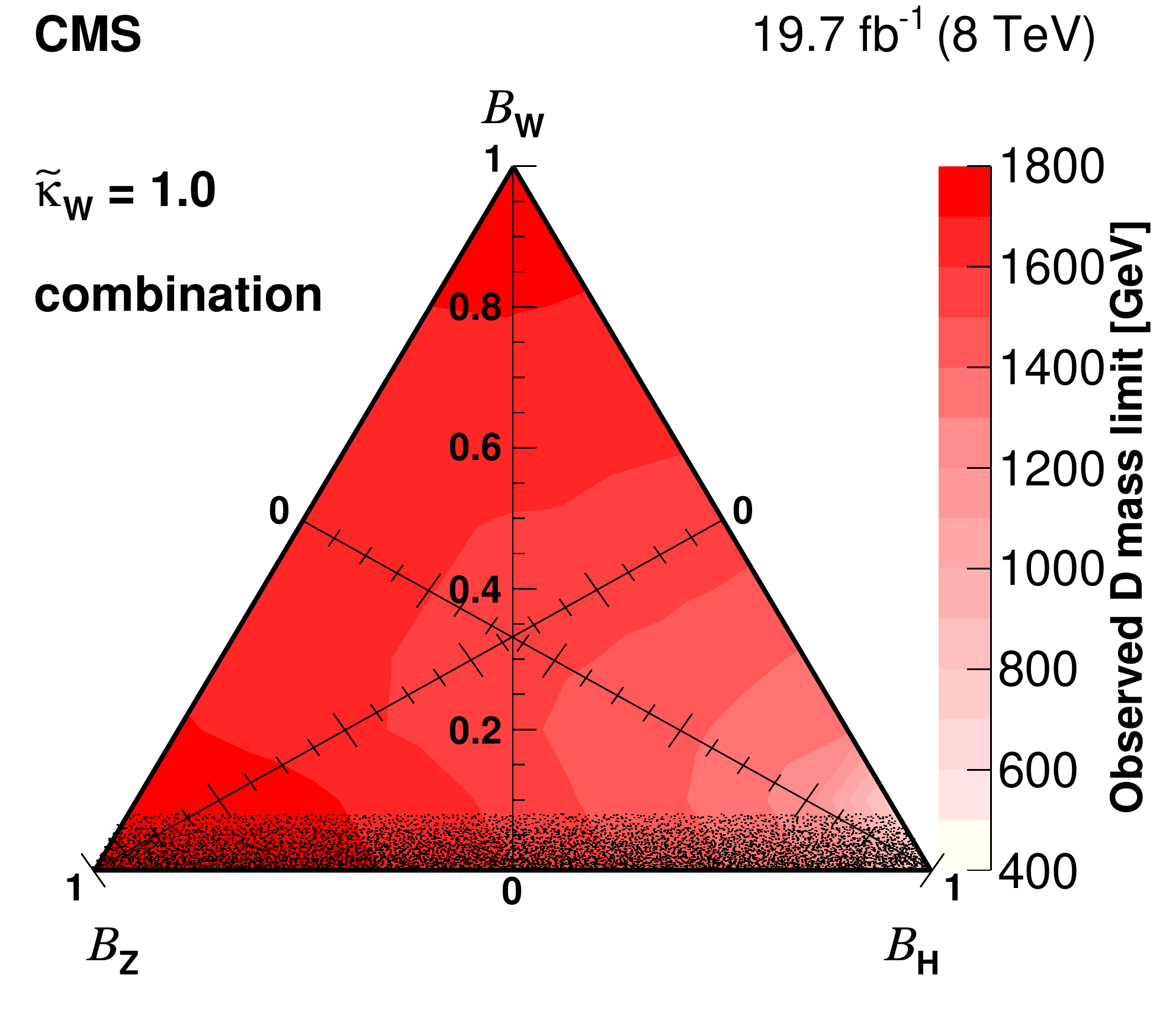}
  \caption{The median expected (\cmsLeft) and observed (\cmsRight) combined lower mass
  limits represented in a triangular form, where each point of the triangle corresponds
  to a given set of branching fractions for the decay of a VLQ into a boson
  and a first-generation quark.
  The limit contours are determined assuming that $\tilde{\kappa}_\PW = 1.0$,
  which means that the signal is dominated by electroweak single production.
  The black shaded band near $\mathcal{B}_\PW = 0$ shows a region where the results cannot
  be reliably interpreted because $\tilde{\kappa}_\PZ$ diverges, as explained in the text.
\label{fig:BFplots_D_kappa1_combination}}
\end{figure}

\begin{figure}[hbtp]
  \centering
 \includegraphics[width=0.48\textwidth]{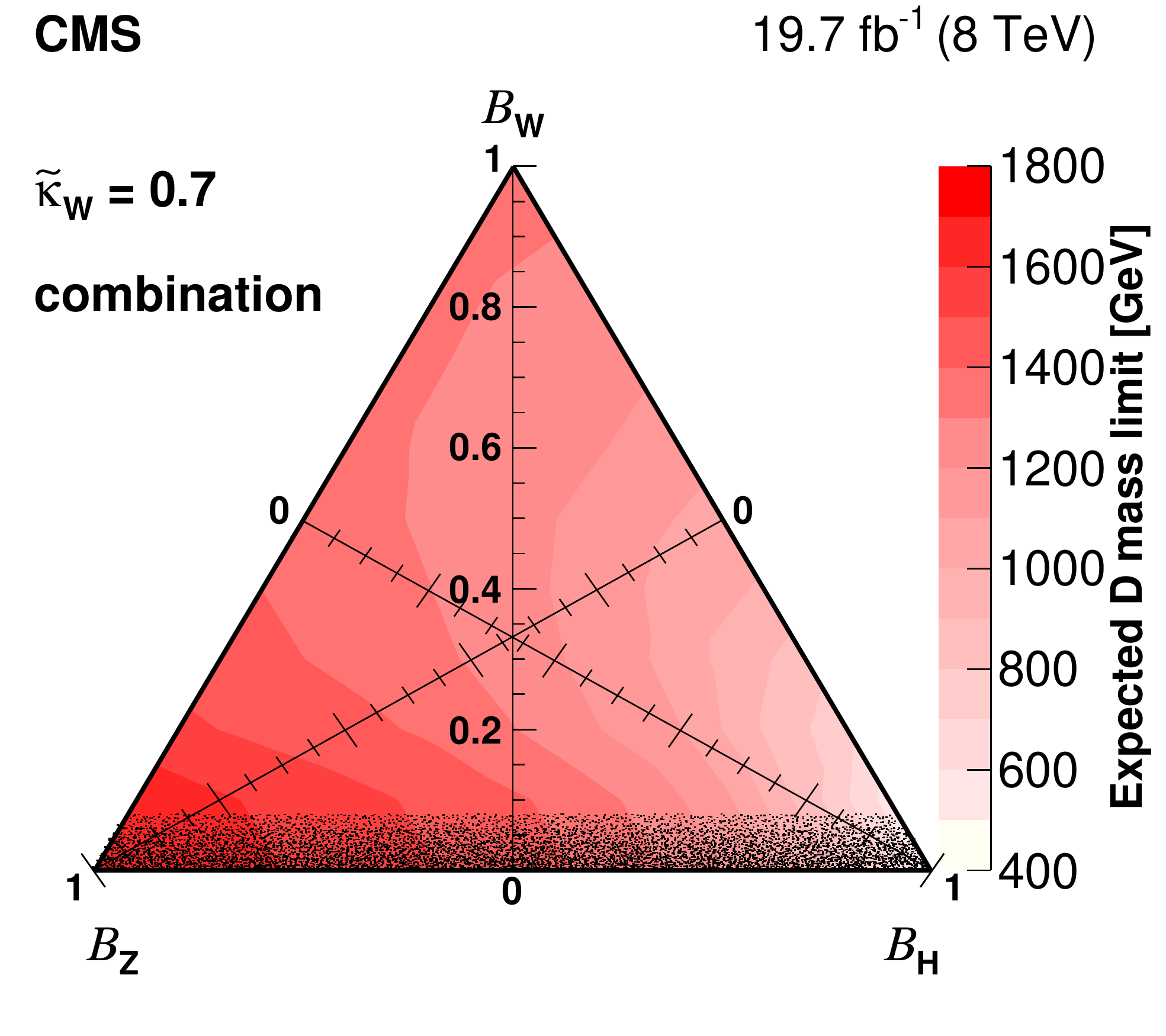}
   \includegraphics[width=0.48\textwidth]{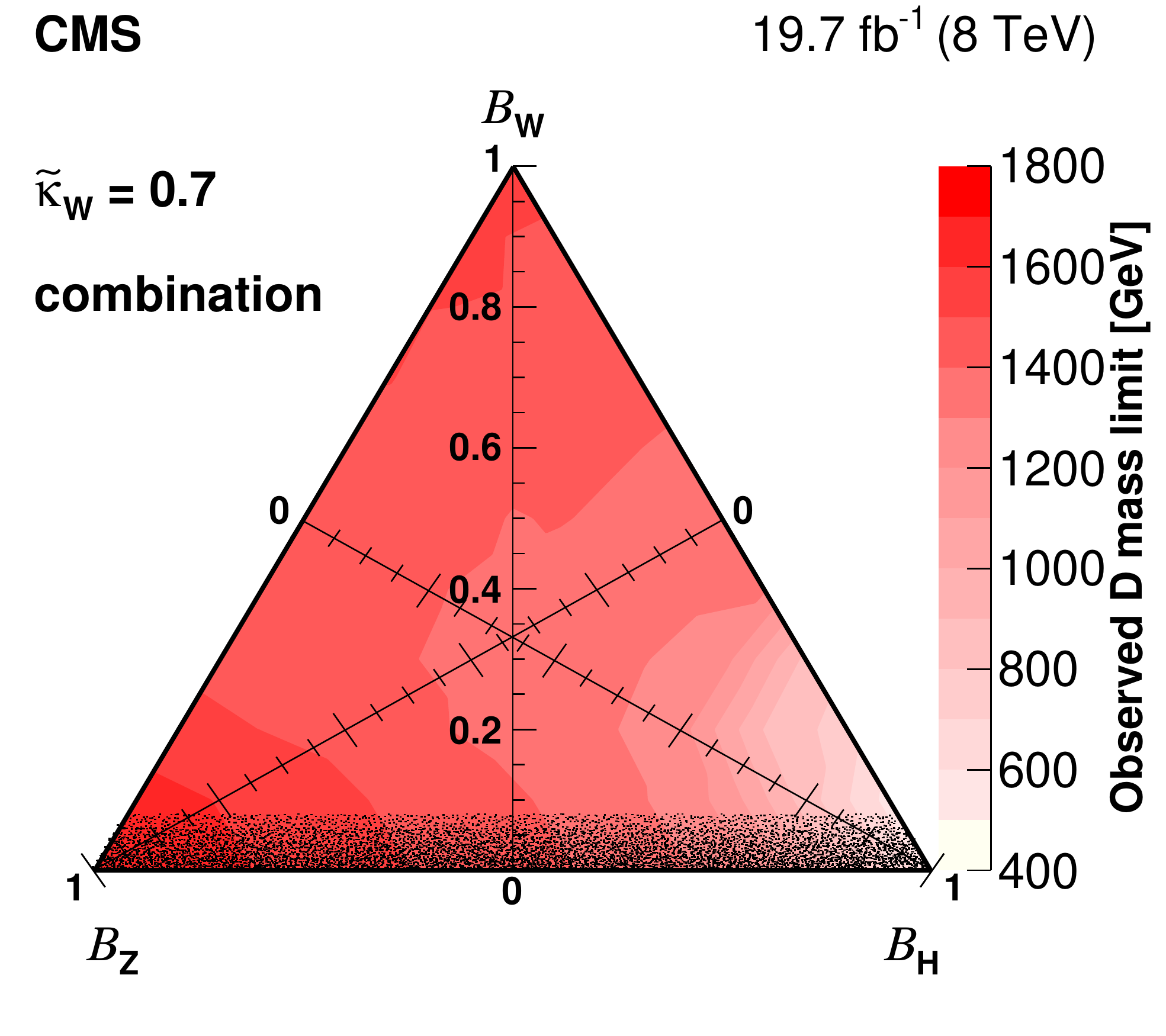}
  \caption{The median expected (\cmsLeft) and observed (\cmsRight) combined lower mass
  limits represented in a triangular form, where each point of the triangle corresponds
  to a given set of branching fractions for a VLQ decaying into a boson and
  a first-generation quark.
  The limit contours are determined assuming $\tilde{\kappa}_\PW = 0.7$,
  which means that the signal will be dominated by electroweak single production
  for most of the parameter space represented by the triangles.
  The black shaded band near $\mathcal{B}_\PW = 0$ represents a region where results cannot
  be reliably interpreted because $\tilde{\kappa}_\PZ$ diverges, as explained in the text.
\label{fig:BFplots_D_kappa0p7_combination}}
\end{figure}

\begin{figure}[hbtp]
  \centering
  \includegraphics[width=0.48\textwidth]{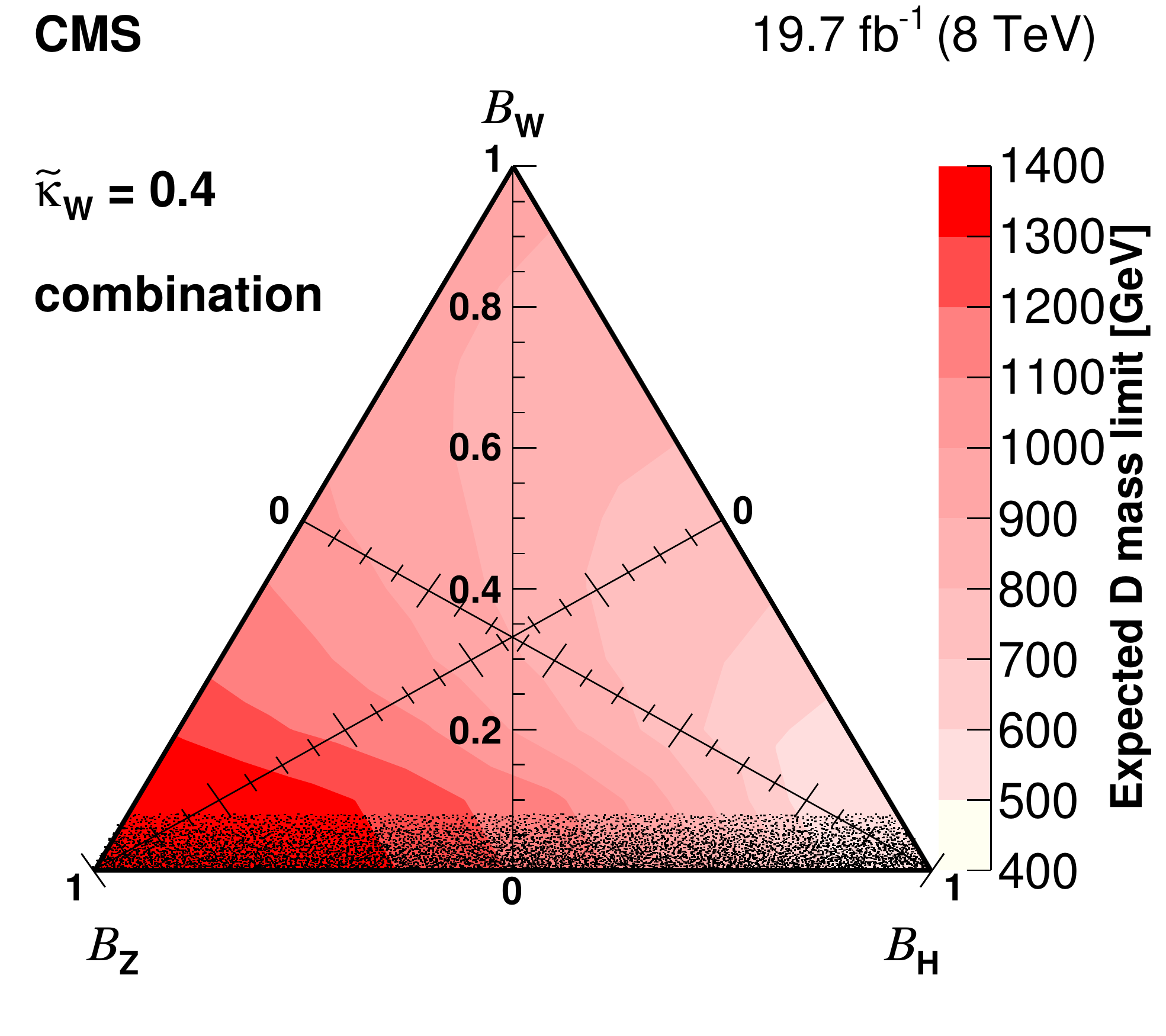}
   \includegraphics[width=0.48\textwidth]{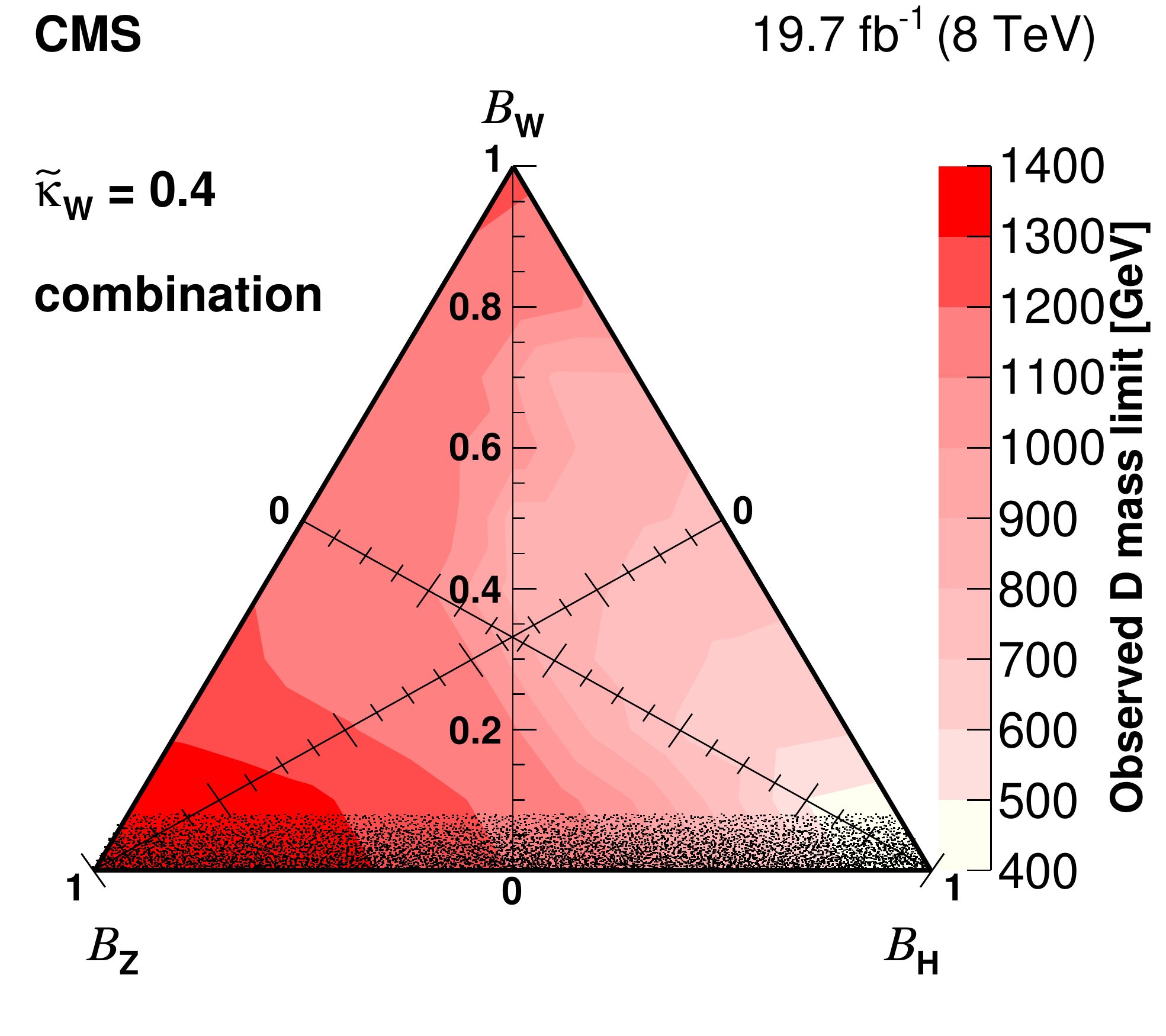}
  \caption{The median expected (\cmsLeft) and observed (\cmsRight) combined lower mass
  limits represented in a triangular form, where each point of the triangle
  corresponds to a given set of branching fractions for a VLQ
  decaying into a boson and a first-generation quark.
  The limit contours are determined assuming $\tilde{\kappa}_\PW = 0.4$,
  which means that the signal is dominated by electroweak single production
  in most of the parameter space represented by the triangles, but in which
  the relative importance of the pair-produced signal has increased.
  The black shaded band near $\mathcal{B}_\PW = 0$ represents a region where
  results cannot be reliably interpreted because $\tilde{\kappa}_\PZ$
  diverges, as explained in the text.
\label{fig:BFplots_D_kappa0p4_combination}}
\end{figure}

\begin{figure}[hbtp]
  \centering
 \includegraphics[width=0.48\textwidth]{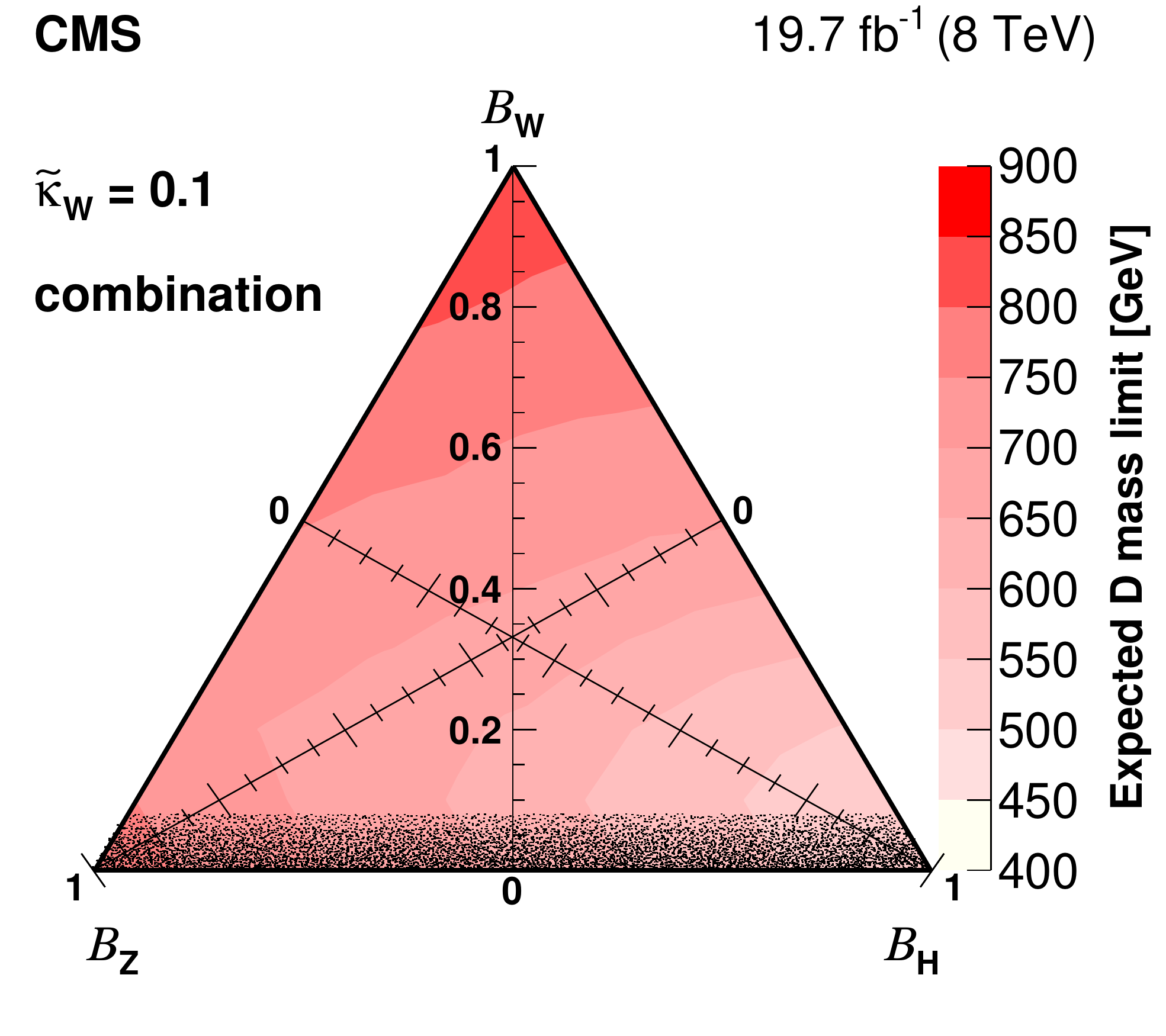}
  \includegraphics[width=0.48\textwidth]{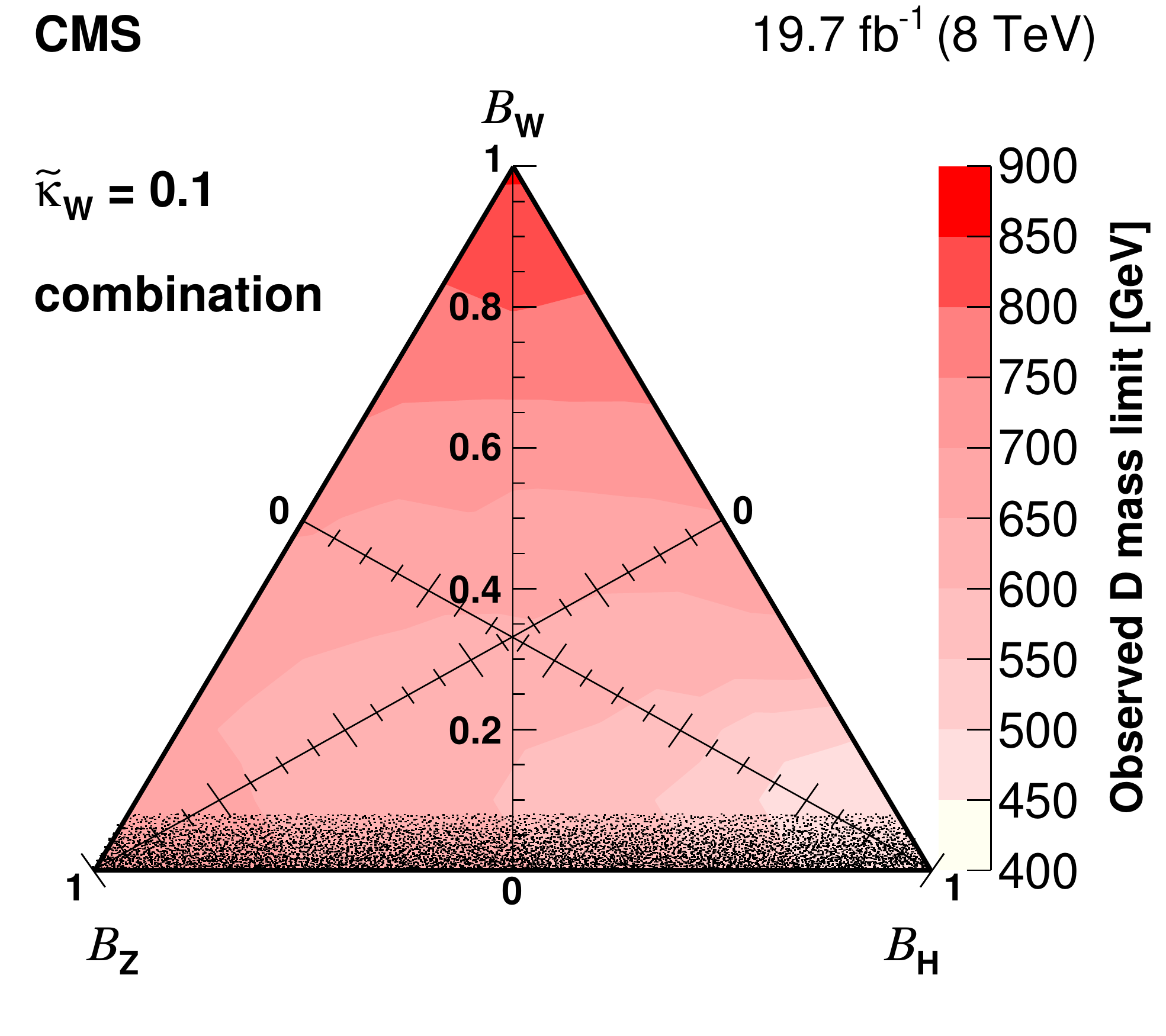}
  \caption{The median expected (\cmsLeft) and observed (\cmsRight) combined lower mass
  limits represented in a triangular form, where each point of the triangle corresponds
  to a given set of branching fractions for a VLQ decaying into a boson and
  a first-generation quark.
  The limit contours are determined assuming $\tilde{\kappa}_\PW = 0.1$,
  which means that the signal is dominated by strong pair production for
  most of the parameter space represented by the triangles.
  The black shaded band near $\mathcal{B}_\PW = 0$ indicates a region where
  results cannot be reliably interpreted because $\tilde{\kappa}_\PZ$ diverges, as explained in the text.
\label{fig:BFplots_D_kappa0p1_combination}}
\end{figure}

\begin{figure}[hbtp]
  \centering
  \includegraphics[width=0.48\textwidth]{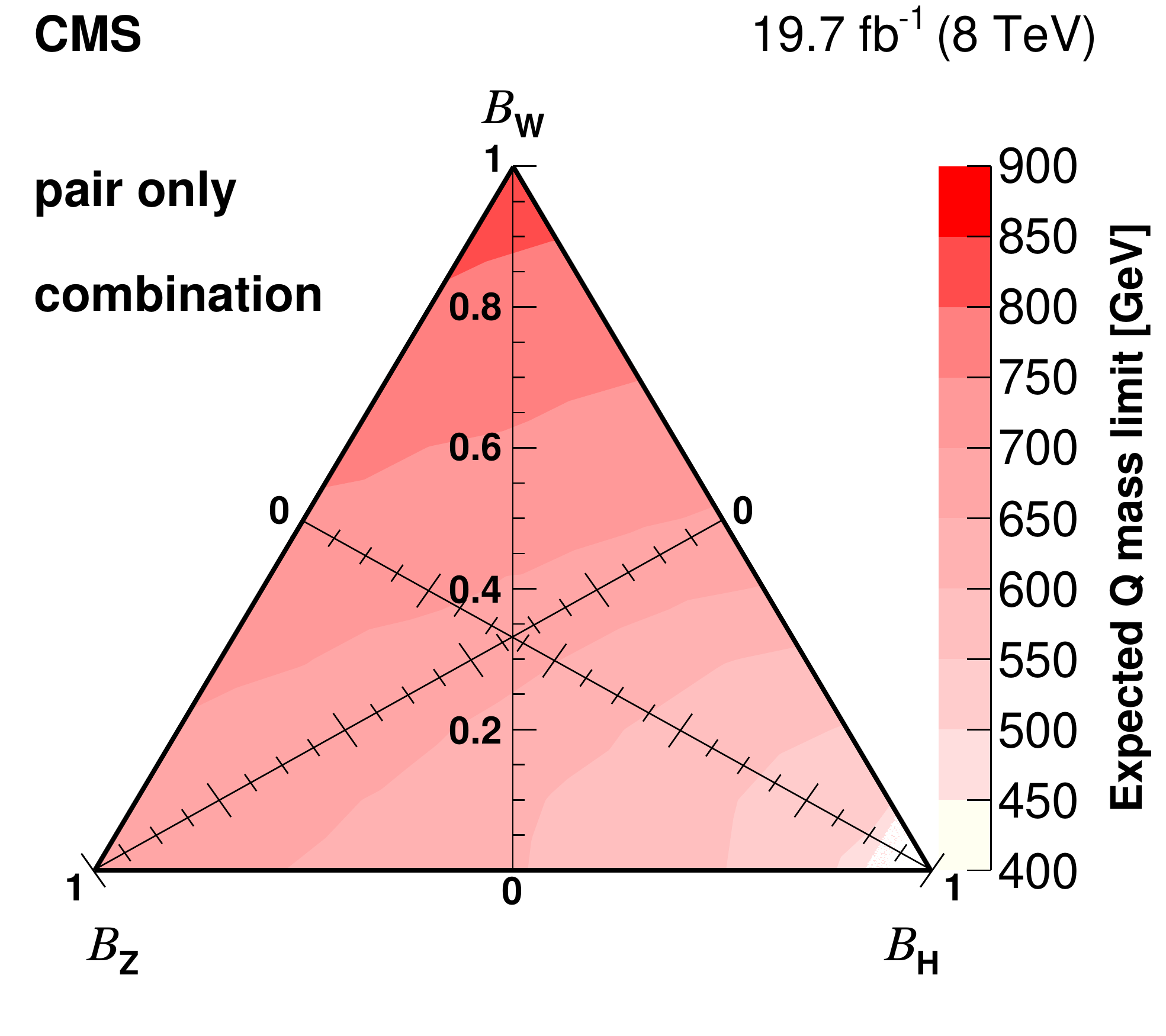}
  \includegraphics[width=0.48\textwidth]{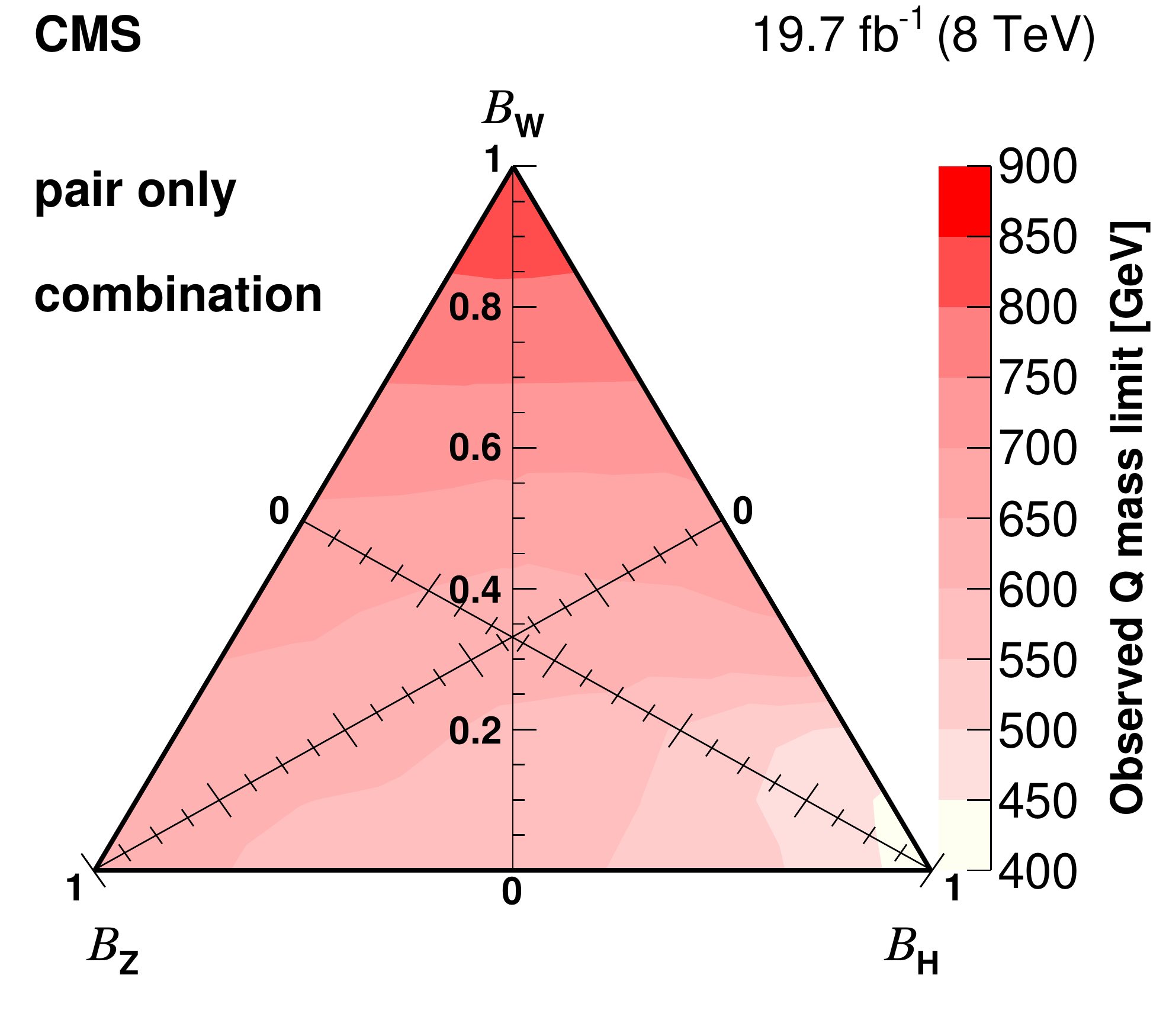}
  \caption{The median expected (\cmsLeft) and observed (\cmsRight) combined lower mass
  limits represented in a triangular form, where each point of the triangle corresponds
  to a given set of branching fractions for a VLQ decaying into a boson and
  a first-generation quark.
  The limit contours are determined assuming that $\tilde{\kappa}_\PW$
  and $\tilde{\kappa}_\PZ$ are so small that the single-production modes
  can be neglected, and therefore that the heavy quarks can only be produced in pairs
  via strong interaction. The white area in the triangle with expected limits indicates mass limits below 400\GeV.
\label{fig:BFplots_D_paironly_combination}}
\end{figure}

The results of the branching fraction scans for the charged-current VLQ single-production
coupling parameters $\tilde{\kappa}_\PW = 1.0$, $\tilde{\kappa}_\PW = 0.7$,
$\tilde{\kappa}_\PW = 0.4$, $\tilde{\kappa}_\PW = 0.1$ are shown
in Figs.~\ref{fig:BFplots_D_kappa1_combination} to~\ref{fig:BFplots_D_kappa0p1_combination}.
For values of $\tilde{\kappa}_\PW = 1.0$ and 0.7, single production is by far the dominant
signal production mode, while the relative importance
of the pair-production mode increases in much of the parameter space
for $\tilde{\kappa}_\PW = 0.4$, and even more so for $\tilde{\kappa}_\PW = 0.1$.
The black shaded region below $\mathcal{B}_\PW \approx 0.1$ in each
branching fraction triangle indicates the region where care should be taken
with the interpretation of the results. In this region, $\mathcal{B}_\PW$ approaches 0,
and as explained in Section~\ref{sec:single_strateg}, the neutral-current single-production
strength parameter $\tilde{\kappa}_\PZ$ diverges and the limits cannot be calculated.
Results for an alternative single-production coupling parametrization that does not exhibit
divergent behavior throughout the scan are available in tabulated form in \suppMaterial.
The results from a branching fraction scan based on the pair-production data
alone are shown in Fig.~\ref{fig:BFplots_D_paironly_combination}.
The lower limits on the mass, together with the uncertainties in the median expected limits,
are presented in Tables~\ref{tab:BFtable_D_kappa1_combination}
to~\ref{tab:BFtable_D_paironly_combination}.

The existence of a heavy vector-like \QD quark with a mass below 1595\GeV is
excluded at 95\% \CL when using the following choice of model parameters:
$\tilde{\kappa}_\PW = 1.0$, $\mathcal{B}_\PW = 0.5$, and
$\mathcal{B}_\PZ = 0.25$. This limit may be compared with the expected value of 1460\GeV.
In the case where the VLQ couples only to the $\PW$ boson,
the observed (expected) limit at 95\% \CL is 1745 (1620)\GeV.

The sensitivity to pair production of VLQs for the event categories in which a kinematic fit is performed becomes
more important for lower $\tilde{\kappa}_\PW$.
In the extreme case where only
pair production is considered (as shown in Fig.~\ref{fig:BFplots_D_paironly_combination}),
the added sensitivity of the combined analysis when compared to the categories that use a kinematic fit or not is
illustrated using some example parameter choices, as shown in
Table~\ref{tab:examplelimitcomparison}. When the branching fraction for the decay to
a $\PW$ boson becomes large, the event categories using the kinematic fit to
the VLQ signal mass become more important.
For lower $\mathcal{B}_\PW$ and relatively large $\mathcal{B}_\PZ$ and $\mathcal{B}_\PH$,
the dilepton $\Z\PQq\PH\PQq$ event category drives the sensitivity.

\begin{table}
\centering
\topcaption{Observed and median expected lower limits on the VLQ mass (in \GeV)
at 95\% \CL, or greater than 95\% \CL when indicated with $*$, for a range of different combinations of decay branching fractions.
The ranges containing 68 and 95\%, respectively, of the distribution of limits expected under the background-only hypothesis, are also given.
The limits are determined assuming $\tilde{\kappa}_\PW = 1.0$.}
\label{tab:BFtable_D_kappa1_combination}
\begin{scotch}{cccrrrr}
 $\mathcal{B}_\PW$  & $\mathcal{B}_\PZ$ & $\mathcal{B}_\PH$  & Observed & Median&  68\% &  95\% \\
 &&&&expected&expected&expected\\
\hline
0.1 & 0.8 & 0.1 & 1760 & 1785 & [1705,$1800^{*}$] & [1615,$1800^{*}$] \\
0.1 & 0.6 & 0.3 & 1660 & 1675 & [1580,1760] & [1505,$1800^{*}$] \\
0.1 & 0.4 & 0.5 & 1520 & 1525 & [1450,1605] & [1375,1690] \\
0.1 & 0.2 & 0.7 & 1365 & 1310 & [1200,1405] & [1125,1470] \\
0.1 & 0.0 & 0.9 & 760 & 700 & [590,830] & [400,965] \\
0.2 & 0.8 & 0.0 & 1710 & 1690 & [1605,1780] & [1515,$1800^{*}$] \\
0.2 & 0.6 & 0.2 & 1620 & 1595 & [1510,1700] & [1435,1770] \\
0.2 & 0.4 & 0.4 & 1520 & 1475 & [1390,1570] & [1305,1660] \\
0.2 & 0.2 & 0.6 & 1420 & 1300 & [1185,1395] & [1105,1500] \\
0.2 & 0.0 & 0.8 & 1305 & 990 & [810,1110] & [710,1260] \\
0.4 & 0.6 & 0.0 & 1660 & 1595 & [1485,1695] & [1395,1790] \\
0.4 & 0.4 & 0.2 & 1605 & 1510 & [1395,1620] & [1305,1730] \\
0.4 & 0.2 & 0.4 & 1530 & 1375 & [1275,1535] & [1165,1635] \\
0.4 & 0.0 & 0.6 & 1480 & 1275 & [1100,1380] & [955,1545] \\
0.6 & 0.4 & 0.0 & 1700 & 1565 & [1445,1690] & [1340,1780] \\
0.6 & 0.2 & 0.2 & 1645 & 1495 & [1355,1630] & [1250,1730] \\
0.6 & 0.0 & 0.4 & 1605 & 1385 & [1270,1565] & [1150,1665] \\
0.8 & 0.2 & 0.0 & 1700 & 1580 & [1435,1715] & [1325,1800] \\
0.8 & 0.0 & 0.2 & 1695 & 1525 & [1365,1675] & [1260,1775] \\
1.0 & 0.0 & 0.0 & 1745 & 1620 & [1450,1730] & [1335,$1800^{*}$] \\
\end{scotch}
\end{table}

\begin{table}
\centering
\topcaption{Observed and median expected lower limits on the VLQ mass (in \GeV)
at 95\% \CL, for a range of different combinations of decay branching fractions.
The ranges containing 68 and 95\%, respectively, of the distribution of limits expected under the background-only hypothesis, are also given.
The limits are determined using $\tilde{\kappa}_\PW = 0.7$.}
\label{tab:BFtable_D_kappa0p7_combination}
\begin{scotch}{cccrrrr}
 $\mathcal{B}_\PW$  & $\mathcal{B}_\PZ$ & $\mathcal{B}_\PH$  & Observed & Median &  68\%&  95\% \\
 &&&&expected&expected&expected\\
\hline
0.1 & 0.8 & 0.1 & 1595 & 1615 & [1535,1705] & [1460,1770] \\
0.1 & 0.6 & 0.3 & 1485 & 1510 & [1435,1595] & [1360,1670] \\
0.1 & 0.4 & 0.5 & 1380 & 1380 & [1300,1450] & [1200,1515] \\
0.1 & 0.2 & 0.7 & 1175 & 1130 & [1005,1215] & [915,1300] \\
0.1 & 0.0 & 0.9 & 560 & 550 & [435,625] & [400,710] \\
0.2 & 0.8 & 0.0 & 1525 & 1525 & [1445,1610] & [1380,1690] \\
0.2 & 0.6 & 0.2 & 1465 & 1435 & [1350,1510] & [1255,1580] \\
0.2 & 0.4 & 0.4 & 1360 & 1305 & [1200,1400] & [1120,1470] \\
0.2 & 0.2 & 0.6 & 1240 & 1105 & [960,1195] & [840,1295] \\
0.2 & 0.0 & 0.8 & 745 & 725 & [600,840] & [505,965] \\
0.4 & 0.6 & 0.0 & 1470 & 1400 & [1300,1495] & [1200,1585] \\
0.4 & 0.4 & 0.2 & 1405 & 1300 & [1190,1400] & [1095,1500] \\
0.4 & 0.2 & 0.4 & 1355 & 1155 & [1025,1280] & [890,1380] \\
0.4 & 0.0 & 0.6 & 1315 & 985 & [820,1120] & [720,1265] \\
0.6 & 0.4 & 0.0 & 1470 & 1335 & [1210,1450] & [1110,1560] \\
0.6 & 0.2 & 0.2 & 1435 & 1245 & [1105,1365] & [985,1505] \\
0.6 & 0.0 & 0.4 & 1385 & 1140 & [1005,1285] & [835,1385] \\
0.8 & 0.2 & 0.0 & 1500 & 1320 & [1205,1445] & [1060,1565] \\
0.8 & 0.0 & 0.2 & 1465 & 1265 & [1090,1380] & [980,1540] \\
1.0 & 0.0 & 0.0 & 1550 & 1335 & [1210,1480] & [1055,1615] \\
\end{scotch}
\end{table}

\begin{table}
\centering
\topcaption{Observed and median expected lower limits on the VLQ mass (in \GeV)
at 95\% \CL, for a range of different combinations of decay branching fractions.
The ranges containing 68 and 95\%, respectively, of the distribution of limits expected under the background-only hypothesis, are also given.
The cases where the limits could not be evaluated because simulated signal samples for VLQ masses below 400\GeV are not available, are indicated with `n.a.'.
The limits are determined assuming $\tilde{\kappa}_\PW = 0.4$.}
\label{tab:BFtable_D_kappa0p4_combination}
\begin{scotch}{cccrrrr}
 $\mathcal{B}_\PW$  & $\mathcal{B}_\PZ$ & $\mathcal{B}_\PH$  & Observed & Median &  68\%  &  95\%   \\
 &&&&expected&expected&expected\\
\hline
0.1 & 0.8 & 0.1 & 1370 & 1400 & [1305,1460] & [1220,1525] \\
0.1 & 0.6 & 0.3 & 1260 & 1275 & [1175,1365] & [1110,1430] \\
0.1 & 0.4 & 0.5 & 1145 & 1120 & [1000,1190] & [890,1290] \\
0.1 & 0.2 & 0.7 & 745 & 765 & [595,905] & [495,990] \\
0.1 & 0.0 & 0.9 & 460 & 505 & [n.a.,555] & [n.a.,595] \\
0.2 & 0.8 & 0.0 & 1280 & 1285 & [1180,1370] & [1115,1435] \\
0.2 & 0.6 & 0.2 & 1205 & 1165 & [1080,1255] & [965,1340] \\
0.2 & 0.4 & 0.4 & 1115 & 995 & [895,1110] & [745,1185] \\
0.2 & 0.2 & 0.6 & 690 & 730 & [590,840] & [510,955] \\
0.2 & 0.0 & 0.8 & 610 & 565 & [500,645] & [n.a.,715] \\
0.4 & 0.6 & 0.0 & 1195 & 1110 & [975,1195] & [880,1280] \\
0.4 & 0.4 & 0.2 & 1110 & 960 & [840,1080] & [730,1165] \\
0.4 & 0.2 & 0.4 & 810 & 790 & [700,895] & [610,995] \\
0.4 & 0.0 & 0.6 & 725 & 715 & [605,780] & [525,850] \\
0.6 & 0.4 & 0.0 & 1160 & 980 & [865,1090] & [770,1200] \\
0.6 & 0.2 & 0.2 & 1065 & 860 & [775,985] & [705,1080] \\
0.6 & 0.0 & 0.4 & 805 & 795 & [720,880] & [635,995] \\
0.8 & 0.2 & 0.0 & 1160 & 930 & [830,1050] & [755,1175] \\
0.8 & 0.0 & 0.2 & 1090 & 870 & [785,980] & [720,1080] \\
1.0 & 0.0 & 0.0 & 1250 & 940 & [845,1055] & [780,1165] \\
\end{scotch}
\end{table}

\begin{table}
\centering
\topcaption{Observed and median expected lower limits on the VLQ mass (in \GeV)
at 95\% \CL, for a range of different combinations of decay branching fractions.
The ranges containing 68 and 95\%, respectively, of the distribution of limits expected under the background-only hypothesis, are also given.
The cases where the limits could not be evaluated because simulated signal samples for VLQ masses below 400\GeV are not available, are indicated with `n.a.'.
The limits are determined assuming $\tilde{\kappa}_\PW = 0.1$.}
\label{tab:BFtable_D_kappa0p1_combination}
\begin{scotch}{cccrrrr}
 $\mathcal{B}_{\PW}$  & $\mathcal{B}_\Z$ & $\mathcal{B}_\PH$  & Observed & Median&  68\%&  95\%\\
 &&&&expected&expected&expected\\
\hline
0.1 & 0.8 & 0.1 & 660 & 720 & [650,795] & [580,885] \\
0.1 & 0.6 & 0.3 & 615 & 665 & [595,730] & [550,785] \\
0.1 & 0.4 & 0.5 & 575 & 610 & [555,680] & [510,725] \\
0.1 & 0.2 & 0.7 & 520 & 560 & [510,605] & [455,660] \\
0.1 & 0.0 & 0.9 & 455 & 505 & [n.a.,550] & [n.a.,585] \\
0.2 & 0.8 & 0.0 & 660 & 715 & [650,770] & [590,825] \\
0.2 & 0.6 & 0.2 & 630 & 690 & [615,740] & [565,790] \\
0.2 & 0.4 & 0.4 & 610 & 645 & [580,705] & [525,755] \\
0.2 & 0.2 & 0.6 & 575 & 585 & [535,660] & [490,715] \\
0.2 & 0.0 & 0.8 & 510 & 545 & [480,605] & [n.a.,675] \\
0.4 & 0.6 & 0.0 & 680 & 735 & [675,795] & [605,840] \\
0.4 & 0.4 & 0.2 & 665 & 715 & [640,770] & [580,820] \\
0.4 & 0.2 & 0.4 & 650 & 685 & [590,745] & [530,795] \\
0.4 & 0.0 & 0.6 & 660 & 655 & [565,725] & [490,765] \\
0.6 & 0.4 & 0.0 & 740 & 770 & [705,830] & [640,885] \\
0.6 & 0.2 & 0.2 & 725 & 745 & [680,805] & [600,865] \\
0.6 & 0.0 & 0.4 & 730 & 735 & [660,790] & [570,840] \\
0.8 & 0.2 & 0.0 & 785 & 805 & [745,860] & [675,915] \\
0.8 & 0.0 & 0.2 & 795 & 785 & [725,845] & [660,900] \\
1.0 & 0.0 & 0.0 & 860 & 835 & [775,890] & [725,940] \\
\end{scotch}
\end{table}

\begin{table}
\centering
\topcaption{Observed and median expected lower limits on the VLQ mass (in \GeV)
at 95\% \CL, for a range of different combinations of decay branching fractions.
The ranges containing 68 and 95\%, respectively, of the distribution of limits expected under the background-only hypothesis, are also given.
The cases where the limits could not be evaluated because simulated signal samples for VLQ masses below 400\GeV are not available, are indicated with `n.a.'.
The limits are determined under the assumption that pair production is the only
available VLQ production mechanism.}
\label{tab:BFtable_D_paironly_combination}
\begin{scotch}{cccrrrr}
 $\mathcal{B}_{\PW}$  & $\mathcal{B}_\Z$ & $\mathcal{B}_\PH$  & Observed & Median&  68\% &  95\% \\
 &&&&expected&expected&expected\\
\hline
0.0 & 1.0 & 0.0 & 605 & 675 & [625,725] & [580,765] \\
0.0 & 0.8 & 0.2 & 590 & 655 & [600,700] & [550,750] \\
0.0 & 0.6 & 0.4 & 580 & 625 & [575,680] & [530,725] \\
0.0 & 0.4 & 0.6 & 550 & 585 & [540,640] & [495,690] \\
0.0 & 0.2 & 0.8 & 510 & 535 & [490,580] & [430,620] \\
0.0 & 0.0 & 1.0 & 430 & n.a. & [n.a.,505] & [n.a.,535] \\
0.2 & 0.8 & 0.0 & 625 & 695 & [645,745] & [595,785] \\
0.2 & 0.6 & 0.2 & 620 & 675 & [610,725] & [560,770] \\
0.2 & 0.4 & 0.4 & 585 & 635 & [575,700] & [525,745] \\
0.2 & 0.2 & 0.6 & 545 & 585 & [530,655] & [475,710] \\
0.2 & 0.0 & 0.8 & 495 & 545 & [470,600] & [n.a.,675] \\
0.4 & 0.6 & 0.0 & 670 & 725 & [670,780] & [610,825] \\
0.4 & 0.4 & 0.2 & 650 & 710 & [635,760] & [575,810] \\
0.4 & 0.2 & 0.4 & 645 & 680 & [590,740] & [535,785] \\
0.4 & 0.0 & 0.6 & 665 & 650 & [565,720] & [490,765] \\
0.6 & 0.4 & 0.0 & 725 & 760 & [700,820] & [625,870] \\
0.6 & 0.2 & 0.2 & 715 & 745 & [670,800] & [585,845] \\
0.6 & 0.0 & 0.4 & 710 & 725 & [650,780] & [560,830] \\
0.8 & 0.2 & 0.0 & 785 & 795 & [730,855] & [660,905] \\
0.8 & 0.0 & 0.2 & 785 & 785 & [715,840] & [640,885] \\
1.0 & 0.0 & 0.0 & 845 & 825 & [765,880] & [710,930] \\
\end{scotch}
\end{table}

\begin{table}
\centering
\topcaption{Comparison of several expected 95\% \CL lower mass limits for signal
pair production only, illustrating the added sensitivity in
the combination of the event categories that use and do not use a kinematic fit.}
\label{tab:examplelimitcomparison}
\begin{scotch}{crrr}
 Signal benchmark  & Dilepton and multilepton                          & Single-lepton channel    &  Combination \\
                   & channels                                          & using kinematic fit       &              \\
\hline
 $\mathcal{B}_\PW = 1.0$, $\mathcal{B}_\PZ = 0.0$ & 725\GeV & 810\GeV & 825\GeV \\
 $\mathcal{B}_\PW = 0.5$, $\mathcal{B}_\PZ = 0.2$ & 585\GeV &  680\GeV & 720\GeV \\
 $\mathcal{B}_\PW = 0.1$, $\mathcal{B}_\PZ = 0.5$ & 600\GeV &  405\GeV & 630\GeV \\
 $\mathcal{B}_\PW = 0.1$, $\mathcal{B}_\PZ = 0.1$ & 420\GeV & $<$400\GeV & 525\GeV \\
\end{scotch}
\end{table}

\begin{figure}
  \centering
  \includegraphics[width=0.48\textwidth]{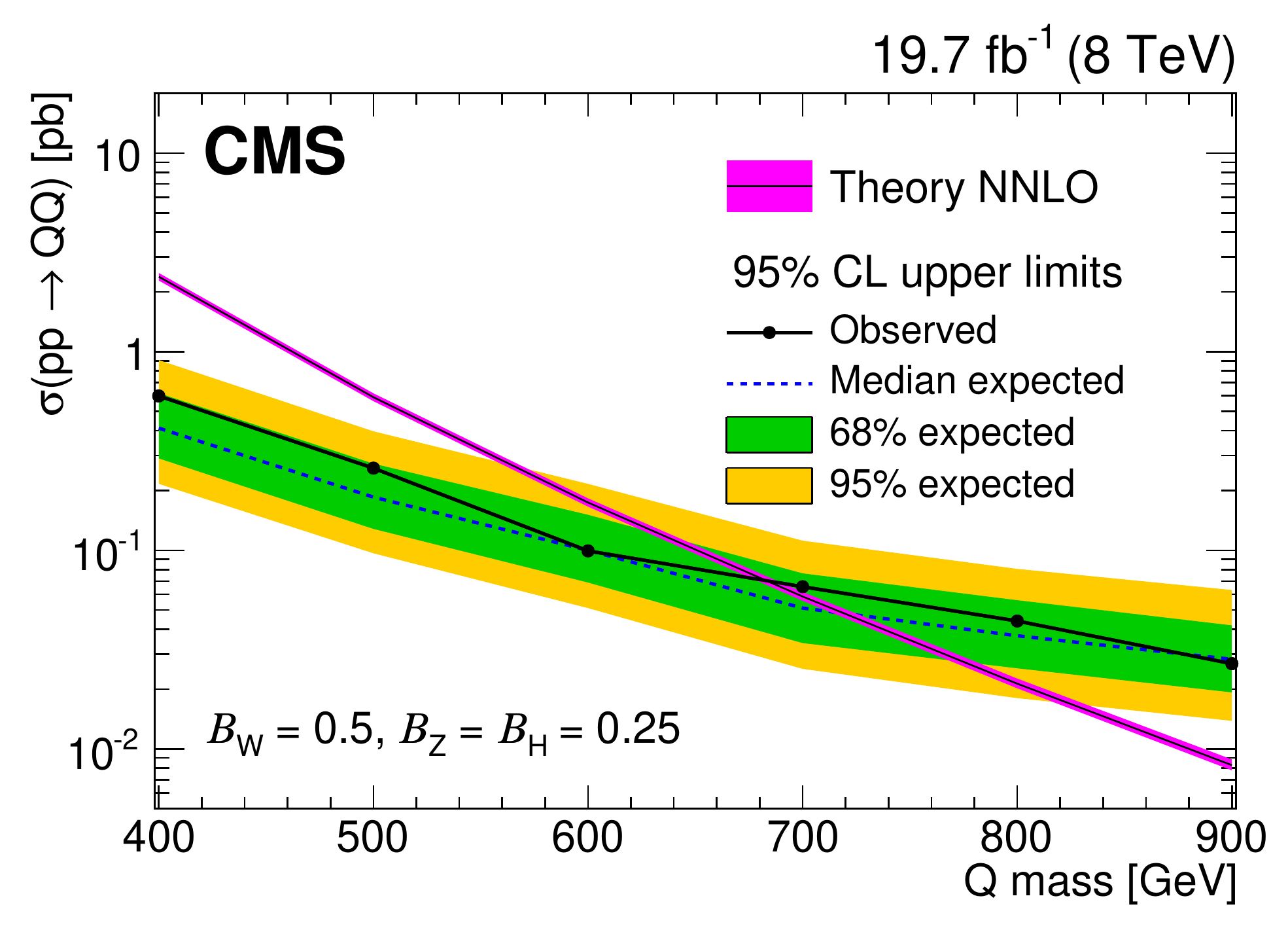}
   \includegraphics[width=0.48\textwidth]{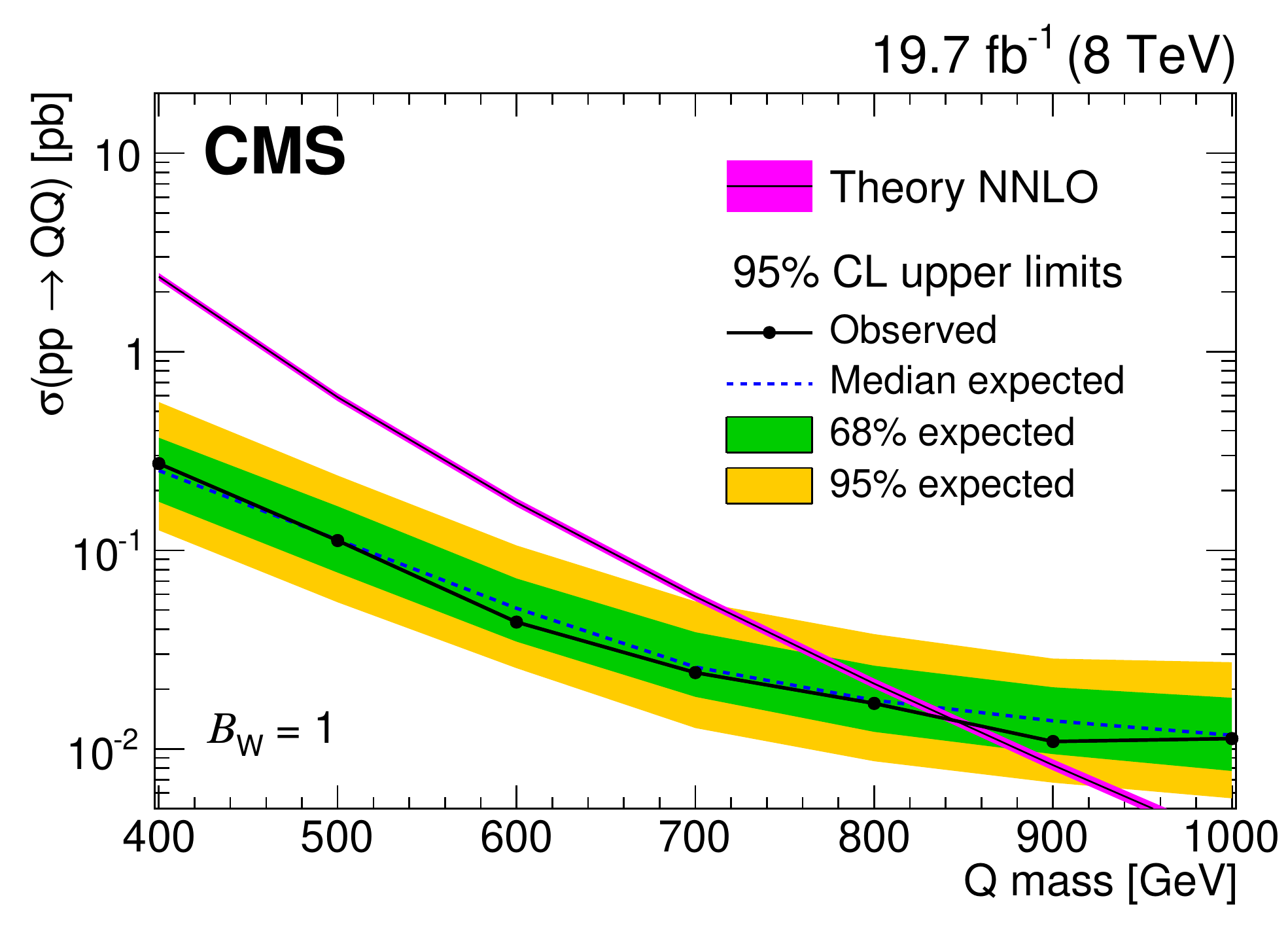}
  \caption{The 95\% \CL exclusion limits on the production cross section
  determined assuming different sets of model parameters ($\mathcal{B}_\PW = 0.5$,
  $\mathcal{B}_\PZ = 0.25$ (\cmsLeft), and $\mathcal{B}_\PW = 1$
 (\cmsRight)) as a function of the hypothetical VLQ mass, and for
  the scenario where only strong pair production of the VLQs
  is considered.
  The median expected and observed exclusion limits are indicated
  with a dashed and a solid line, respectively. The inner (green) band and the outer (yellow) band
  indicate the regions containing 68 and 95\%, respectively, of the distribution of limits expected under the background-only hypothesis.
  The cross section from a full NNLO
  calculation~\cite{Czakon_Mitov}, including uncertainties in the PDF description and the
  renormalization and factorization scales, is shown by the magenta band.
}
  \label{fig:xsection_vs_mass_paironly_SuperCombination_refsignal}
\end{figure}

Figure~\ref{fig:xsection_vs_mass_paironly_SuperCombination_refsignal} shows
the 95\% \CL limit on the production cross section as a function
of the VLQ mass, for the scenario where only pair production of
the VLQs is considered, and for two different parameter choices. In Fig.~\ref{fig:xsection_vs_mass_paironly_SuperCombination_refsignal} (\cmsLeft)
the result is shown for
$\mathcal{B}_\PW = 0.5$ and $\mathcal{B}_\PZ = 0.25$. For this
set of parameters, we exclude VLQs with masses
below 685\GeV at 95\% \CL, compared to an expected exclusion limit of 720\GeV.
In Fig.~\ref{fig:xsection_vs_mass_paironly_SuperCombination_refsignal}(\cmsRight),
the exclusion limits are shown for $\mathcal{B}_\PW = 1$. In
this case we exclude VLQs with masses below
845\GeV at 95\% \CL, compared to an expected lower limit of 825\GeV.

In this search we use signal mass distributions simulated using the narrow-width
approximation, where the decay width is about 1\% of the mass of the VLQ and is
significantly less than the experimental resolution.
We have verified that this approximation does not affect the results. At
smaller mass values ($\sim$700\GeV) and for a
parameter space with an exclusion limit close to this mass, the theoretically
calculated width reaches up to about 4\%, which is still well below the experimental
resolution (about 9\% in the $\PW\PQq\PQq$ category, for example). For the highest mass values
probed in the analysis ($\sim$1800\GeV), the width approaches the experimental resolution
in part of the parameter space. This does not change the results, as the
width of the signal mass distributions remains smaller than the bin size at these
high masses.

The results tighten the constraints on the masses, cross sections and decay branching fractions
of VLQs coupled to light-flavor quarks. In the scenario where the VLQ couples to first-generation quarks only via the $\PW$ boson,
the results can be compared to those obtained previously.
The presented exclusion limits in this paper are more stringent than those obtained by the ATLAS experiment,
when considering single production of VLQs alone at $\sqrt{s} = 7$\TeV~\cite{ATLASlightVLQ2011}
and pair production of VLQs alone at 8\TeV~\cite{ATLASlightVLQ_pair_8}.

\section{Summary}
\label{sec:summary}
A search has been performed for vector-like quarks coupled to light quarks,
produced in either single-production or pair-production processes, in proton-proton collisions at
$\sqrt{s} = 8$\TeV at the LHC. In the single-production mode the search
has been performed for down-type quarks
(electric charge of magnitude $1/3$), while in the pair-production mode the search is
sensitive to decays of vector-like quarks into up, down and strange quarks.
Approaches with and without kinematic fits have been used to perform this search.
No significant excess over standard model expectations has been observed.
Lower limits on the mass of the vector-like quarks have been determined by
combining the results from both the single- and pair-production searches.
Limits have also been extracted using the data from the pair-production
search alone. For all processes considered, the lower
mass limits range from 400 to 1800\GeV, depending on the vector-like
quark branching fractions for decays to $\PW$, $\PZ$, and Higgs bosons and the assumed
value of the electroweak single-production strength.
When considering pair production alone, vector-like quarks with masses below
845\GeV\,(825\GeV expected) are excluded
for $\mathcal{B}(\PW) = 1.0$, and with masses below 685\GeV (720\GeV expected),
for the widely adopted benchmark with
$\mathcal{B}(\PW) = 0.5$, $\mathcal{B}(\PZ) = \mathcal{B}(\PH) = 0.25$.
These results provide the most stringent mass limits to date on vector-like quarks
that couple to light quarks and that are produced either singly or in pairs.

\begin{acknowledgments}
\hyphenation{Bundes-ministerium Forschungs-gemeinschaft Forschungs-zentren Rachada-pisek} We congratulate our colleagues in the CERN accelerator departments for the excellent performance of the LHC and thank the technical and administrative staffs at CERN and at other CMS institutes for their contributions to the success of the CMS effort. In addition, we gratefully acknowledge the computing centers and personnel of the Worldwide LHC Computing Grid for delivering so effectively the computing infrastructure essential to our analyses. Finally, we acknowledge the enduring support for the construction and operation of the LHC and the CMS detector provided by the following funding agencies: the Austrian Federal Ministry of Science, Research and Economy and the Austrian Science Fund; the Belgian Fonds de la Recherche Scientifique, and Fonds voor Wetenschappelijk Onderzoek; the Brazilian Funding Agencies (CNPq, CAPES, FAPERJ, and FAPESP); the Bulgarian Ministry of Education and Science; CERN; the Chinese Academy of Sciences, Ministry of Science and Technology, and National Natural Science Foundation of China; the Colombian Funding Agency (COLCIENCIAS); the Croatian Ministry of Science, Education and Sport, and the Croatian Science Foundation; the Research Promotion Foundation, Cyprus; the Secretariat for Higher Education, Science, Technology and Innovation, Ecuador; the Ministry of Education and Research, Estonian Research Council via IUT23-4 and IUT23-6 and European Regional Development Fund, Estonia; the Academy of Finland, Finnish Ministry of Education and Culture, and Helsinki Institute of Physics; the Institut National de Physique Nucl\'eaire et de Physique des Particules~/~CNRS, and Commissariat \`a l'\'Energie Atomique et aux \'Energies Alternatives~/~CEA, France; the Bundesministerium f\"ur Bildung und Forschung, Deutsche Forschungsgemeinschaft, and Helmholtz-Gemeinschaft Deutscher Forschungszentren, Germany; the General Secretariat for Research and Technology, Greece; the National Scientific Research Foundation, and National Innovation Office, Hungary; the Department of Atomic Energy and the Department of Science and Technology, India; the Institute for Studies in Theoretical Physics and Mathematics, Iran; the Science Foundation, Ireland; the Istituto Nazionale di Fisica Nucleare, Italy; the Ministry of Science, ICT and Future Planning, and National Research Foundation (NRF), Republic of Korea; the Lithuanian Academy of Sciences; the Ministry of Education, and University of Malaya (Malaysia); the Mexican Funding Agencies (BUAP, CINVESTAV, CONACYT, LNS, SEP, and UASLP-FAI); the Ministry of Business, Innovation and Employment, New Zealand; the Pakistan Atomic Energy Commission; the Ministry of Science and Higher Education and the National Science Centre, Poland; the Funda\c{c}\~ao para a Ci\^encia e a Tecnologia, Portugal; JINR, Dubna; the Ministry of Education and Science of the Russian Federation, the Federal Agency of Atomic Energy of the Russian Federation, Russian Academy of Sciences, the Russian Foundation for Basic Research and the Russian Competitiveness Program of NRNU ``MEPhI"; the Ministry of Education, Science and Technological Development of Serbia; the Secretar\'{\i}a de Estado de Investigaci\'on, Desarrollo e Innovaci\'on, Programa Consolider-Ingenio 2010, Plan de Ciencia, Tecnolog\'{i}a e Innovaci\'on 2013-2017 del Principado de Asturias and Fondo Europeo de Desarrollo Regional, Spain; the Swiss Funding Agencies (ETH Board, ETH Zurich, PSI, SNF, UniZH, Canton Zurich, and SER); the Ministry of Science and Technology, Taipei; the Thailand Center of Excellence in Physics, the Institute for the Promotion of Teaching Science and Technology of Thailand, Special Task Force for Activating Research and the National Science and Technology Development Agency of Thailand; the Scientific and Technical Research Council of Turkey, and Turkish Atomic Energy Authority; the National Academy of Sciences of Ukraine, and State Fund for Fundamental Researches, Ukraine; the Science and Technology Facilities Council, UK; the US Department of Energy, and the US National Science Foundation.

Individuals have received support from the Marie-Curie program and the European Research Council and Horizon 2020 Grant, contract No. 675440 (European Union); the Leventis Foundation; the A. P. Sloan Foundation; the Alexander von Humboldt Foundation; the Belgian Federal Science Policy Office; the Fonds pour la Formation \`a la Recherche dans l'Industrie et dans l'Agriculture (FRIA-Belgium); the Agentschap voor Innovatie door Wetenschap en Technologie (IWT-Belgium); the Ministry of Education, Youth and Sports (MEYS) of the Czech Republic; the Council of Scientific and Industrial Research, India; the HOMING PLUS program of the Foundation for Polish Science, cofinanced from European Union, Regional Development Fund, the Mobility Plus program of the Ministry of Science and Higher Education, the National Science Center (Poland), contracts Harmonia 2014/14/M/ST2/00428, Opus 2014/13/B/ST2/02543, 2014/15/B/ST2/03998, and 2015/19/B/ST2/02861, Sonata-bis 2012/07/E/ST2/01406; the National Priorities Research Program by Qatar National Research Fund; the Programa Clar\'in-COFUND del Principado de Asturias; the Thalis and Aristeia programs cofinanced by EU-ESF and the Greek NSRF; the Rachadapisek Sompot Fund for Postdoctoral Fellowship, Chulalongkorn University and the Chulalongkorn Academic into Its 2nd Century Project Advancement Project (Thailand); and the Welch Foundation, contract C-1845.
\end{acknowledgments}
\clearpage
\numberwithin{table}{section}
\bibliography{auto_generated}
\ifthenelse{\boolean{cms@external}}{}{
\clearpage
\appendix
\section{Results using an alternative parametrization of charged and neutral coupling strengths\label{app:suppMat}}
In Section~\ifthenelse{\boolean{cms@external}}{8 in the main document}{\ref{sec:results}} results are presented for a scan over the branching fractions of the VLQ, while keeping the value of $\tilde{\kappa}_{\PW}$ fixed.
As noted in Section~\ifthenelse{\boolean{cms@external}}{2.1}{\ref{sec:single_strateg}}, for non-zero $\tilde{\kappa}_\PW$ the exclusion limits on the VLQ mass cannot be evaluated for $\mathcal{B}_{\PW} = 0$,
as Eq.~(\ifthenelse{\boolean{cms@external}}{2}{\ref{eq:kappaWZrelation}}) implies that the neutral-current single-production strength
parameter $\tilde{\kappa}_\PZ$ diverges in this limit.
This is indicated by the black shaded region below $\mathcal{B}_\PW \approx 0.1$ in Figs.~\ifthenelse{\boolean{cms@external}}{11}{\ref{fig:BFplots_D_kappa1_combination}} to~\ifthenelse{\boolean{cms@external}}{14}{\ref{fig:BFplots_D_kappa0p1_combination}}.

However, from Ref.~\cite{vlqbuchkremer} a parametrization can be chosen that does not exhibit this divergent behavior.
This involves fixing one generic single-production strength parameter $\kappa_\QD$ and scanning over the branching fractions as before.
The single parameter $\kappa_\QD$ contains information from the charged-current, and $\PZ$ and $\PH$ neutral-current interactions, because it can be expressed as
\begin{equation}
\label{eq:kappaD}
  \kappa^2_\QD= \kappa^2_{\PW} + \frac{\kappa^2_\PZ}{2} + \kappa^2_\PH\left(\frac{1}{2} - \frac{m^2_{\PH}}{m_{\Q}}\right)
\end{equation}
with $m_\PH$ the mass of the Higgs boson.
Since $\kappa$ is to be interpreted as a mixing angle, the range of $\kappa_{\QD}$ is physically restricted between 0 and 1.

The following relations between the default and alternative parametrization can be deduced:
\begin{align}
\tilde{\kappa}_\PW &= \frac{\sqrt{2 \mathcal{B}_\PW} m_{\Q} }{v} \kappa_\QD, \\
\tilde{\kappa}_\PZ &= \frac{2 \sqrt{\mathcal{B}_\PZ} m_{\Q}}{v} \kappa_\QD\end{align}

From these relations it is seen that Eq.~(\ifthenelse{\boolean{cms@external}}{2}{\ref{eq:kappaWZrelation}}) still holds, but fixing $\kappa_\QD$ in the scan instead of $\tilde{\kappa}_\PW$
provides a more consistent behavior throughout the scan.
In particular, the combination $\kappa_\QD\neq 0$ and $\mathcal{B}_\PW = 0$ does not automatically lead to a divergence of $\tilde{\kappa}_\PZ$.
Results derived in this parametrization are especially useful for scenarios where the VLQ only couples to Z or Higgs bosons;
such scenarios have only been covered in the default results in Section~\ifthenelse{\boolean{cms@external}}{8}{\ref{sec:results}} when considering VLQ pair production alone, but not including single production.

When fixing values of $\kappa_\QD$ and scanning over the branching fractions, results are obtained for the combination of all channels
in Tables~\ref{tab:reparam_BFtable_D_kappaD0p05_combination} to~\ref{tab:reparam_BFtable_D_kappaD1_combination}.
The scan in $\kappa_\QD$ is performed from 0.05 to 1, initially in steps of 0.05, but in larger steps of 0.1 from $\kappa_\QD= 0.2$ onwards.
Even for relatively small $\kappa_\QD$ values, the mass limits become larger than 1800\GeV and cannot be evaluated with the produced VLQ signal MC samples.
The reason for these high mass limits is that the single-production strengths governed by $\tilde{\kappa}_\PW$ and $\tilde{\kappa}_\PZ$
may become large even for relatively small $\kappa_\QD$ values.

\begin{table*}
\centering
\topcaption{Observed and median expected lower limits on the VLQ mass (in \GeV)
at 95\% CL for a range of different combinations of decay branching fractions.
The ranges containing 68 and 95\%, respectively, of the distribution of limits expected under the background-only hypothesis, are also given. 
The cases where the limits could not be evaluated because simulated signal samples for VLQ masses below 400\GeV are not available, are indicated with `n.a.'.
The limits are determined assuming $\kappa_\QD= 0.05$.}
\label{tab:reparam_BFtable_D_kappaD0p05_combination}
\begin{scotch}{cccrrrr}
 $\mathcal{B}_\PW$  & $\mathcal{B}_\cPZ$ & $\mathcal{B}_\PH$  & Observed & Median expected &  68\% expected &  95\% expected  \\
\hline
0.0 & 1.0 & 0.0 & 635 & 690 & [630,745] & [580,815] \\
0.0 & 0.8 & 0.2 & 610 & 660 & [600,715] & [555,765] \\
0.0 & 0.6 & 0.4 & 585 & 625 & [575,680] & [530,730] \\
0.0 & 0.4 & 0.6 & 555 & 585 & [540,640] & [495,690] \\
0.0 & 0.2 & 0.8 & 500 & 535 & [485,575] & [425,620] \\
0.0 & 0.0 & 1.0 & 430 & n.a. & [n.a.,505] & [n.a.,535] \\
0.2 & 0.8 & 0.0 & 645 & 710 & [650,775] & [590,850] \\
0.2 & 0.6 & 0.2 & 620 & 685 & [610,740] & [565,785] \\
0.2 & 0.4 & 0.4 & 605 & 640 & [575,705] & [530,755] \\
0.2 & 0.2 & 0.6 & 560 & 585 & [535,655] & [475,715] \\
0.2 & 0.0 & 0.8 & 550 & 545 & [480,605] & [400,685] \\
0.4 & 0.6 & 0.0 & 690 & 745 & [685,810] & [610,880] \\
0.4 & 0.4 & 0.2 & 665 & 715 & [645,780] & [580,835] \\
0.4 & 0.2 & 0.4 & 655 & 685 & [590,750] & [530,800] \\
0.4 & 0.0 & 0.6 & 660 & 655 & [565,725] & [500,770] \\
0.6 & 0.4 & 0.0 & 750 & 775 & [715,845] & [645,895] \\
0.6 & 0.2 & 0.2 & 735 & 755 & [695,820] & [600,875] \\
0.6 & 0.0 & 0.4 & 725 & 735 & [665,790] & [580,850] \\
0.8 & 0.2 & 0.0 & 820 & 820 & [750,880] & [685,945] \\
0.8 & 0.0 & 0.2 & 810 & 795 & [730,860] & [660,915] \\
1.0 & 0.0 & 0.0 & 890 & 850 & [785,925] & [725,1010] \\
\end{scotch}

\end{table*}

\begin{table*}
\centering
\topcaption{Observed and median expected lower limits on the VLQ mass (in \GeV)
at 95\% CL, or greater than 95\% CL when indicated with $*$, for a range of different combinations of decay branching fractions.
The ranges containing 68 and 95\%, respectively, of the distribution of limits expected under the background-only hypothesis, are also given. 
The cases where the limits could not be evaluated because simulated signal samples for VLQ masses below 400\GeV are not available, are indicated with `n.a.'.
The limits are determined assuming $\kappa_\QD= 0.1$.}
\label{tab:reparam_BFtable_D_kappaD0p1_combination}
\begin{scotch}{cccrrrr}
 $\mathcal{B}_\PW$  & $\mathcal{B}_\cPZ$ & $\mathcal{B}_\PH$  & Observed & Median expected &  68\% expected &  95\% expected  \\
\hline
0.0 & 1.0 & 0.0 & 1140 & 1145 & [775,1265] & [620,1385] \\
0.0 & 0.8 & 0.2 & 665 & 780 & [645,1130] & [570,1215] \\
0.0 & 0.6 & 0.4 & 615 & 660 & [580,750] & [535,960] \\
0.0 & 0.4 & 0.6 & 555 & 585 & [540,655] & [495,710] \\
0.0 & 0.2 & 0.8 & 505 & 535 & [485,575] & [425,615] \\
0.0 & 0.0 & 1.0 & 430 & n.a. & [n.a.,505] & [n.a.,535] \\
0.2 & 0.8 & 0.0 & 1160 & 1135 & [785,1265] & [650,1385] \\
0.2 & 0.6 & 0.2 & 675 & 780 & [655,1100] & [575,1195] \\
0.2 & 0.4 & 0.4 & 630 & 665 & [580,755] & [525,875] \\
0.2 & 0.2 & 0.6 & 600 & 590 & [530,670] & [470,735] \\
0.2 & 0.0 & 0.8 & 495 & 550 & [470,600] & [400,690] \\
0.4 & 0.6 & 0.0 & 1290 & 1110 & [790,1285] & [660,1400] \\
0.4 & 0.4 & 0.2 & 730 & 785 & [685,1035] & [580,1215] \\
0.4 & 0.2 & 0.4 & 685 & 710 & [600,795] & [535,895] \\
0.4 & 0.0 & 0.6 & 675 & 660 & [570,740] & [495,795] \\
0.6 & 0.4 & 0.0 & 1420 & 1120 & [810,1340] & [705,1540] \\
0.6 & 0.2 & 0.2 & 1360 & 835 & [735,1130] & [625,1370] \\
0.6 & 0.0 & 0.4 & 805 & 770 & [685,870] & [565,1090] \\
0.8 & 0.2 & 0.0 & 1620 & 1280 & [870,1565] & [755,1750] \\
0.8 & 0.0 & 0.2 & 1555 & 1055 & [800,1385] & [695,1685] \\
1.0 & 0.0 & 0.0 & 1765 & 1475 & [1215,1730] & [835,$1800^{*}$] \\
\end{scotch}
\end{table*}

\begin{table*}
\centering
\topcaption{Observed and median expected lower limits on the VLQ mass (in \GeV)
at 95\% CL, or greater than 95\% CL when indicated with $*$, for a range of different combinations of decay branching fractions.
The ranges containing 68 and 95\%, respectively, of the distribution of limits expected under the background-only hypothesis, are also given. 
The cases where the limits could not be evaluated because simulated signal samples for VLQ masses below 400\GeV are not available, are indicated with `n.a.'.
The limits are determined assuming $\kappa_\QD= 0.15$.}
\label{tab:reparam_BFtable_D_kappaD0p15_combination}
\begin{scotch}{cccrrrr}
 $\mathcal{B}_\PW$  & $\mathcal{B}_\cPZ$ & $\mathcal{B}_\PH$  & Observed & Median expected &  68\% expected &  95\% expected  \\
\hline
0.0 & 1.0 & 0.0 & 1355 & 1420 & [1300,1510] & [1165,1605] \\
0.0 & 0.8 & 0.2 & 1190 & 1275 & [1125,1400] & [775,1490] \\
0.0 & 0.6 & 0.4 & 950 & 1070 & [685,1190] & [550,1325] \\
0.0 & 0.4 & 0.6 & 575 & 610 & [550,720] & [500,990] \\
0.0 & 0.2 & 0.8 & 505 & 535 & [485,580] & [430,620] \\
0.0 & 0.0 & 1.0 & 430 & n.a. & [n.a.,505] & [n.a.,535] \\
0.2 & 0.8 & 0.0 & 1425 & 1425 & [1310,1530] & [1150,1630] \\
0.2 & 0.6 & 0.2 & 1325 & 1250 & [1115,1400] & [720,1495] \\
0.2 & 0.4 & 0.4 & 690 & 955 & [625,1175] & [540,1300] \\
0.2 & 0.2 & 0.6 & 610 & 600 & [540,710] & [470,820] \\
0.2 & 0.0 & 0.8 & 500 & 550 & [485,610] & [400,685] \\
0.4 & 0.6 & 0.0 & 1575 & 1465 & [1320,1635] & [1150,1765] \\
0.4 & 0.4 & 0.2 & 1495 & 1310 & [1120,1495] & [730,1655] \\
0.4 & 0.2 & 0.4 & 1400 & 895 & [685,1275] & [560,1505] \\
0.4 & 0.0 & 0.6 & 705 & 710 & [585,825] & [495,1255] \\
0.6 & 0.4 & 0.0 & 1770 & 1630 & [1385,1790] & [1200,$1800^{*}$] \\
0.6 & 0.2 & 0.2 & 1735 & 1510 & [1250,1715] & [810,$1800^{*}$] \\
0.6 & 0.0 & 0.4 & 1675 & 1320 & [805,1635] & [675,1775] \\
0.8 & 0.2 & 0.0 & $1800^{*}$ & 1785 & [1615,$1800^{*}$] & [1335,$1800^{*}$] \\
0.8 & 0.0 & 0.2 & $1800^{*}$ & 1725 & [1505,$1800^{*}$] & [1205,$1800^{*}$] \\
1.0 & 0.0 & 0.0 & $1800^{*}$ & $1800^{*}$ & [1750,$1800^{*}$] & [1560,$1800^{*}$] \\
\end{scotch}
\end{table*}

\begin{table*}
\centering
\topcaption{Observed and median expected lower limits on the VLQ mass (in \GeV)
at 95\% CL, or greater than 95\% CL when indicated with $*$, for a range of different combinations of decay branching fractions.
The ranges containing 68 and 95\%, respectively, of the distribution of limits expected under the background-only hypothesis, are also given. 
The cases where the limits could not be evaluated because simulated signal samples for VLQ masses below 400\GeV are not available, are indicated with `n.a.'.
The limits are determined assuming $\kappa_\QD= 0.2$.}
\label{tab:reparam_BFtable_D_kappaD0p2_combination}
\begin{scotch}{cccrrrr}
 $\mathcal{B}_\PW$  & $\mathcal{B}_\cPZ$ & $\mathcal{B}_\PH$  & Observed & Median expected &  68\% expected &  95\% expected   \\
\hline
0.0 & 1.0 & 0.0 & 1500 & 1565 & [1470,1710] & [1380,1785] \\
0.0 & 0.8 & 0.2 & 1380 & 1455 & [1350,1555] & [1200,1660] \\
0.0 & 0.6 & 0.4 & 1210 & 1280 & [1140,1410] & [780,1485] \\
0.0 & 0.4 & 0.6 & 655 & 900 & [565,1130] & [495,1225] \\
0.0 & 0.2 & 0.8 & 505 & 540 & [485,585] & [420,645] \\
0.0 & 0.0 & 1.0 & 430 & n.a. & [n.a.,505] & [n.a.,535] \\
0.2 & 0.8 & 0.0 & 1605 & 1590 & [1480,1715] & [1370,$1800^{*}$] \\
0.2 & 0.6 & 0.2 & 1495 & 1460 & [1340,1590] & [1180,1710] \\
0.2 & 0.4 & 0.4 & 1350 & 1265 & [1120,1410] & [595,1530] \\
0.2 & 0.2 & 0.6 & 665 & 695 & [555,990] & [480,1210] \\
0.2 & 0.0 & 0.8 & 605 & 555 & [480,625] & [400,710] \\
0.4 & 0.6 & 0.0 & $1800^{*}$ & 1725 & [1555,$1800^{*}$] & [1405,$1800^{*}$] \\
0.4 & 0.4 & 0.2 & 1745 & 1585 & [1400,1780] & [1230,$1800^{*}$] \\
0.4 & 0.2 & 0.4 & 1635 & 1395 & [1155,1640] & [740,1785] \\
0.4 & 0.0 & 0.6 & 1540 & 1035 & [670,1385] & [525,1700] \\
0.6 & 0.4 & 0.0 & $1800^{*}$ & $1800^{*}$ & [1720,$1800^{*}$] & [1540,$1800^{*}$] \\
0.6 & 0.2 & 0.2 & $1800^{*}$ & $1800^{*}$ & [1615,$1800^{*}$] & [1355,$1800^{*}$] \\
0.6 & 0.0 & 0.4 & $1800^{*}$ & 1725 & [1425,$1800^{*}$] & [1170,$1800^{*}$] \\
0.8 & 0.2 & 0.0 & $1800^{*}$ & $1800^{*}$ & [$1800^{*}$,$1800^{*}$] & [1720,$1800^{*}$] \\
0.8 & 0.0 & 0.2 & $1800^{*}$ & $1800^{*}$ & [1795,$1800^{*}$] & [1620,$1800^{*}$] \\
1.0 & 0.0 & 0.0 & $1800^{*}$ & $1800^{*}$ & [$1800^{*}$,$1800^{*}$] & [$1800^{*}$,$1800^{*}$] \\
\end{scotch}
\end{table*}

\begin{table*}
\centering
\topcaption{Observed and median expected lower limits on the VLQ mass (in \GeV)
at 95\% CL, or greater than 95\% CL when indicated with $*$, for a range of different combinations of decay branching fractions.
The ranges containing 68 and 95\%, respectively, of the distribution of limits expected under the background-only hypothesis, are also given. 
The cases where the limits could not be evaluated because simulated signal samples for VLQ masses below 400\GeV are not available, are indicated with `n.a.'.
The limits are determined assuming $\kappa_\QD= 0.3$.}
\label{tab:reparam_BFtable_D_kappaD0p3_combination}
\begin{scotch}{cccrrrr}
 $\mathcal{B}_\PW$  & $\mathcal{B}_\cPZ$ & $\mathcal{B}_\PH$  & Observed & Median expected &  68\% expected &  95\% expected  \\
\hline
0.0 & 1.0 & 0.0 & 1760 & $1800^{*}$ & [1720,$1800^{*}$] & [1585,$1800^{*}$] \\
0.0 & 0.8 & 0.2 & 1600 & 1700 & [1565,1790] & [1465,$1800^{*}$] \\
0.0 & 0.6 & 0.4 & 1455 & 1515 & [1420,1615] & [1300,1730] \\
0.0 & 0.4 & 0.6 & 1205 & 1275 & [1145,1405] & [550,1490] \\
0.0 & 0.2 & 0.8 & 490 & 555 & [495,645] & [430,955] \\
0.0 & 0.0 & 1.0 & 430 & n.a. & [n.a.,505] & [n.a.,535] \\
0.2 & 0.8 & 0.0 & $1800^{*}$ & $1800^{*}$ & [1750,$1800^{*}$] & [1615,$1800^{*}$] \\
0.2 & 0.6 & 0.2 & 1785 & 1755 & [1615,$1800^{*}$] & [1485,$1800^{*}$] \\
0.2 & 0.4 & 0.4 & 1655 & 1590 & [1440,1730] & [1320,$1800^{*}$] \\
0.2 & 0.2 & 0.6 & 1505 & 1300 & [1065,1500] & [520,1665] \\
0.2 & 0.0 & 0.8 & 665 & 570 & [495,735] & [n.a.,1255] \\
0.4 & 0.6 & 0.0 & $1800^{*}$ & $1800^{*}$ & [$1800^{*}$,$1800^{*}$] & [1750,$1800^{*}$] \\
0.4 & 0.4 & 0.2 & $1800^{*}$ & $1800^{*}$ & [1795,$1800^{*}$] & [1650,$1800^{*}$] \\
0.4 & 0.2 & 0.4 & $1800^{*}$ & $1800^{*}$ & [1675,$1800^{*}$] & [1495,$1800^{*}$] \\
0.4 & 0.0 & 0.6 & $1800^{*}$ & 1720 & [1470,$1800^{*}$] & [1215,$1800^{*}$] \\
0.6 & 0.4 & 0.0 & $1800^{*}$ & $1800^{*}$ & [$1800^{*}$,$1800^{*}$] & [$1800^{*}$,$1800^{*}$] \\
0.6 & 0.2 & 0.2 & $1800^{*}$ & $1800^{*}$ & [$1800^{*}$,$1800^{*}$] & [$1800^{*}$,$1800^{*}$] \\
0.6 & 0.0 & 0.4 & $1800^{*}$ & $1800^{*}$ & [$1800^{*}$,$1800^{*}$] & [1745,$1800^{*}$] \\
0.8 & 0.2 & 0.0 & $1800^{*}$ & $1800^{*}$ & [$1800^{*}$,$1800^{*}$] & [$1800^{*}$,$1800^{*}$] \\
0.8 & 0.0 & 0.2 & $1800^{*}$ & $1800^{*}$ & [$1800^{*}$,$1800^{*}$] & [$1800^{*}$,$1800^{*}$] \\
1.0 & 0.0 & 0.0 & $1800^{*}$ & $1800^{*}$ & [$1800^{*}$,$1800^{*}$] & [$1800^{*}$,$1800^{*}$] \\
\end{scotch}
\end{table*}

\begin{table*}
\centering
\topcaption{Observed and median expected lower limits on the VLQ mass (in \GeV)
at 95\% CL, or greater than 95\% CL when indicated with $*$, for a range of different combinations of decay branching fractions.
The ranges containing 68 and 95\%, respectively, of the distribution of limits expected under the background-only hypothesis, are also given. 
The cases where the limits could not be evaluated because simulated signal samples for VLQ masses below 400\GeV are not available, are indicated with `n.a.'.
The limits are determined assuming $\kappa_\QD= 0.4$.}
\label{tab:reparam_BFtable_D_kappaD0p4_combination}
\begin{scotch}{cccrrrr}
 $\mathcal{B}_\PW$  & $\mathcal{B}_\cPZ$ & $\mathcal{B}_\PH$  & Observed & Median expected &  68\% expected &  95\% expected  \\
\hline
0.0 & 1.0 & 0.0 & $1800^{*}$ & $1800^{*}$ & [$1800^{*}$,$1800^{*}$] & [1775,$1800^{*}$] \\
0.0 & 0.8 & 0.2 & 1770 & $1800^{*}$ & [1750,$1800^{*}$] & [1620,$1800^{*}$] \\
0.0 & 0.6 & 0.4 & 1590 & 1695 & [1560,1790] & [1470,$1800^{*}$] \\
0.0 & 0.4 & 0.6 & 1405 & 1450 & [1345,1545] & [1205,1645] \\
0.0 & 0.2 & 0.8 & 650 & 710 & [505,1120] & [430,1225] \\
0.0 & 0.0 & 1.0 & 430 & n.a. & [n.a.,505] & [n.a.,535] \\
0.2 & 0.8 & 0.0 & $1800^{*}$ & $1800^{*}$ & [$1800^{*}$,$1800^{*}$] & [$1800^{*}$,$1800^{*}$] \\
0.2 & 0.6 & 0.2 & $1800^{*}$ & $1800^{*}$ & [$1800^{*}$,$1800^{*}$] & [1715,$1800^{*}$] \\
0.2 & 0.4 & 0.4 & $1800^{*}$ & 1800 & [1660,$1800^{*}$] & [1530,$1800^{*}$] \\
0.2 & 0.2 & 0.6 & 1725 & 1610 & [1400,1775] & [1275,$1800^{*}$] \\
0.2 & 0.0 & 0.8 & 1520 & 1030 & [530,1385] & [n.a.,1690] \\
0.4 & 0.6 & 0.0 & $1800^{*}$ & $1800^{*}$ & [$1800^{*}$,$1800^{*}$] & [$1800^{*}$,$1800^{*}$] \\
0.4 & 0.4 & 0.2 & $1800^{*}$ & $1800^{*}$ & [$1800^{*}$,$1800^{*}$] & [$1800^{*}$,$1800^{*}$] \\
0.4 & 0.2 & 0.4 & $1800^{*}$ & $1800^{*}$ & [$1800^{*}$,$1800^{*}$] & [1795,$1800^{*}$] \\
0.4 & 0.0 & 0.6 & $1800^{*}$ & $1800^{*}$ & [$1800^{*}$,$1800^{*}$] & [1645,$1800^{*}$] \\
0.6 & 0.4 & 0.0 & $1800^{*}$ & $1800^{*}$ & [$1800^{*}$,$1800^{*}$] & [$1800^{*}$,$1800^{*}$] \\
0.6 & 0.2 & 0.2 & $1800^{*}$ & $1800^{*}$ & [$1800^{*}$,$1800^{*}$] & [$1800^{*}$,$1800^{*}$] \\
0.6 & 0.0 & 0.4 & $1800^{*}$ & $1800^{*}$ & [$1800^{*}$,$1800^{*}$] & [$1800^{*}$,$1800^{*}$] \\
0.8 & 0.2 & 0.0 & $1800^{*}$ & $1800^{*}$ & [$1800^{*}$,$1800^{*}$] & [$1800^{*}$,$1800^{*}$] \\
0.8 & 0.0 & 0.2 & $1800^{*}$ & $1800^{*}$ & [$1800^{*}$,$1800^{*}$] & [$1800^{*}$,$1800^{*}$] \\
1.0 & 0.0 & 0.0 & $1800^{*}$ & $1800^{*}$ & [$1800^{*}$,$1800^{*}$] & [$1800^{*}$,$1800^{*}$] \\
\end{scotch}
\end{table*}

\begin{table*}
\centering
\topcaption{Observed and median expected lower limits on the VLQ mass (in \GeV)
at 95\% CL, or greater than 95\% CL when indicated with $*$, for a range of different combinations of decay branching fractions.
The ranges containing 68 and 95\%, respectively, of the distribution of limits expected under the background-only hypothesis, are also given. 
The cases where the limits could not be evaluated because simulated signal samples for VLQ masses below 400\GeV are not available, are indicated with `n.a.'.
The limits are determined assuming $\kappa_\QD= 0.5$.}
\label{tab:reparam_BFtable_D_kappaD0p5_combination}

\begin{scotch}{cccrrrr}
 $\mathcal{B}_\PW$  & $\mathcal{B}_\cPZ$ & $\mathcal{B}_\PH$  & Observed & Median expected &  68\% expected &  95\% expected  \\
\hline
0.0 & 1.0 & 0.0 & $1800^{*}$ & $1800^{*}$ & [$1800^{*}$,$1800^{*}$] & [$1800^{*}$,$1800^{*}$] \\
0.0 & 0.8 & 0.2 & $1800^{*}$ & $1800^{*}$ & [$1800^{*}$,$1800^{*}$] & [1770,$1800^{*}$] \\
0.0 & 0.6 & 0.4 & 1735 & $1800^{*}$ & [1720,$1800^{*}$] & [1585,$1800^{*}$] \\
0.0 & 0.4 & 0.6 & 1485 & 1570 & [1470,1705] & [1370,1780] \\
0.0 & 0.2 & 0.8 & 1135 & 1150 & [545,1295] & [435,1405] \\
0.0 & 0.0 & 1.0 & 430 & n.a. & [n.a.,505] & [n.a.,535] \\
0.2 & 0.8 & 0.0 & $1800^{*}$ & $1800^{*}$ & [$1800^{*}$,$1800^{*}$] & [$1800^{*}$,$1800^{*}$] \\
0.2 & 0.6 & 0.2 & $1800^{*}$ & $1800^{*}$ & [$1800^{*}$,$1800^{*}$] & [$1800^{*}$,$1800^{*}$] \\
0.2 & 0.4 & 0.4 & $1800^{*}$ & $1800^{*}$ & [$1800^{*}$,$1800^{*}$] & [1710,$1800^{*}$] \\
0.2 & 0.2 & 0.6 & $1800^{*}$ & $1800^{*}$ & [1640,$1800^{*}$] & [1460,$1800^{*}$] \\
0.2 & 0.0 & 0.8 & 1755 & 1530 & [895,1730] & [n.a.,$1800^{*}$] \\
0.4 & 0.6 & 0.0 & $1800^{*}$ & $1800^{*}$ & [$1800^{*}$,$1800^{*}$] & [$1800^{*}$,$1800^{*}$] \\
0.4 & 0.4 & 0.2 & $1800^{*}$ & $1800^{*}$ & [$1800^{*}$,$1800^{*}$] & [$1800^{*}$,$1800^{*}$] \\
0.4 & 0.2 & 0.4 & $1800^{*}$ & $1800^{*}$ & [$1800^{*}$,$1800^{*}$] & [$1800^{*}$,$1800^{*}$] \\
0.4 & 0.0 & 0.6 & $1800^{*}$ & $1800^{*}$ & [$1800^{*}$,$1800^{*}$] & [$1800^{*}$,$1800^{*}$] \\
0.6 & 0.4 & 0.0 & $1800^{*}$ & $1800^{*}$ & [$1800^{*}$,$1800^{*}$] & [$1800^{*}$,$1800^{*}$] \\
0.6 & 0.2 & 0.2 & $1800^{*}$ & $1800^{*}$ & [$1800^{*}$,$1800^{*}$] & [$1800^{*}$,$1800^{*}$] \\
0.6 & 0.0 & 0.4 & $1800^{*}$ & $1800^{*}$ & [$1800^{*}$,$1800^{*}$] & [$1800^{*}$,$1800^{*}$] \\
0.8 & 0.2 & 0.0 & $1800^{*}$ & $1800^{*}$ & [$1800^{*}$,$1800^{*}$] & [$1800^{*}$,$1800^{*}$] \\
0.8 & 0.0 & 0.2 & $1800^{*}$ & $1800^{*}$ & [$1800^{*}$,$1800^{*}$] & [$1800^{*}$,$1800^{*}$] \\
1.0 & 0.0 & 0.0 & $1800^{*}$ & $1800^{*}$ & [$1800^{*}$,$1800^{*}$] & [$1800^{*}$,$1800^{*}$] \\
\end{scotch}
\end{table*}

\begin{table*}
\centering
\topcaption{Observed and median expected lower limits on the VLQ mass (in \GeV)
at 95\% CL, or greater than 95\% CL when indicated with $*$, for a range of different combinations of decay branching fractions.
The ranges containing 68 and 95\%, respectively, of the distribution of limits expected under the background-only hypothesis, are also given. 
The cases where the limits could not be evaluated because simulated signal samples for VLQ masses below 400\GeV are not available, are indicated with `n.a.'.
The limits are determined assuming $\kappa_\QD= 0.6$.}
\label{tab:reparam_BFtable_D_kappaD0p6_combination}

\begin{scotch}{cccrrrr}
 $\mathcal{B}_\PW$  & $\mathcal{B}_\cPZ$ & $\mathcal{B}_\PH$  & Observed & Median expected &  68\% expected &  95\% expected  \\
\hline
0.0 & 1.0 & 0.0 & $1800^{*}$ & $1800^{*}$ & [$1800^{*}$,$1800^{*}$] & [$1800^{*}$,$1800^{*}$] \\
0.0 & 0.8 & 0.2 & $1800^{*}$ & $1800^{*}$ & [$1800^{*}$,$1800^{*}$] & [$1800^{*}$,$1800^{*}$] \\
0.0 & 0.6 & 0.4 & $1800^{*}$ & $1800^{*}$ & [$1800^{*}$,$1800^{*}$] & [1720,$1800^{*}$] \\
0.0 & 0.4 & 0.6 & 1600 & 1700 & [1560,1790] & [1475,$1800^{*}$] \\
0.0 & 0.2 & 0.8 & 1205 & 1280 & [1145,1405] & [450,1485] \\
0.0 & 0.0 & 1.0 & 430 & n.a. & [n.a.,505] & [n.a.,535] \\
0.2 & 0.8 & 0.0 & $1800^{*}$ & $1800^{*}$ & [$1800^{*}$,$1800^{*}$] & [$1800^{*}$,$1800^{*}$] \\
0.2 & 0.6 & 0.2 & $1800^{*}$ & $1800^{*}$ & [$1800^{*}$,$1800^{*}$] & [$1800^{*}$,$1800^{*}$] \\
0.2 & 0.4 & 0.4 & $1800^{*}$ & $1800^{*}$ & [$1800^{*}$,$1800^{*}$] & [$1800^{*}$,$1800^{*}$] \\
0.2 & 0.2 & 0.6 & $1800^{*}$ & $1800^{*}$ & [1800,$1800^{*}$] & [1650,$1800^{*}$] \\
0.2 & 0.0 & 0.8 & $1800^{*}$ & 1720 & [1510,$1800^{*}$] & [890,$1800^{*}$] \\
0.4 & 0.6 & 0.0 & $1800^{*}$ & $1800^{*}$ & [$1800^{*}$,$1800^{*}$] & [$1800^{*}$,$1800^{*}$] \\
0.4 & 0.4 & 0.2 & $1800^{*}$ & $1800^{*}$ & [$1800^{*}$,$1800^{*}$] & [$1800^{*}$,$1800^{*}$] \\
0.4 & 0.2 & 0.4 & $1800^{*}$ & $1800^{*}$ & [$1800^{*}$,$1800^{*}$] & [$1800^{*}$,$1800^{*}$] \\
0.4 & 0.0 & 0.6 & $1800^{*}$ & $1800^{*}$ & [$1800^{*}$,$1800^{*}$] & [$1800^{*}$,$1800^{*}$] \\
0.6 & 0.4 & 0.0 & $1800^{*}$ & $1800^{*}$ & [$1800^{*}$,$1800^{*}$] & [$1800^{*}$,$1800^{*}$] \\
0.6 & 0.2 & 0.2 & $1800^{*}$ & $1800^{*}$ & [$1800^{*}$,$1800^{*}$] & [$1800^{*}$,$1800^{*}$] \\
0.6 & 0.0 & 0.4 & $1800^{*}$ & $1800^{*}$ & [$1800^{*}$,$1800^{*}$] & [$1800^{*}$,$1800^{*}$] \\
0.8 & 0.2 & 0.0 & $1800^{*}$ & $1800^{*}$ & [$1800^{*}$,$1800^{*}$] & [$1800^{*}$,$1800^{*}$] \\
0.8 & 0.0 & 0.2 & $1800^{*}$ & $1800^{*}$ & [$1800^{*}$,$1800^{*}$] & [$1800^{*}$,$1800^{*}$] \\
1.0 & 0.0 & 0.0 & $1800^{*}$ & $1800^{*}$ & [$1800^{*}$,$1800^{*}$] & [$1800^{*}$,$1800^{*}$] \\
\end{scotch}
\end{table*}

\begin{table*}
\centering
\topcaption{Observed and median expected lower limits on the VLQ mass (in \GeV)
at 95\% CL, or greater than 95\% CL when indicated with $*$, for a range of different combinations of decay branching fractions.
The ranges containing 68 and 95\%, respectively, of the distribution of limits expected under the background-only hypothesis, are also given. 
The cases where the limits could not be evaluated because simulated signal samples for VLQ masses below 400\GeV are not available, are indicated with `n.a.'.
The limits are determined assuming $\kappa_\QD= 0.7$.}
\label{tab:reparam_BFtable_D_kappaD0p7_combination}
\begin{scotch}{cccrrrr}
 $\mathcal{B}_\PW$  & $\mathcal{B}_\cPZ$ & $\mathcal{B}_\PH$  & Observed & Median expected &  68\% expected &  95\% expected \\
\hline
0.0 & 1.0 & 0.0 & $1800^{*}$ & $1800^{*}$ & [$1800^{*}$,$1800^{*}$] & [$1800^{*}$,$1800^{*}$] \\
0.0 & 0.8 & 0.2 & $1800^{*}$ & $1800^{*}$ & [$1800^{*}$,$1800^{*}$] & [$1800^{*}$,$1800^{*}$] \\
0.0 & 0.6 & 0.4 & $1800^{*}$ & $1800^{*}$ & [$1800^{*}$,$1800^{*}$] & [1785,$1800^{*}$] \\
0.0 & 0.4 & 0.6 & 1720 & 1785 & [1680,$1800^{*}$] & [1550,$1800^{*}$] \\
0.0 & 0.2 & 0.8 & 1290 & 1390 & [1240,1480] & [495,1560] \\
0.0 & 0.0 & 1.0 & 430 & n.a. & [n.a.,505] & [n.a.,535] \\
0.2 & 0.8 & 0.0 & $1800^{*}$ & $1800^{*}$ & [$1800^{*}$,$1800^{*}$] & [$1800^{*}$,$1800^{*}$] \\
0.2 & 0.6 & 0.2 & $1800^{*}$ & $1800^{*}$ & [$1800^{*}$,$1800^{*}$] & [$1800^{*}$,$1800^{*}$] \\
0.2 & 0.4 & 0.4 & $1800^{*}$ & $1800^{*}$ & [$1800^{*}$,$1800^{*}$] & [$1800^{*}$,$1800^{*}$] \\
0.2 & 0.2 & 0.6 & $1800^{*}$ & $1800^{*}$ & [$1800^{*}$,$1800^{*}$] & [1770,$1800^{*}$] \\
0.2 & 0.0 & 0.8 & $1800^{*}$ & $1800^{*}$ & [1695,$1800^{*}$] & [1415,$1800^{*}$] \\
0.4 & 0.6 & 0.0 & $1800^{*}$ & $1800^{*}$ & [$1800^{*}$,$1800^{*}$] & [$1800^{*}$,$1800^{*}$] \\
0.4 & 0.4 & 0.2 & $1800^{*}$ & $1800^{*}$ & [$1800^{*}$,$1800^{*}$] & [$1800^{*}$,$1800^{*}$] \\
0.4 & 0.2 & 0.4 & $1800^{*}$ & $1800^{*}$ & [$1800^{*}$,$1800^{*}$] & [$1800^{*}$,$1800^{*}$] \\
0.4 & 0.0 & 0.6 & $1800^{*}$ & $1800^{*}$ & [$1800^{*}$,$1800^{*}$] & [$1800^{*}$,$1800^{*}$] \\
0.6 & 0.4 & 0.0 & $1800^{*}$ & $1800^{*}$ & [$1800^{*}$,$1800^{*}$] & [$1800^{*}$,$1800^{*}$] \\
0.6 & 0.2 & 0.2 & $1800^{*}$ & $1800^{*}$ & [$1800^{*}$,$1800^{*}$] & [$1800^{*}$,$1800^{*}$] \\
0.6 & 0.0 & 0.4 & $1800^{*}$ & $1800^{*}$ & [$1800^{*}$,$1800^{*}$] & [$1800^{*}$,$1800^{*}$] \\
0.8 & 0.2 & 0.0 & $1800^{*}$ & $1800^{*}$ & [$1800^{*}$,$1800^{*}$] & [$1800^{*}$,$1800^{*}$] \\
0.8 & 0.0 & 0.2 & $1800^{*}$ & $1800^{*}$ & [$1800^{*}$,$1800^{*}$] & [$1800^{*}$,$1800^{*}$] \\
1.0 & 0.0 & 0.0 & $1800^{*}$ & $1800^{*}$ & [$1800^{*}$,$1800^{*}$] & [$1800^{*}$,$1800^{*}$] \\
\end{scotch}
\end{table*}

\begin{table*}
\centering
\topcaption{Observed and median expected lower limits on the VLQ mass (in \GeV)
at 95\% CL, or greater than 95\% CL when indicated with $*$, for a range of different combinations of decay branching fractions.
The ranges containing 68 and 95\%, respectively, of the distribution of limits expected under the background-only hypothesis, are also given. 
The cases where the limits could not be evaluated because simulated signal samples for VLQ masses below 400\GeV are not available, are indicated with `n.a.'.
The limits are determined assuming $\kappa_\QD= 0.8$.}
\label{tab:reparam_BFtable_D_kappaD0p8_combination}

\begin{scotch}{cccrrrr}
 $\mathcal{B}_\PW$  & $\mathcal{B}_\cPZ$ & $\mathcal{B}_\PH$  & Observed & Median expected &  68\% expected &  95\% expected  \\
\hline
0.0 & 1.0 & 0.0 & $1800^{*}$ & $1800^{*}$ & [$1800^{*}$,$1800^{*}$] & [$1800^{*}$,$1800^{*}$] \\
0.0 & 0.8 & 0.2 & $1800^{*}$ & $1800^{*}$ & [$1800^{*}$,$1800^{*}$] & [$1800^{*}$,$1800^{*}$] \\
0.0 & 0.6 & 0.4 & $1800^{*}$ & $1800^{*}$ & [$1800^{*}$,$1800^{*}$] & [$1800^{*}$,$1800^{*}$] \\
0.0 & 0.4 & 0.6 & 1775 & $1800^{*}$ & [1750,$1800^{*}$] & [1635,$1800^{*}$] \\
0.0 & 0.2 & 0.8 & 1390 & 1450 & [1350,1545] & [1195,1670] \\
0.0 & 0.0 & 1.0 & 430 & n.a. & [n.a.,505] & [n.a.,535] \\
0.2 & 0.8 & 0.0 & $1800^{*}$ & $1800^{*}$ & [$1800^{*}$,$1800^{*}$] & [$1800^{*}$,$1800^{*}$] \\
0.2 & 0.6 & 0.2 & $1800^{*}$ & $1800^{*}$ & [$1800^{*}$,$1800^{*}$] & [$1800^{*}$,$1800^{*}$] \\
0.2 & 0.4 & 0.4 & $1800^{*}$ & $1800^{*}$ & [$1800^{*}$,$1800^{*}$] & [$1800^{*}$,$1800^{*}$] \\
0.2 & 0.2 & 0.6 & $1800^{*}$ & $1800^{*}$ & [$1800^{*}$,$1800^{*}$] & [$1800^{*}$,$1800^{*}$] \\
0.2 & 0.0 & 0.8 & $1800^{*}$ & $1800^{*}$ & [1795,$1800^{*}$] & [1640,$1800^{*}$] \\
0.4 & 0.6 & 0.0 & $1800^{*}$ & $1800^{*}$ & [$1800^{*}$,$1800^{*}$] & [$1800^{*}$,$1800^{*}$] \\
0.4 & 0.4 & 0.2 & $1800^{*}$ & $1800^{*}$ & [$1800^{*}$,$1800^{*}$] & [$1800^{*}$,$1800^{*}$] \\
0.4 & 0.2 & 0.4 & $1800^{*}$ & $1800^{*}$ & [$1800^{*}$,$1800^{*}$] & [$1800^{*}$,$1800^{*}$] \\
0.4 & 0.0 & 0.6 & $1800^{*}$ & $1800^{*}$ & [$1800^{*}$,$1800^{*}$] & [$1800^{*}$,$1800^{*}$] \\
0.6 & 0.4 & 0.0 & $1800^{*}$ & $1800^{*}$ & [$1800^{*}$,$1800^{*}$] & [$1800^{*}$,$1800^{*}$] \\
0.6 & 0.2 & 0.2 & $1800^{*}$ & $1800^{*}$ & [$1800^{*}$,$1800^{*}$] & [$1800^{*}$,$1800^{*}$] \\
0.6 & 0.0 & 0.4 & $1800^{*}$ & $1800^{*}$ & [$1800^{*}$,$1800^{*}$] & [$1800^{*}$,$1800^{*}$] \\
0.8 & 0.2 & 0.0 & $1800^{*}$ & $1800^{*}$ & [$1800^{*}$,$1800^{*}$] & [$1800^{*}$,$1800^{*}$] \\
0.8 & 0.0 & 0.2 & $1800^{*}$ & $1800^{*}$ & [$1800^{*}$,$1800^{*}$] & [$1800^{*}$,$1800^{*}$] \\
1.0 & 0.0 & 0.0 & $1800^{*}$ & $1800^{*}$ & [$1800^{*}$,$1800^{*}$] & [$1800^{*}$,$1800^{*}$] \\
\end{scotch}
\end{table*}

\begin{table*}
\centering
\topcaption{Observed and median expected lower limits on the VLQ mass (in \GeV)
at 95\% CL, or greater than 95\% CL when indicated with $*$, for a range of different combinations of decay branching fractions.
The ranges containing 68 and 95\%, respectively, of the distribution of limits expected under the background-only hypothesis, are also given. 
The cases where the limits could not be evaluated because simulated signal samples for VLQ masses below 400\GeV are not available, are indicated with `n.a.'.
The limits are determined assuming $\kappa_\QD= 0.9$.}
\label{tab:reparam_BFtable_D_kappaD0p9_combination}
\begin{scotch}{cccrrrr}
 $\mathcal{B}_\PW$  & $\mathcal{B}_\cPZ$ & $\mathcal{B}_\PH$  & Observed & Median expected &  68\% expected &  95\% expected  \\
\hline
0.0 & 1.0 & 0.0 & $1800^{*}$ & $1800^{*}$ & [$1800^{*}$,$1800^{*}$] & [$1800^{*}$,$1800^{*}$] \\
0.0 & 0.8 & 0.2 & $1800^{*}$ & $1800^{*}$ & [$1800^{*}$,$1800^{*}$] & [$1800^{*}$,$1800^{*}$] \\
0.0 & 0.6 & 0.4 & $1800^{*}$ & $1800^{*}$ & [$1800^{*}$,$1800^{*}$] & [$1800^{*}$,$1800^{*}$] \\
0.0 & 0.4 & 0.6 & $1800^{*}$ & $1800^{*}$ & [$1800^{*}$,$1800^{*}$] & [1720,$1800^{*}$] \\
0.0 & 0.2 & 0.8 & 1450 & 1510 & [1415,1620] & [1315,1730] \\
0.0 & 0.0 & 1.0 & 430 & n.a. & [n.a.,505] & [n.a.,535] \\
0.2 & 0.8 & 0.0 & $1800^{*}$ & $1800^{*}$ & [$1800^{*}$,$1800^{*}$] & [$1800^{*}$,$1800^{*}$] \\
0.2 & 0.6 & 0.2 & $1800^{*}$ & $1800^{*}$ & [$1800^{*}$,$1800^{*}$] & [$1800^{*}$,$1800^{*}$] \\
0.2 & 0.4 & 0.4 & $1800^{*}$ & $1800^{*}$ & [$1800^{*}$,$1800^{*}$] & [$1800^{*}$,$1800^{*}$] \\
0.2 & 0.2 & 0.6 & $1800^{*}$ & $1800^{*}$ & [$1800^{*}$,$1800^{*}$] & [$1800^{*}$,$1800^{*}$] \\
0.2 & 0.0 & 0.8 & $1800^{*}$ & $1800^{*}$ & [$1800^{*}$,$1800^{*}$] & [1750,$1800^{*}$] \\
0.4 & 0.6 & 0.0 & $1800^{*}$ & $1800^{*}$ & [$1800^{*}$,$1800^{*}$] & [$1800^{*}$,$1800^{*}$] \\
0.4 & 0.4 & 0.2 & $1800^{*}$ & $1800^{*}$ & [$1800^{*}$,$1800^{*}$] & [$1800^{*}$,$1800^{*}$] \\
0.4 & 0.2 & 0.4 & $1800^{*}$ & $1800^{*}$ & [$1800^{*}$,$1800^{*}$] & [$1800^{*}$,$1800^{*}$] \\
0.4 & 0.0 & 0.6 & $1800^{*}$ & $1800^{*}$ & [$1800^{*}$,$1800^{*}$] & [$1800^{*}$,$1800^{*}$] \\
0.6 & 0.4 & 0.0 & $1800^{*}$ & $1800^{*}$ & [$1800^{*}$,$1800^{*}$] & [$1800^{*}$,$1800^{*}$] \\
0.6 & 0.2 & 0.2 & $1800^{*}$ & $1800^{*}$ & [$1800^{*}$,$1800^{*}$] & [$1800^{*}$,$1800^{*}$] \\
0.6 & 0.0 & 0.4 & $1800^{*}$ & $1800^{*}$ & [$1800^{*}$,$1800^{*}$] & [$1800^{*}$,$1800^{*}$] \\
0.8 & 0.2 & 0.0 & $1800^{*}$ & $1800^{*}$ & [$1800^{*}$,$1800^{*}$] & [$1800^{*}$,$1800^{*}$] \\
0.8 & 0.0 & 0.2 & $1800^{*}$ & $1800^{*}$ & [$1800^{*}$,$1800^{*}$] & [$1800^{*}$,$1800^{*}$] \\
1.0 & 0.0 & 0.0 & $1800^{*}$ & $1800^{*}$ & [$1800^{*}$,$1800^{*}$] & [$1800^{*}$,$1800^{*}$] \\
\end{scotch}
\end{table*}

\begin{table*}
\centering
\topcaption{Observed and median expected lower limits on the VLQ mass (in \GeV)
at 95\% CL, or greater than 95\% CL when indicated with $*$, for a range of different combinations of decay branching fractions.
The ranges containing 68 and 95\%, respectively, of the distribution of limits expected under the background-only hypothesis, are also given. 
The cases where the limits could not be evaluated because simulated signal samples for VLQ masses below 400\GeV are not available, are indicated with `n.a.'.
The limits are determined assuming $\kappa_\QD= 1.0$.}
\label{tab:reparam_BFtable_D_kappaD1_combination}
\begin{scotch}{cccrrrr}
 $\mathcal{B}_\PW$  & $\mathcal{B}_\cPZ$ & $\mathcal{B}_\PH$  & Observed & Median expected &  68\% expected &  95\% expected  \\
\hline
0.0 & 1.0 & 0.0 & $1800^{*}$ & $1800^{*}$ & [$1800^{*}$,$1800^{*}$] & [$1800^{*}$,$1800^{*}$] \\
0.0 & 0.8 & 0.2 & $1800^{*}$ & $1800^{*}$ & [$1800^{*}$,$1800^{*}$] & [$1800^{*}$,$1800^{*}$] \\
0.0 & 0.6 & 0.4 & $1800^{*}$ & $1800^{*}$ & [$1800^{*}$,$1800^{*}$] & [$1800^{*}$,$1800^{*}$] \\
0.0 & 0.4 & 0.6 & $1800^{*}$ & $1800^{*}$ & [$1800^{*}$,$1800^{*}$] & [1760,$1800^{*}$] \\
0.0 & 0.2 & 0.8 & 1490 & 1565 & [1475,1705] & [1380,1780] \\
0.0 & 0.0 & 1.0 & 430 & n.a. & [n.a.,505] & [n.a.,535] \\
0.2 & 0.8 & 0.0 & $1800^{*}$ & $1800^{*}$ & [$1800^{*}$,$1800^{*}$] & [$1800^{*}$,$1800^{*}$] \\
0.2 & 0.6 & 0.2 & $1800^{*}$ & $1800^{*}$ & [$1800^{*}$,$1800^{*}$] & [$1800^{*}$,$1800^{*}$] \\
0.2 & 0.4 & 0.4 & $1800^{*}$ & $1800^{*}$ & [$1800^{*}$,$1800^{*}$] & [$1800^{*}$,$1800^{*}$] \\
0.2 & 0.2 & 0.6 & $1800^{*}$ & $1800^{*}$ & [$1800^{*}$,$1800^{*}$] & [$1800^{*}$,$1800^{*}$] \\
0.2 & 0.0 & 0.8 & $1800^{*}$ & $1800^{*}$ & [$1800^{*}$,$1800^{*}$] & [$1800^{*}$,$1800^{*}$] \\
0.4 & 0.6 & 0.0 & $1800^{*}$ & $1800^{*}$ & [$1800^{*}$,$1800^{*}$] & [$1800^{*}$,$1800^{*}$] \\
0.4 & 0.4 & 0.2 & $1800^{*}$ & $1800^{*}$ & [$1800^{*}$,$1800^{*}$] & [$1800^{*}$,$1800^{*}$] \\
0.4 & 0.2 & 0.4 & $1800^{*}$ & $1800^{*}$ & [$1800^{*}$,$1800^{*}$] & [$1800^{*}$,$1800^{*}$] \\
0.4 & 0.0 & 0.6 & $1800^{*}$ & $1800^{*}$ & [$1800^{*}$,$1800^{*}$] & [$1800^{*}$,$1800^{*}$] \\
0.6 & 0.4 & 0.0 & $1800^{*}$ & $1800^{*}$ & [$1800^{*}$,$1800^{*}$] & [$1800^{*}$,$1800^{*}$] \\
0.6 & 0.2 & 0.2 & $1800^{*}$ & $1800^{*}$ & [$1800^{*}$,$1800^{*}$] & [$1800^{*}$,$1800^{*}$] \\
0.6 & 0.0 & 0.4 & $1800^{*}$ & $1800^{*}$ & [$1800^{*}$,$1800^{*}$] & [$1800^{*}$,$1800^{*}$] \\
0.8 & 0.2 & 0.0 & $1800^{*}$ & $1800^{*}$ & [$1800^{*}$,$1800^{*}$] & [$1800^{*}$,$1800^{*}$] \\
0.8 & 0.0 & 0.2 & $1800^{*}$ & $1800^{*}$ & [$1800^{*}$,$1800^{*}$] & [$1800^{*}$,$1800^{*}$] \\
1.0 & 0.0 & 0.0 & $1800^{*}$ & $1800^{*}$ & [$1800^{*}$,$1800^{*}$] & [$1800^{*}$,$1800^{*}$] \\
\end{scotch}
\end{table*}

}
\cleardoublepage \section{The CMS Collaboration \label{app:collab}}\begin{sloppypar}\hyphenpenalty=5000\widowpenalty=500\clubpenalty=5000\textbf{Yerevan Physics Institute,  Yerevan,  Armenia}\\*[0pt]
A.M.~Sirunyan, A.~Tumasyan
\vskip\cmsinstskip
\textbf{Institut f\"{u}r Hochenergiephysik,  Wien,  Austria}\\*[0pt]
W.~Adam, E.~Asilar, T.~Bergauer, J.~Brandstetter, E.~Brondolin, M.~Dragicevic, J.~Er\"{o}, M.~Flechl, M.~Friedl, R.~Fr\"{u}hwirth\cmsAuthorMark{1}, V.M.~Ghete, C.~Hartl, N.~H\"{o}rmann, J.~Hrubec, M.~Jeitler\cmsAuthorMark{1}, A.~K\"{o}nig, I.~Kr\"{a}tschmer, D.~Liko, T.~Matsushita, I.~Mikulec, D.~Rabady, N.~Rad, B.~Rahbaran, H.~Rohringer, J.~Schieck\cmsAuthorMark{1}, J.~Strauss, W.~Waltenberger, C.-E.~Wulz\cmsAuthorMark{1}
% ...
\vskip\cmsinstskip
% ...
1:~~Also at Vienna University of Technology, Vienna, Austria\\
% ...

\end{sloppypar}
\end{document}